\documentclass[12pt]{article}
\usepackage{epsfig}
\usepackage{amssymb}
\usepackage{lscape}
\usepackage{cleveref}
\usepackage{subfig}
\usepackage{multirow}

\begin{document}

\title{Modeling the transmission of new coronavirus in S\~{a}o Paulo State,
Brazil -- Assessing epidemiological impacts of isolating young and elder
persons\\
\bigskip }
\author{Hyun Mo Yang{$^1$}\thanks{%
Corresponding author -- email: hyunyang@ime.unicamp.br; tel: + 55 19
3521-6031}, Luis Pedro Lombardi Junior{$^1$} and Ariana Campos Yang{$^2$}
\bigskip \bigskip \\
{$^1$}UNICAMP -- IMECC -- DMA; Pra\c{c}a S\'{e}rgio Buarque de Holanda, 651;%
\\
CEP: 13083-859, Campinas, SP, Brazil\\
\bigskip {$^2$}HC-FMUSP and HC-UNICAMP}
\date{ }
\maketitle

\begin{abstract}
We developed a mathematical model to describe the transmission of new
coronavirus in the S\~{a}o Paulo State, Brazil. The model divided a
community in subpopulations comprised by young and elder persons, in order
to take into account higher risk of fatality among elder persons with severe
CoViD-19. From data collected in the S\~{a}o Paulo State, we estimated the
transmission and additional mortality rates, from which we calculated the
basic reproduction number $R_0$. From estimated parameters, estimation of
the deaths due to CiViD-19 was three times lower than those found in
literature. Considering isolation as a control mechanism, we varied
isolation rates of young and elder persons in order to assess their
epidemiological impacts. The epidemiological scenarios focused mainly on
evaluating the number of severe CoViD-19 cases and deaths due to this
disease when isolation is introduced in a population.

\bigskip

\end{abstract}

\section{Introduction}

Coronavirus disease 2019 (CoViD-19) is caused by severe acute respiratory
syndrome coronavirus 2 (SARS-CoV-2), a strain of the SARS-CoV-1 (pandemic in
2002/2003), originated in Wuhan, China, in December 2019, and spread out
worldwide. World Health Organization (WHO) declared CoViD-19 pandemic on 11
March, based on its own definition: \textquotedblleft A pandemic is the
worldwide spread of a new disease. An influenza pandemic occurs when a new
influenza virus emerges and spreads around the world, and most people do not
have immunity\textquotedblright .

Coronavirus (RNA virus) can be transmitted by droplets that escape lungs
through coughing or sneezing and infects humans (direct transmission), or
they are deposited in surfaces and infects humans when in contact with this
contaminated surface (indirect transmission). This virus enters in
susceptible persons through nose, mouth or eyes, and infects cells in the
respiratory tract, being capable of releasing millions of new virus. In
serious cases, immune cells overreact and attack lung cell causing acute
respiratory disease syndrome and possibly death. In general, the fatality
rate in elder patients (60 years or more) is much higher than the average,
and under 40 years seems to be around $0.2\%$. Currently, there is not
vaccine, neither efficient treatment, even many drugs (cloroquine, for
instance) are under clinical trial. Like all RNA-based viruses, coronavirus
tends to mutate faster than DNA-viruses, but lower than influenza viruses.

Many mathematical and computational models are being used to describe
current new coronavirus pandemics. In mathematical model, there is a
fundamental threshold (see \cite{anderson}) called the basic reproduction
number, which is defined as the secondary cases produced by one case
introduced in a completely susceptible population, and is denoted by $R_{0}$%
. When a control mechanisms is introduced, this number is reduced, and is
called as the reduced reproduction number $R_{r}$. Ferguson \textit{et al}. 
\cite{ferguson} proposed a model in order to investigate the effects of
isolation of susceptible persons. They analyzed two scenarios called by them
as mitigation and suppression. Roughly, mitigation reduces the basic
reproduction number $R_{0}$, but not lower than one ($1<R_{r}<R_{0}$), while
suppression reduces the basic reproduction number lower than one ($R_{r}<1$%
). They predicted the numbers of severe cases and deaths due to CoViD-19
without control measure, and compared them with those numbers when
isolations (mitigation and suppression) are introduced as control measures.
Li \textit{et al}. discussed the role of undocumented infections \cite{li}.

In this paper we formulate a mathematical model based on ordinary
differential equations aiming firstly to understand the dynamics of CoViD-19
transmission, and, using the data from S\~{a}o Paulo State, Brazil, estimate
model parameters, and, then, study potential scenarios introducing isolation
as a control mechanism.

The paper is structured as follows. In Section 2, we introduce a model,
which is numerically studied in Section 3. Discussions are presented in
Section 4, and conclusions, in Section 5.

\section{Material and methods}

In a community where SARS-CoV-2 (new coronavirus) is circulating, the risk
of infection is greater in elder than young persons, as well as under
increased probability of being symptomatic and higher CoViD-19 induced
mortality. Hence, a community is divided in two groups, comprised by young
(under 60 years old, denoted by subscript $y$), and elder (above 60 years
old, denoted by subscript $o$) persons. The vital dynamics of this community
is given by per-capita rates of birth ($\phi $) and mortality ($\mu $).

For each sub-population $j$ ($j=y,o$), the persons are divided in seven
classes: susceptible $S_{j}$, susceptible persons who are isolated $Q_{j}$,
exposed $E_{j}$, asymptomatic $A_{j}$, asymptomatic persons who are caught
by test and then isolated $Q_{1j}$, symptomatic persons at initial phase of
CoViD-19 (or pre-diseased) $D_{1j}$, pre-diseased persons caught by test and
then isolated, plus mild CoViD-19 (or non-hospitalized) $Q_{2j}$, and
symptomatic persons with severe CoViD-19 (hospitalized) $D_{2j}$. However,
all persons in young and elder classes enter to same immunized class $I$,
after experiencing infection.

With respect to new coronavirus transmission, the history of natural
infection is the same in young ($j=y$) and elder ($j=o$) classes. We assume
that only persons in asymptomatic ($A_{j}$) and pre-diseased ($D_{1j}$)
classes are transmitting the virus, and other infected classes ($Q_{1j}$, $%
Q_{2j}$ and $D_{2j}$) are under voluntary or forced isolation. Susceptible
persons are infected according to $\lambda _{j}S_{j}/N$ and enter to classes 
$E_{j}$, where $\lambda _{j}$ is the per-capita incidence rate (or force of
infection) defined by $\lambda _{j}=\lambda \left( \delta _{jy}+\psi \delta
_{jo}\right) $, with $\lambda $ being%
\begin{equation}
\lambda =\beta _{1y}A_{y}+\beta _{2y}D_{1y}+\beta _{1o}A_{o}+\beta
_{2o}D_{1o},  \label{force}
\end{equation}%
where $\delta _{ij}$ is Kronecker delta, with $\delta _{ij}=1$ if $i=j$, and 
$0$, if $i\neq j$, and $S_{j}/N$ is the probability of virus encountering
susceptible persons. After an average period of time $1/\sigma _{j}$ in
classes $E_{j}$, where $\sigma _{j}$ is the incubation rate, exposed persons
enter to asymptomatic $A_{j}$ (with probability $p_{j}$) or pre-diseased $%
D_{1j}$ (with probability $1-p_{j}$) classes. After an average period of
time $1/\gamma _{j}$ in class $A_{j}$, where $\gamma _{j}$ is the infection
rate of asymptomatic persons, symptomatic persons acquire immunity
(recovered) and enter to immunized class $I$. Another route of exit from
class $A_{j}$ is being caught by a test at a rate $\eta _{j}$ and enters to
class $Q_{1j}$, and, then, after a period of time $1/\gamma _{j}$, enters to
class $I$. With very low intensity, asymptomatic persons are in voluntary
isolation, which is described by voluntary isolation rate $\chi _{j}$. With
respect to symptomatic persons, after an average period of time $1/\gamma
_{1j}$ in class $D_{1j}$, where $\gamma _{1j}$ is the infection rate of
pre-diseased persons, pre-diseased persons enter to non-hospitalized $Q_{2j}$
(with probability $m_{j}$) or hospitalized $D_{2j}$ (with probability $%
1-m_{j}$) classes. Hospitalized persons acquire immunity after a period of
time $1/\gamma _{2j}$, where $\gamma _{2j}$ is recovery rate of severe
CoViD-19, and enter to immunized class $I$, or die under disease induced
(additional) mortality rate $\alpha $. After an average period of time $%
1/\gamma _{j}$ in class $Q_{2j}$, non-hospitalized persons acquire immunity
and enter to immunized class $I$, or enter to class $D_{2j}$ at a relapsing
rate of pre-diseased persons $\xi _{j}$.

Figure 1 shows the flowchart of new coronavirus transmission model.

\begin{figure}[!h]
\centering
\includegraphics[scale=0.7]{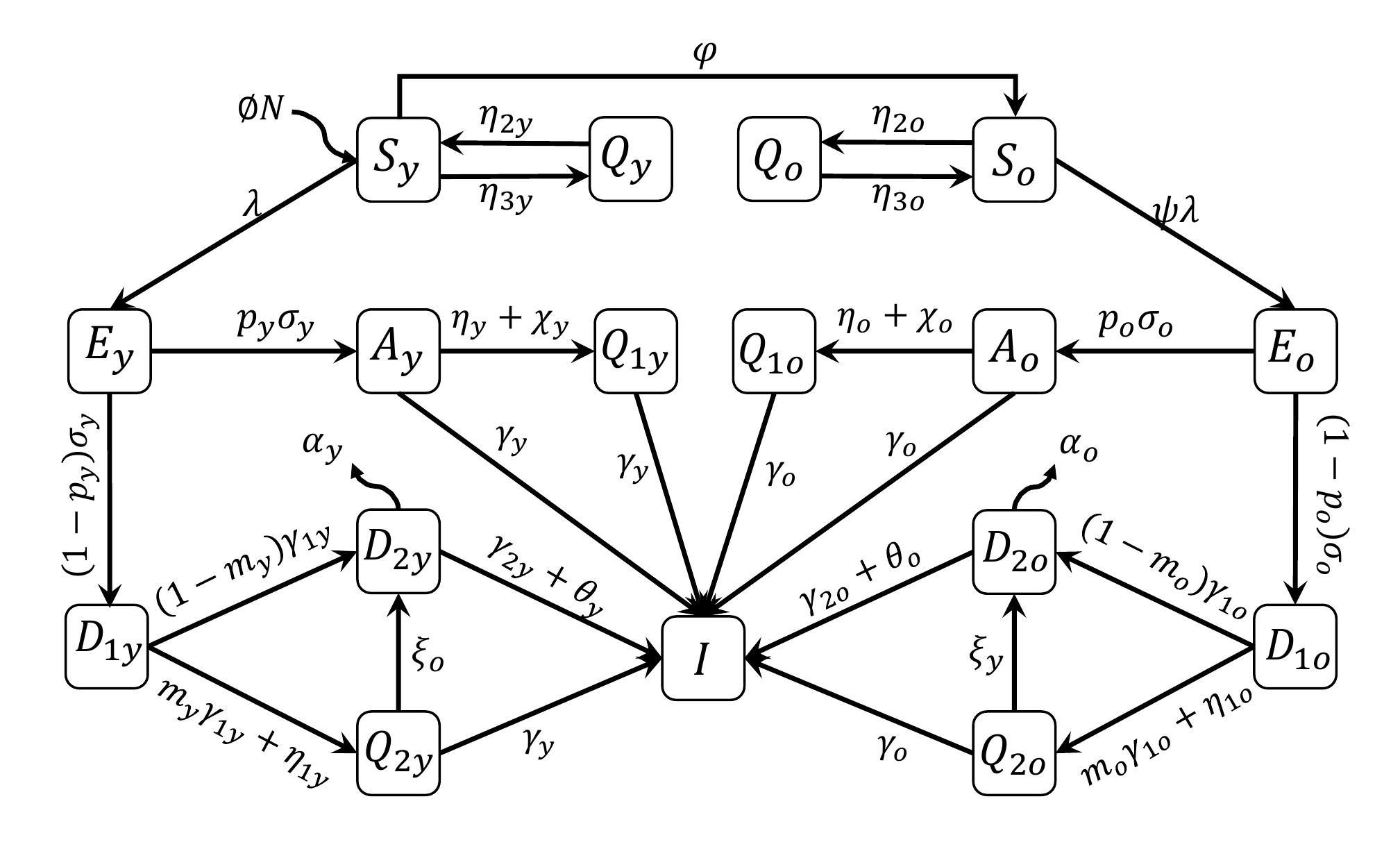}
\caption{The flowchart of new coronavirus transmission model with variables
and parameters.}%
\end{figure}

The new coronavirus transmission model, based on above descriptions
summarized in Figure 1, is described by system of ordinary differential
equations, with $j=y,o$. Equations for susceptible persons are%
\begin{equation}
\left\{ 
\begin{array}{rll}
\displaystyle\frac{d}{dt}S_{y} & = & \phi N-\left( \eta _{2y}+\varphi +\mu
\right) S_{y}-\lambda \frac{S_{y}}{N}+\eta _{3y}Q_{y} \\ 
\displaystyle\frac{d}{dt}S_{o} & = & \varphi S_{y}-\left( \eta _{2o}+\mu
\right) S_{o}-\lambda \psi \frac{S_{o}}{N}+\eta _{3o}Q_{o},%
\end{array}%
\right.  \label{system2a}
\end{equation}%
for infectious persons,%
\begin{equation}
\left\{ 
\begin{array}{rll}
\displaystyle\frac{d}{dt}Q_{y} & = & \eta _{2j}S_{j}-\left( \eta _{3j}+\mu
\right) Q_{j} \\ 
\displaystyle\frac{d}{dt}E_{j} & = & \lambda \left( \delta _{jy}+\psi \delta
_{jo}\right) \frac{S_{j}}{N}-\left( \sigma _{j}+\mu \right) E_{j} \\ 
\displaystyle\frac{d}{dt}A_{j} & = & p_{j}\sigma _{j}E_{j}-\left( \gamma
_{j}+\eta _{j}+\chi _{j}+\mu \right) A_{j} \\ 
\displaystyle\frac{d}{dt}Q_{1j} & = & \left( \eta _{j}+\chi _{j}\right)
A_{j}-\left( \gamma _{j}+\mu \right) Q_{1j} \\ 
\displaystyle\frac{d}{dt}D_{1j} & = & \left( 1-p_{j}\right) \sigma
_{j}E_{j}-\left( \gamma _{1j}+\eta _{1j}+\mu \right) D_{1j} \\ 
\displaystyle\frac{d}{dt}Q_{2j} & = & \left( m_{j}\gamma _{1j}+\eta
_{1j}\right) D_{1j}-\left( \gamma _{j}+\xi _{j}+\mu \right) Q_{2j}, \\ 
\displaystyle\frac{d}{dt}D_{2j} & = & \left( 1-m_{j}\right) \gamma
_{1j}D_{1j}+\xi _{j}Q_{2j}-\left( \gamma _{2j}+\theta _{j}+\mu +\alpha
_{j}\right) D_{2j},%
\end{array}%
\right.  \label{system1c}
\end{equation}%
and for immune persons,%
\begin{equation}
\begin{array}{rll}
\displaystyle\frac{d}{dt}I & = & \gamma _{y}A_{y}+\gamma _{y}Q_{1y}+\gamma
_{y}Q_{2y}+\left( \gamma _{2y}+\theta _{y}\right) D_{2y}+\gamma
_{o}A_{o}+\gamma _{o}Q_{1o}+\gamma _{o}Q_{2o}+ \\ 
&  & \left( \gamma _{2o}+\theta _{o}\right) D_{2o}-\mu I,%
\end{array}
\label{system2b}
\end{equation}%
with $N_{j}=S_{j}+Q_{j}+E_{j}+A_{j}+Q_{1j}+D_{1j}+Q_{2j}+D_{2j}$ obeying,
with $N=N_{y}+N_{o}+I$,%
\begin{equation}
\frac{d}{dt}N=\left( \phi -\mu \right) N-\alpha _{y}D_{2y}-\alpha _{o}D_{2o},
\label{nvar}
\end{equation}%
where, the initial number of population at $t=0$ is $N(0)=N_{0}$. If $\phi
=\mu +\left( \alpha _{y}D_{2y}+\alpha _{o}D_{2o}\right) /N$, the total size
of population is constant. The initial conditions (at $t=0$) supplied to
equations (\ref{system2a}), (\ref{system1c}) and (\ref{system2b}) are%
\[
\begin{array}{cccccc}
S_{j}\left( 0\right) =N_{0j}, & Q_{j}\left( 0\right) =0, & \mathrm{and} & 
X_{j}\left( 0\right) =n_{X_{j}}, & \mathrm{where} & 
X_{j}=E_{j},A_{j},Q_{1j},D_{1j},Q_{2j},D_{2j},I,%
\end{array}%
\]%
where $n_{X_{j}}$ is a non-negative number. For instance, $%
n_{E_{y}}=n_{E_{o}}=0$ means that there is not any exposed persons in the
beginning of epidemics.

Table 1 summarizes model variables.

\begin{table}[!h]
\centering
\caption{Summary of the model variables ($j=y,o$).}
\begin{tabular}{lll}
\hline
Symbol &  & Meaning \\ \hline
$S_{j}$ &  & Susceptible persons \\ 
$Q_{j}$ &  & Isolated among susceptible persons \\ 
$E_{j}$ &  & Exposed \\ 
$A_{j}$ &  & Asymptomatic \\ 
$Q_{1j}$ &  & Isolated among asymptomatic by test \\ 
$D_{1j}$ &  & Initial symptomatic (pre-diseased persons) \\ 
$Q_{2j}$ &  & Isolated among pre-diseased by test \\ 
$D_{2j}$ &  & Symptomatic (diseased persons) \\ 
$I_{j}$ &  & Immune persons (recovered persons) \\ \hline
\end{tabular}
\end{table}

Table 2 summarizes model parameters and values (values for elder classes are
between parentheses).

\begin{table}[!h]
\centering
\caption{Summary of the model parameters ($j=y,o$) and values (rates in $%
days^{-1}$, time in $days$ and proportions are dimensionless). Some values
are calculated ($^{\&}$), or varied ($^{\#}$), or assumed ($^{\ast }$), or
estimated ($^{\ast \ast }$) or not available ($^{\ast \ast \ast }$).}
\begin{tabular}{lllll}
\hline
Symbol &  & Meaning &  & Value \\ \hline
$\mu $ &  & Natural mortality rate &  & $1/(75\times 360)$\cite{seade} \\ 
$\phi $ &  & Birth rate &  & $1/(75\times 360)^{\ast }$ \\ 
$\varphi $ &  & Aging rate &  & $6.7\times 10^{-6}$ \\ 
$\sigma _{y}\left( \sigma _{o}\right) $ &  & Incubation rate &  & $1/6\left(
1/5\right) $\cite{who} \\ 
$\gamma _{y}\left( \gamma _{o}\right) $ &  & Infection rate of asymptomatic
persons &  & $1/10\left( 1/12\right) $\cite{who} \\ 
$\gamma _{1y}\left( \gamma _{1o}\right) $ &  & Infection rate of
pre-diseased persons &  & $1/3\left( 1/2\right) $\cite{who} \\ 
$\gamma _{2y}\left( \gamma _{2o}\right) $ &  & Recovery rate of severe
CoViD-19 &  & $1/10\left( 1/14\right) $\cite{who} \\ 
$\xi _{y}\left( \xi _{yo}\right) $ &  & Relapsing rate of pre-diseased
persons &  & $0.005\left( 0.01\right) ^{\ast }$ \\ 
$\alpha _{y}\left( \alpha _{o}\right) $ &  & Additional mortality rate &  & $%
0.0009\left( 0.009\right) ^{\ast \ast }$ \\ 
$\eta _{y}\left( \eta _{o}\right) $ &  & Testing rate among asymptomatic
persons &  & $0\left( 0\right) ^{\ast \ast \ast }$ \\ 
$\chi _{y}\left( \chi _{o}\right) $ &  & Voluntary isolation rate of
asymptomatic persons &  & $0\left( 0\right) ^{\ast }$ \\ 
$\eta _{1y}\left( \eta _{1o}\right) $ &  & Testing rate among pre-diseased
persons &  & $0\left( 0\right) ^{\ast \ast \ast }$ \\ 
$\eta _{2y}\left( \eta _{2o}\right) $ &  & Isolation rate of susceptible
persons &  & $0.035\left( 0.035\right) ^{\#}$ \\ 
$\eta _{3y}\left( \eta _{3o}\right) $ &  & Releasing rate of isolated persons
&  & $0.035\left( 0.035\right) ^{\#}$ \\ 
$\theta _{y}\left( \theta _{o}\right) $ &  & Treatment rate &  & $0(0)^{\ast
\ast \ast }$ \\ 
$\beta _{1y}\left( \beta _{1o}\right) $ &  & Transmission rate due to
asymptomatic persons &  & $0.77\left( 0.77\right) ^{\ast \ast }$ \\ 
$\beta _{2y}\left( \beta _{2o}\right) $ &  & Transmission rate due to
rpe-diseased persons &  & $0.77\left( 0.77\right) ^{\ast \ast }$ \\ 
$\psi $ &  & Scaling factor of transmission among elder persons &  & $%
1.17^{\&}$ \\ 
$p_{y}\left( p_{o}\right) $ &  & Proportion of asymptomatic persons &  & $%
0.8(0.75)^{\ast }$ \\ 
$m_{y}\left( m_{o}\right) $ &  & Proportion of mild (non-hospitalized)
CoViD-19 &  & $0.8\left( 0.75\right) $\cite{bepi8} \\ \hline
\end{tabular}
\end{table}

Isolation of persons deserves some words. In the modeling, the isolation is
applied to susceptible persons, which are known only at exact time of the
introduction of new virus, that is, $S(0)=N_{0}$. However, as time passes,
susceptible persons are decreased and become immunized persons, and, due to
asymptomatic persons, susceptible and immunized persons are
indistinguishable (except caught by test and hospitalized persons). For this
reason, if isolation of persons is not done at the time of virus
introduction, it is probable that virus should be circulating among them,
but at very lower transmission rate (virus circulates only among household
and neighborhood persons).

From the system of equations (\ref{system2a}), (\ref{system1c}) and (\ref%
{system2b}) we can derive some epidemiological parameters: new cases, new
CoViD-19 cases, severe CoViD-19 cases, number of deaths due to CoViD-19, and
isolated persons.

The number of persons infected with new coronavirus are given by $%
E_{y}+A_{y}+Q_{1y}+D_{1y}+Q_{2y}+D_{2y}$ for young persons, and $%
E_{o}+A_{o}+Q_{1o}+D_{1o}+Q_{2o}+D_{2o}$ for elder persons. The incidence
rates are 
\begin{equation}
\begin{array}{ccccc}
\Lambda _{y}=\lambda \frac{S_{y}}{N} &  & \mathrm{and} &  & \Lambda
_{o}=\lambda \psi \frac{S_{o}}{N},%
\end{array}
\label{incide}
\end{equation}%
where the per-capita incidence rate $\lambda $ is given by equation (\ref%
{force}), and the numbers of new cases $C_{y}$ and $C_{o}$ are%
\[
\begin{array}{ccccc}
\displaystyle\frac{d}{dt}C_{y}=\Lambda _{y}dt &  & \mathrm{and} &  & %
\displaystyle\frac{d}{dt}C_{o}=\Lambda _{o}dt,%
\end{array}%
\]%
with $C_{y}(0)=0$ and $C_{o}(0)=0$, and the numbers of new cases in a day is%
\[
\begin{array}{ccccc}
C_{y}^{i}=\int\limits_{T_{i}}^{T_{i+1}}\Lambda _{y}dt=C_{y}\left(
T_{i+1}\right) -C_{y}\left( T_{i}\right) &  & \mathrm{and} &  & 
C_{o}^{i}=\int\limits_{T_{i}}^{T_{i+1}}\Lambda _{o}dt=C_{o}\left(
T_{i+1}\right) -C_{o}\left( T_{i}\right) ,%
\end{array}%
\]%
where $T_{i}=i\tau $, $\tau =T_{i+1}-T_{i}=1$ $day$, for $i=1,\cdots $, with 
$T_{0}=0$. Notice that $C_{y}^{i}$ and $C_{o}^{i}$ are entering in exposed
classes at each day.

The numbers of CoViD-19 cases $\Delta _{y}$ and $\Delta _{o}$ are given by
outflux of $A_{y}$, $D_{1y}$, $A_{o}$ and $D_{1o}$, that is,%
\[
\begin{array}{ccc}
\displaystyle\frac{d}{dt}\Delta _{y}=\eta _{y}A_{y}+\left( \gamma _{1j}+\eta
_{1j}\right) D_{1y} & \mathrm{and} & \displaystyle\frac{d}{dt}\Delta
_{o}=\eta _{o}A_{o}+\left( \gamma _{1o}+\eta _{1o}\right) D_{1o},%
\end{array}%
\]%
with $\Delta _{y}(0)=0$ and $\Delta _{o}(0)=0$, and the numbers of CoViD-19
cases in a day are%
\[
\left\{ 
\begin{array}{l}
\Delta _{y}^{i}=\int\limits_{T_{i}}^{T_{i+1}}\left[ \eta _{y}A_{y}+\left(
\gamma _{1j}+\eta _{1j}\right) D_{1y}\right] dt=\Delta _{y}\left(
T_{i+1}\right) -\Delta _{y}\left( T_{i}\right) \\ 
\Delta _{o}^{i}=\int\limits_{T_{i}}^{T_{i+1}}\left[ \eta _{o}A_{o}+\left(
\gamma _{1o}+\eta _{1o}\right) D_{1o}\right] dy=\Delta _{o}\left(
T_{i+1}\right) -\Delta _{o}\left( T_{i}\right) ,%
\end{array}%
\right. 
\]%
which are entering in classes $Q_{1y}$, $D_{2y}$, $Q_{2y}$, $Q_{1o}$, $%
D_{2o} $ and $Q_{2o}$ at each day.

The numbers of severe CoViD-19 (hospitalized) cases $\Omega _{y}$ and $%
\Omega _{o}$ are given by outflux of $D_{1y}$, $Q_{2o}$, $D_{2o}$ and $%
Q_{2y} $, that is,%
\begin{equation}
\begin{array}{ccc}
\displaystyle\frac{d}{dt}\Omega _{y}=\left( 1-m_{y}\right) \gamma
_{1y}D_{1y}+\xi _{y}Q_{2y} & \mathrm{and} & \displaystyle\frac{d}{dt}\Omega
_{o}=\left( 1-m_{o}\right) \gamma _{1o}D_{1o}+\xi _{o}Q_{2o},%
\end{array}
\label{severe}
\end{equation}%
with $\Omega _{y}(0)=0$ and $\Omega _{o}(0)=0$, and the numbers of
hospitalized cases in a day are%
\[
\left\{ 
\begin{array}{l}
\Omega _{y}^{i}=\int\limits_{T_{i}}^{T_{i+1}}\left[ \left( 1-m_{y}\right)
\gamma _{1y}D_{1y}+\xi _{y}Q_{2y}\right] dt=\Omega _{y}(T_{i+1})-\Omega
_{y}(T_{i}) \\ 
\Omega _{o}^{i}=\int\limits_{T_{i}}^{T_{i+1}}\left[ \left( 1-m_{o}\right)
\gamma _{1o}D_{1o}+\xi _{o}Q_{2o}\right] dt=\Omega _{o}(T_{i+1})-\Omega
_{o}(T_{i}),%
\end{array}%
\right. 
\]%
which are entering in classes $D_{2y}$ and $D_{2o}$ at each day.

The number of deaths caused by severe CoViD-19 cases $\Pi $ can be
calculated from hospitalized cases. This number of deaths is 
\begin{equation}
\frac{d}{dt}\Pi =\alpha _{y}D_{2y}+\alpha _{o}D_{2o},  \label{mort}
\end{equation}%
with $\Pi (0)=0$. The number of died persons in a day is%
\[
\begin{array}{ccccc}
\pi =\pi _{y}+\pi _{o} &  & \mathrm{with} &  & \left\{ 
\begin{array}{l}
\pi _{y}=\int\limits_{T_{i}}^{T_{i+1}}\alpha _{y}D_{2y}dt \\ 
\pi _{o}=\int\limits_{T_{i}}^{T_{i+1}}\alpha _{o}D_{2o}dt,%
\end{array}%
\right.%
\end{array}%
\]%
where $\pi _{y}$ and $\pi _{0}$ are the numbers of deaths of young and elder
persons at each day.

The number of susceptible persons in isolation in the absence of releasing
is obtained from%
\begin{equation}
\begin{array}{ccc}
S^{is}=S_{y}^{is}+S_{o}^{is}, & \mathrm{where} & \left\{ 
\begin{array}{lll}
\displaystyle\frac{d}{dt}S_{y}^{is}=\eta _{2y}S_{y}, & \mathrm{with} & 
S_{y}^{is}(0)=0 \\ 
\displaystyle\frac{d}{dt}S_{o}^{is}=\eta _{2o}S_{o}, & \mathrm{with} & 
S_{o}^{is}(0)=0,%
\end{array}%
\right.%
\end{array}
\label{propotion}
\end{equation}%
where the corresponding fractions of isolated susceptible persons are $%
f_{y}^{is}=S_{y}^{is}/N_{y}$ and $f_{y}^{is}=S_{o}^{is}/N_{o}$.

The system of equations (\ref{system2a}), (\ref{system1c}) and (\ref%
{system2b}) is non-autonomous. Nevertheless the fractions of persons in each
compartment approach to the steady state (see Appendix A), hence, by using
equations (\ref{Rr}) and (\ref{Rr1}), the reduced reproduction number $R_{r}$
is given by%
\begin{equation}
\begin{array}{c}
R_{r}=R_{ry}+R_{ro}=\left[ p_{y}R_{0y}^{1}+\left( 1-p_{y}\right) R_{0y}^{2}%
\right] \frac{S_{y}^{0}}{N_{0}}+\left[ p_{o}R_{0o}^{1}+\left( 1-p_{o}\right)
R_{0o}^{2}\right] \frac{S_{o}^{0}}{N_{0}},%
\end{array}
\label{Rred}
\end{equation}%
where $s_{y}^{0}$ and $s_{o}^{0}$ are substituted by $S_{y}^{0}/N_{0}$ and $%
S_{o}^{0}/N_{0}$.

Given $N$ and $R_{0}$, let us evaluate the number of susceptible persons in
order to trigger and maintain epidemics, but in a special case. Assume that
all model parameters for young and elder classes and all transmission rates
are equal, then $R_{0}=\sigma \beta /\left[ \left( \sigma +\phi \right)
\left( \gamma +\phi \right) \right] $ and $R_{e}=R_{0}S/N$, using
approximated $R_{e}$ given by equation (\ref{Refe}). Letting $R_{e}=1$, the
critical number of susceptible persons $S^{th}$ at equilibrium is%
\begin{equation}
S^{th}\approx \frac{N}{R_{0}}.  \label{scrit}
\end{equation}%
If $S>S^{th}$, epidemics occurs and persists ($R_{e}>1$, non-trivial
equilibrium point $P^{\ast }$), and the fraction of susceptible individuals
is $s^{\ast }=1/R_{e}$, where $s^{\ast }=s_{y}^{\ast }+s_{o}^{\ast }$; but
if $S<S^{th}$, epidemics occurs but fades out ($R_{e}<1$, trivial
equilibrium point $P^{0}$), and the fractions of susceptible individuals $%
s_{y}$ and $s_{o}$ at equilibrium are given by equation (\ref{trivial}), or (%
\ref{trivial0}) if there is not any control.

Let us now evaluate the critical isolation rate of susceptible persons $\eta
_{2}$ assuming that all model parameters for young and elder classes and all
transmission rates are equal. In this special case, $R_{r}=R_{0}\left( \eta
_{3}+\phi \right) /\left( \eta _{2}+\eta _{3}+\phi \right) $, where $%
R_{0}=\sigma \beta /\left[ \left( \sigma +\phi \right) \left( \gamma +\phi
\right) \right] $, and letting $R_{r}=1$, we obtain%
\begin{equation}
\eta _{2}^{th}\approx \left( \eta _{3}+\phi \right) \left( R_{0}-1\right) .
\label{etacrit}
\end{equation}%
If $\eta _{2}<\eta _{2}^{th}$, epidemics occurs and persists ($R_{e}>1$,
non-trivial equilibrium point $P^{\ast }$); but if $\eta _{2}>\eta _{2}^{th}$%
, epidemics occurs but fades out ($R_{e}<1$, trivial equilibrium point $%
P^{0} $).

We apply above results to study the introduction and establishment of new
coronavirus in the S\~{a}o Paulo State, Brazil. From data collected in the S%
\~{a}o Paulo State from March 14, 2020 until April 5, 2020, we estimate
transmission and additional mortality rates, and, then, study potential
scenarios introducing isolation as control mechanisms.

\section{Results}

Results obtained in foregoing section is applied to describe new coronavirus
infection in the S\~{a}o Paulo State, Brazil. The first confirmed case of
CoViD-19, occurred in February 26, 2020, was from a traveler returning from
Italy in February 21, and being hospitalized in February 24. The first death
due to CoViD-19 was a 62 years old male with comorbidity who never travelled
to abroad, hence considered as autochthonous transmission. He manifested
first symptoms in March 10, was hospitalized in March 14, and died in March
16. In March 24, the S\~{a}o Paulo State authorities ordered isolation of
persons acting in non-essential activities, as well as students of all level
until April 6, further the isolation was extended to April 22.

Let us determine the initial conditions. In the S\~{a}o Paulo State, the
number of inhabitants is $N\left( 0\right) =N_{0}=44.6\times 10^{6}$
according to SEADE \cite{seade}. The value of parameter $\varphi $ given in
Table 1 was calculated by equation (\ref{trivial0}), $\varphi =b\phi /\left(
1-b\right) $, where $b$ is the proportion of elder persons. Using $b=0.153$
in the S\~{a}o Paulo State \cite{seade}, we obtained $\varphi =6.7\times
10^{-6}$ $days^{-1}$, hence, $N_{y}\left( 0\right) =N_{0y}=37.8\times 10^{6}$
($\bar{s}_{y}^{0}=N_{0y}/N_{y}\left( 0\right) =0.8475$) and $N_{o}\left(
0\right) =N_{0o}=6.8\times 10^{6}$ ($\bar{s}_{o}^{0}=N_{0o}/N_{o}\left(
0\right) =0.1525$). The initial conditions for susceptible persons are let
to be $S_{y}\left( 0\right) =N_{y}\left( 0\right) $ and $S_{o}\left(
0\right) =N_{o}\left( 0\right) $. For other variables, from Table 2, $%
p_{y}=0.8$ and $m_{y}=0.8$, the ratio asymptomatic:symptomatic is $4:1$, and
the ratio mild:severe (non-hospitalized:hospitalized) CoViD-19 is $4:1$. We
use these ratios for elder persons, even $p_{o}$ and $m_{o}$ are slightly
different. Hence, if we assume that there is 1 person in $D_{2j}$ (the first
confirmed case), then there are 4 persons in $Q_{2j}$. The sum (5) is the
number of persons in class $D_{1j}$, implying that there are 20 in class $%
A_{j}$, hence, the sum (25) is the number of persons in class $E_{j}$.
Finally, we suppose that no one is isolated or tested, and also immunized.
(Probably the first confirmed COViD-19 person transmitted the virus (since
February 21 when returned infected from Italy), as well as other
asymptomatic travelers returning from abroad.)

Therefore, the initial conditions supplied to the dynamic system (\ref%
{system2a}), (\ref{system1c}) and (\ref{system2b}) are%
\[
\left\{ 
\begin{array}{l}
\begin{array}{ccc}
S_{j}\left( 0\right) =N_{0j}, & Q_{j}\left( 0\right) =Q_{1j}(0)=0, & 
E_{j}\left( 0\right) =25,%
\end{array}
\\ 
\begin{array}{lllll}
A_{j}(0)=20, & D_{1j}(0)=5, & Q_{2j}(0)=4 & D_{2j}(0)=1, & I(0)=0,%
\end{array}%
\end{array}%
\right. 
\]%
where the initial simulation time $t=0$ corresponds to calendar time
February 26, 2020, when the first case was confirmed. The system of
equations (\ref{system2a}), (\ref{system1c}) and (\ref{system2b}) is
evaluated numerically using $4^{th}$ order Runge-Kutta method.

This section presents parameters estimation and epidemiological scenarios
considering isolation as control measure. In estimation and epidemiological
scenarios, we assume that all transmission rates in young persons are equal,
as well as in elder persons, that is, we assume that%
\[
\begin{array}{lllll}
\beta _{y}=\beta _{1y}=\beta _{2y}=\beta _{1o}=\beta _{2o}, &  & \mathrm{and}
&  & \beta _{o}=\psi \beta _{y},%
\end{array}%
\]%
hence the forces of infection are $\lambda _{y}=\left(
A_{y}+D_{1y}+A_{o}+D_{1o}\right) \beta _{y}$ and $\lambda _{o}=\psi \lambda
_{y}$.

\subsection{Parameters estimation}

Reliable estimation of both transmission and additional mortality rates are
crucial aiming the prediction of new cases (to adequate the number of beds
in hospital, for instance) and deaths. When the estimation is based on few
number of data, that is, in the beginning of epidemics, some cautions must
be taken, because the rates maybe over or under estimated. The reason is
that in the very beginning phase of epidemics, the spreading out of
infection and deaths increase exponentially without bound.

Currently, there is not sufficient number of kits to detect infection by new
coronavirus. For this reason, tests to confirm infection by this virus is
done only in hospitalized persons, and, also, in persons who died
manifesting symptoms of CoViD-19. Hence, we have only data of hospitalized
persons ($D_{2y}$ and $D_{2o}$) and those who died ($\Pi _{y}$ and $\Pi _{o}$%
). Taking into account hospitalized persons with CoViD-19, we estimate the
transmission rates, and from persons died due to CoViD-19, we estimate the
additional mortality rates. These rates are estimated applying the least
square method (see \cite{yang7}).

The introduction of quarantine is $t=27$, corresponding to calendar time
March 24, but the effects are expected to appear later. Hence, we will
estimate taking into account confirmed cases and deaths from February 26 ($%
t=0$) to April 5 ($t=39$),\footnote{%
Simulations were done in April 6.} hence $n=40$ observations. Notice that
the sum of incubation and recovery periods (see Table 2) is around 16 days,
hence it is expected that at around simulation time $t=43$ (April 10) the
effects of isolation appear.

To estimate the transmission rates $\beta _{y}$ and $\beta _{o}$, we let $%
\alpha _{y}=\alpha _{o}=0$ and the system of equations (\ref{system2a}), (%
\ref{system1c}) and (\ref{system2b}) is evaluated and calculate 
\begin{equation}
\mathrm{\min }\sum_{i=1}^{n}\left\{ \Omega _{y}\left( t_{i}\right) +\Omega
_{o}\left( t_{i}\right) -\left[ D_{2y}^{ob}\left( t_{i}\right)
+D_{2o}^{ob}\left( t_{i}\right) \right] \right\} ^{2},  \label{ls}
\end{equation}%
where $\mathrm{\min }$ stands for minimum value, $n$ is the number of
observations, $t_{i}$ is $i$-th observation time, $\Omega _{y}$ and $\Omega
_{o}$ are given by equation (\ref{severe}), and $D_{2y}^{ob}$ and $%
D_{2o}^{ob}$ are observed number of hospitalized persons. The better
transmission rates are those minimizing the square difference

To estimate the mortality rates $\alpha _{y}$ and $\alpha _{o}$, we fix
previously transmission rates $\beta _{y}$ and $\beta _{o}$ and the system
of equations (\ref{system2a}), (\ref{system1c}) and (\ref{system2b}) is
evaluated and calculate 
\begin{equation}
\mathrm{\min }\sum_{i=1}^{n}\left\{ \Pi _{y}\left( t_{i}\right) +\Pi
_{o}\left( t_{i}\right) -\left[ P_{y}^{ob}\left( t_{i}\right)
+P_{o}^{ob}\left( t_{i}\right) \right] \right\} ^{2},  \label{ls1}
\end{equation}%
where $\mathrm{\min }$ stands for minimum value, $n$ is the number of
observations, $t_{i}$ is $i$-th observation time, $\Pi _{y}$ and $\Pi _{o}$
are given by equation (\ref{mort}), and $P_{y}^{ob}$ and $P_{o}^{ob}$ are
observed number of died persons. The better mortality rates are those
minimizing the square difference.

Instead of using equations (\ref{ls}) and (\ref{ls1}), the least square
estimation method, we vary transmission or additional mortality rates and
choose better fittings by evaluating the sum of squared distances between
curve and data.

\subsubsection{Estimation of transmission and additional mortality rates}

Firstly, letting additional mortality rates equal to zero ($\alpha
_{y}=\alpha _{o}=0$), we estimate a unique $\beta =\beta _{y}=\beta _{o}$,
with $\psi =1$, against hospitalized CoViD-19 cases ($D_{2}$) data from the S%
\~{a}o Paulo State. The estimated value is $\beta =0.8$ $days^{-1}$,
resulting, for the basic reproduction number, $R_{0}=6.99$ (partials $%
R_{0y}=5.83$ and $R_{0o}=1.16$). Around this value, we vary $\beta _{y}$ and 
$\beta _{o}$ and choose better fitted values comparing curves of $%
D_{2}=D_{2y}+D_{2o}$ with observed data. The estimated values are $\beta
_{y}=0.77$ and $\beta _{o}=\psi \beta _{y}=0.9009$ ($days^{-1}$), where $%
\Psi =1.17$, resulting in the basic reproduction number $R_{0}=6.915$
(partials $R_{0y}=5.606$ and $R_{0o}=1.309$). Figure 2 shows the estimated
curve of $D_{2}$ and observed data. This estimated curve is quite the same
as the curve fitted using a unique $\beta $.

\begin{figure}[!h]
\centering
\includegraphics[scale=0.55]{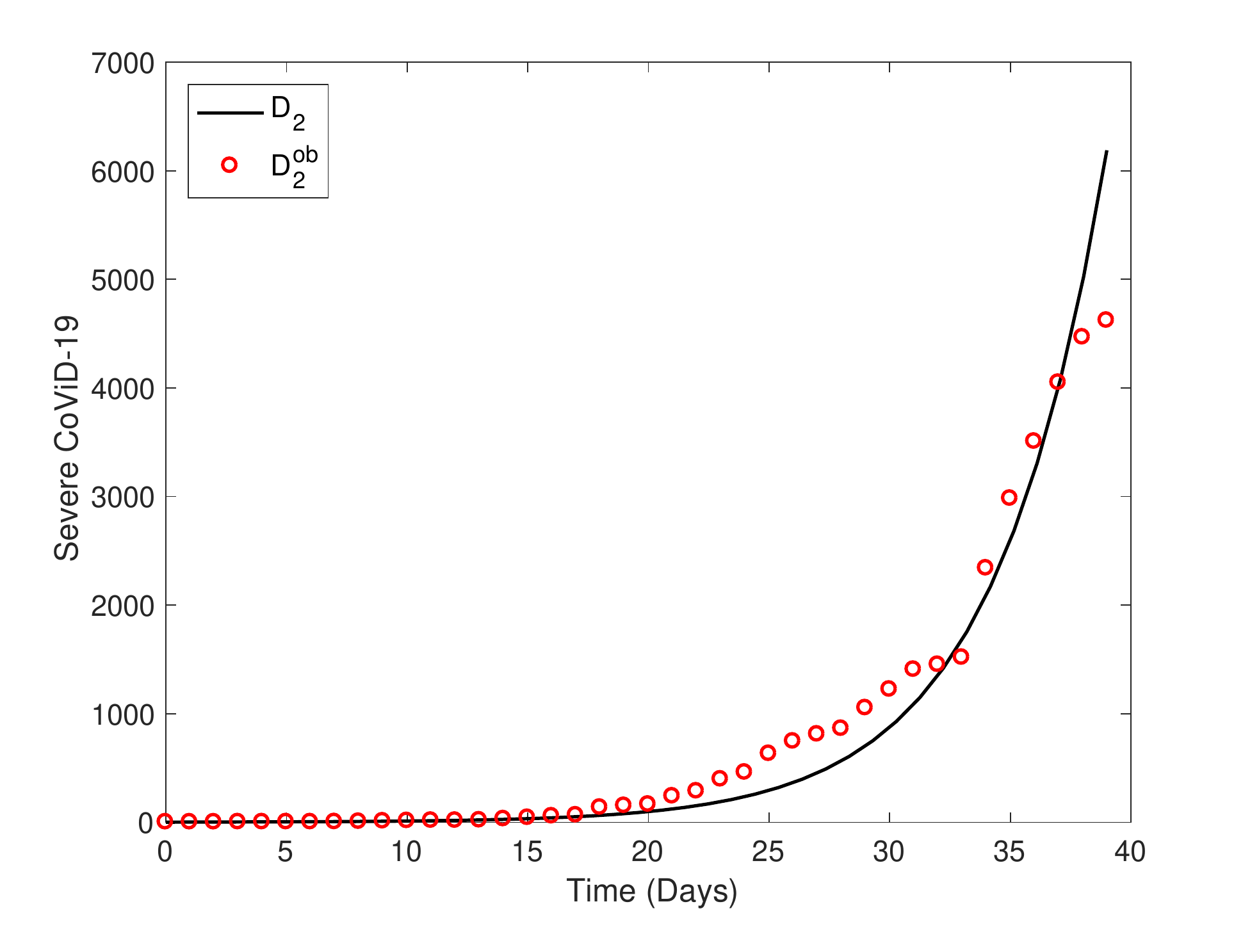}
\caption{The estimated curve of severe CoViD-19 cases $D_{2}$ and observed
data. Estimation of transmission parameters $\beta _{y}=0.77$ and $\beta
_{o}=0.9009$ ($days^{-1}$).}%
\end{figure}

Fixing previously estimated transmission rates $\beta _{y}=0.77$ and $\beta
_{o}=0.9009$ (both $days^{-1}$), we estimate additional mortality rates $%
\alpha _{y}$ and $\alpha _{o}$. We vary $\alpha _{y}$ and $\alpha _{o}$ and
choose better fitted values comparing curves of deaths due to CoViD-19 $\Pi
=\Pi _{y}+\Pi _{o}$ with observed data. By the fact that lethality among
young persons is much lower than elder persons, we let $\alpha
_{y}=0.1\alpha _{o}$ \cite{who}, and fit only one variable $\alpha _{o}$.
The estimated rates are $\alpha _{y}=0.0036$ and $\alpha _{o}=0.036$ ($%
days^{-1}$). Figure 3 shows the estimated curve of $\Pi =\Pi _{y}+\Pi _{o}$
and observed data. We call this as the first estimation method

\begin{figure}[!h]
\centering
\includegraphics[scale=0.55]{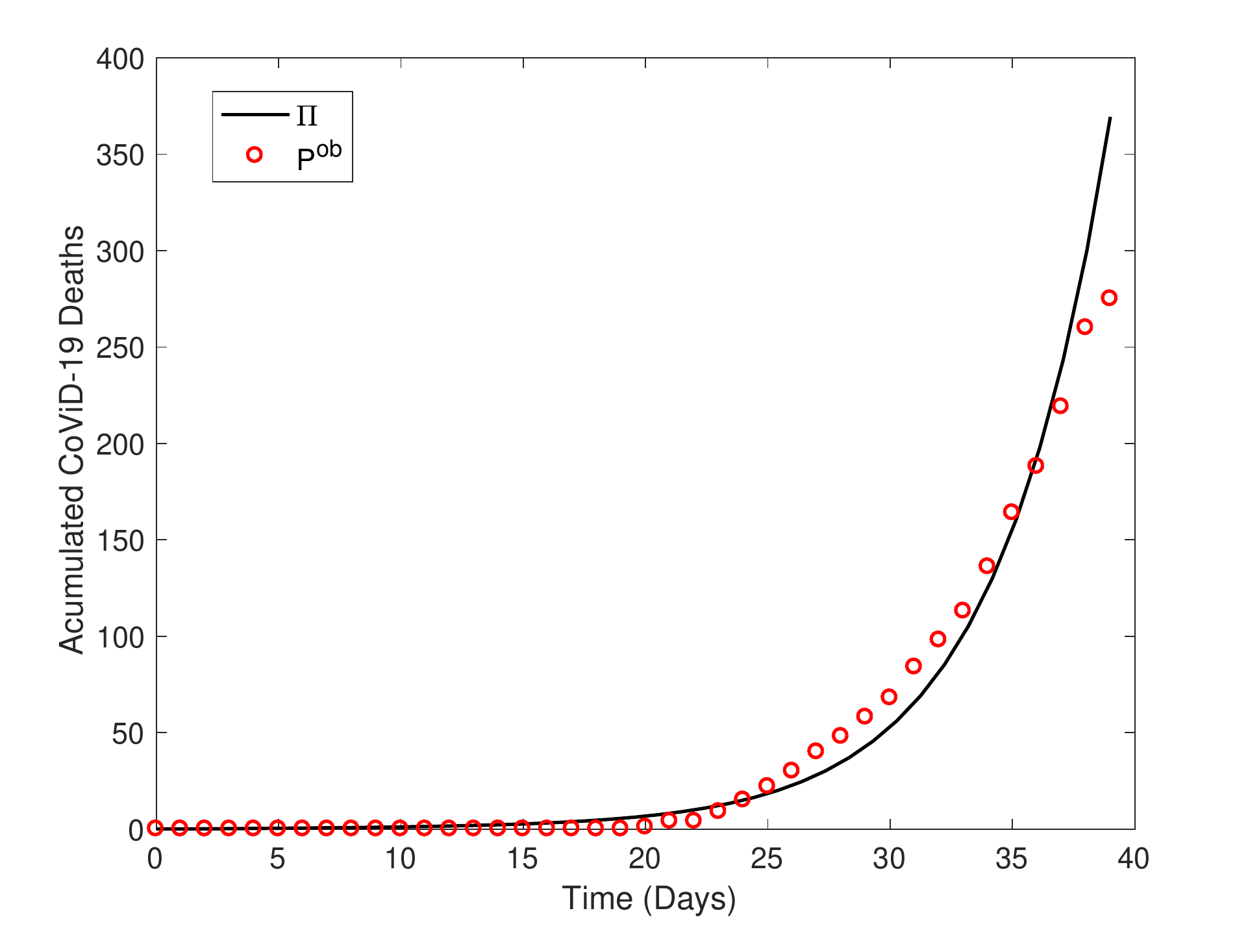}
\caption{The estimated curve of deaths due to CoViD-19 $\Pi $ and observed
data. First estimation method for additional mortality rates $\alpha
_{y}=0.0036$ and $\alpha _{o}=0.036$ ($days^{-1}$).
}%
\end{figure}

The first estimation method used only one information: the risk of death is
higher among elder than young persons (we used $\alpha _{y}=0.1\alpha _{o}$%
). However, the lethality among hospitalized elder persons is $10\%$ \cite%
{bepi8}. Combining both findings, we assume that the numbers of deaths for
young and elder persons are, respectively, $10\%$ and $1\%$ of accumulated
cases when $\Omega _{y}$ and $\Omega _{o}$ approach plateaus (see Figure 6
below). This is called as second estimation method, which takes into account
a second information besides the one used in the first estimation method. In
this procedure, the estimated rates are $\alpha _{y}=0.0009$ and $\alpha
_{o}=0.009$ ($days^{-1}$). Figure 4 shows this estimated curve $\Pi =\Pi
_{y}+\Pi _{o}$ and observe data, which fits very badly in the initial phase
of epidemics, but portraits current epidemiological findings.

\begin{figure}[!h]
\centering
\includegraphics[scale=0.55]{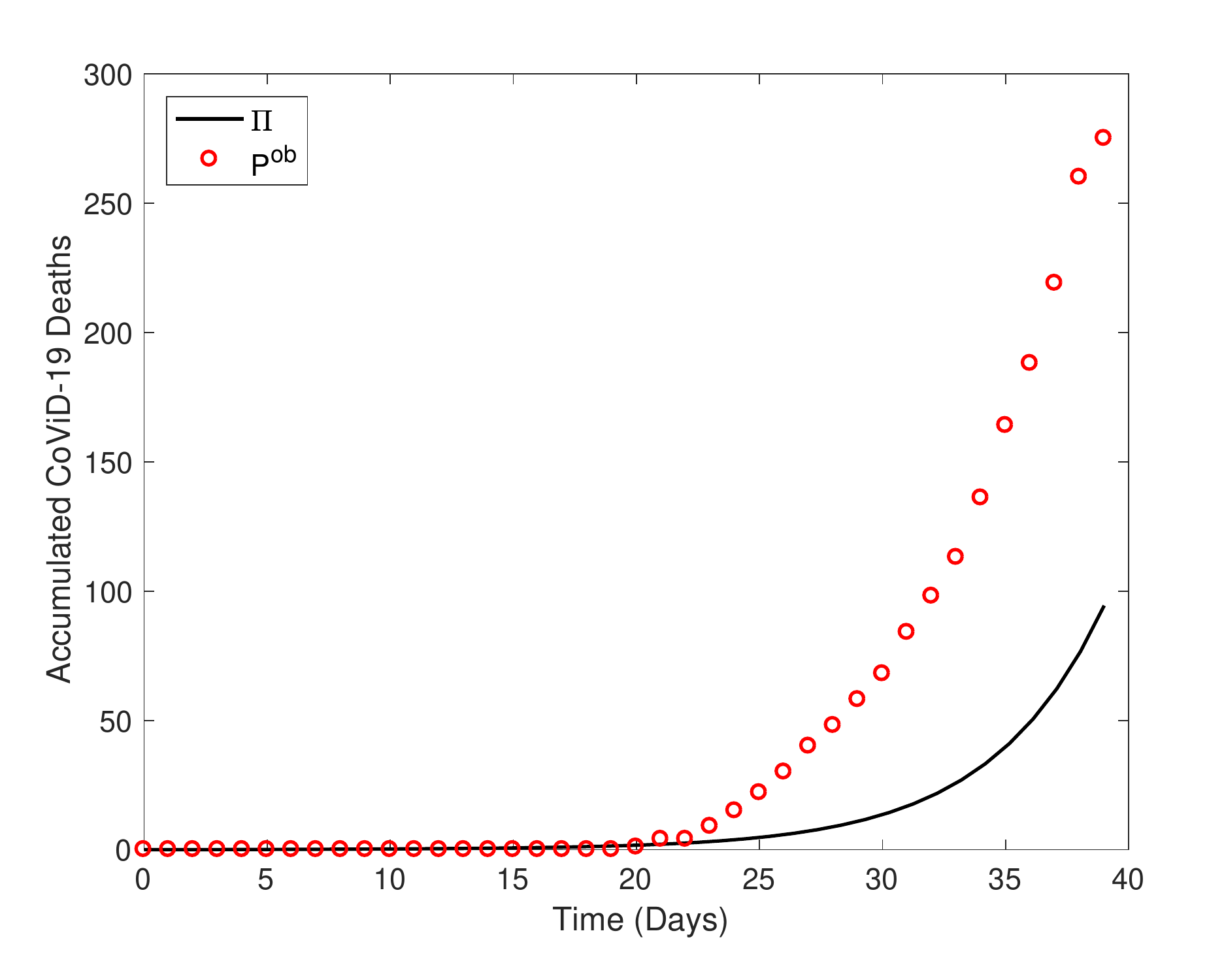}
\caption{The estimated curve of deaths due to CoViD-19 $\Pi $ and observe
data. Second estimation method for additional mortality rates $\alpha
_{y}=0.0009$ and $\alpha _{o}=0.009$ ($days^{-1}$).}%
\end{figure}

The fitted $\beta _{y}$, $\beta _{o}$, $\alpha _{y}$ and $\alpha _{o}$ (two
estimation methods) are fixed, and control variables $\eta _{2y}$ and $\eta
_{2o}$ are varied aiming to obtain of epidemiological scenarios. In general,
the epidemic period of infection by viruses $\tau $ is around 2 years, and
depending on the value of $R_{0}$, a second epidemics occurs after elapsed
many years \cite{yang9}. For this reason, we analyze epidemiological
scenarios of CoViD-19 restricted during the first wave of epidemics letting $%
\tau =140$ days.

Remembering that human population is varying due to the additional mortality
(fatality) of severe CoViD-19, we have, at $t=0$, $N_{0y}=3.780\times 10^{7}$%
, $N_{0o}=0.680\times 10^{7}$ and $N_{0}=N_{0y}+N_{0o}=4.460\times 10^{7}$,
and at $t=140$ days, $N_{y}=3.773\times 10^{7}$ ($0.185\%$), $%
N_{o}=0.662\times 10^{7}$ ($2,647\%$) and $N=4.435\times 10^{7}$ ($0.56\%$)
for the first estimation method, and $N_{y}=3.778\times 10^{7}$ ($0.052\%$), 
$N_{o}=0.674\times 10^{7}$ ($0.882\%$) and $N=4.452\times 10^{7}$ ($0.179\%$%
) for the second estimation method. The percentage of deaths ($100\left(
N_{0j}-N_{j}\right) /N_{0j}$) is given between parentheses. The first
estimation method for $\alpha _{y}$ and $\alpha _{o}$ yielded higher number
of deaths than the second method.

\subsubsection{Epidemiological scenario without any control mechanisms}

All effects of isolation will be compared with new coronavirus transmission
without any control. Initially, estimated curves will be extended until $%
\tau =140$ days, when disease attains low values.

Figure 5 shows the estimated curves of the number of hospitalized (severe)
CoViD-19 ($D_{2y}$, $D_{2o}$ and $D_{2}=D_{2y}+D_{2o}$). We observe that the
peaks of severe CoViD-19 are for elder, young and all persons are,
respectively, $2.061\times 10^{5}$, $5.532\times 10^{5}$ and $7.582\times
10^{5}$, which occur at same time $t=72$ days.

\begin{figure}[!h]
\centering
\includegraphics[scale=0.55]{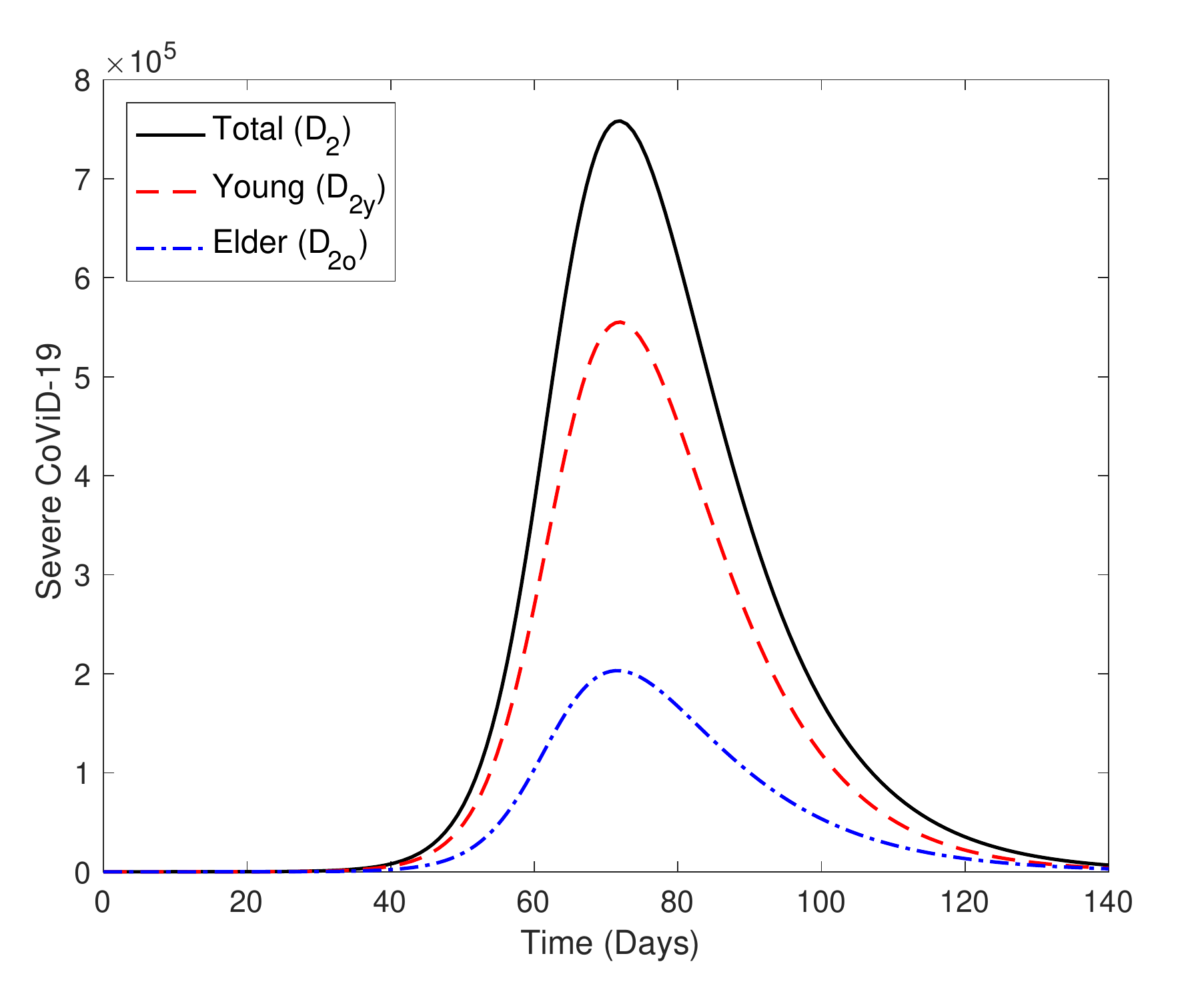}
\caption{The estimated curves of the number of hospitalized (severe)
CoViD-19 ($D_{2y}$, $D_{2o}$ and $D_{2}=D_{2y}+D_{2o}$) during the first
wave of epidemics.}%
\end{figure}

Figure 6 shows the estimated curves of accumulated number of severe CoViD-19
($\Omega _{y}$, $\Omega _{o}$ and $\Omega =\Omega _{y}+\Omega _{o}$), from
equation (\ref{severe}). At $t=140$ days, $\Omega $ is approaching to
asymptote (or plateau), which can be understood as the time when the first
wave of epidemics ends. The curves $\Omega _{y}$, $\Omega _{o}$ and $\Omega $
attain values at $t=140$, respectively, $1.798\times 10^{6}$, $0.563\times
10^{6}$ and $2.361\times 10^{6}$.

\begin{figure}[!h]
\centering
\includegraphics[scale=0.55]{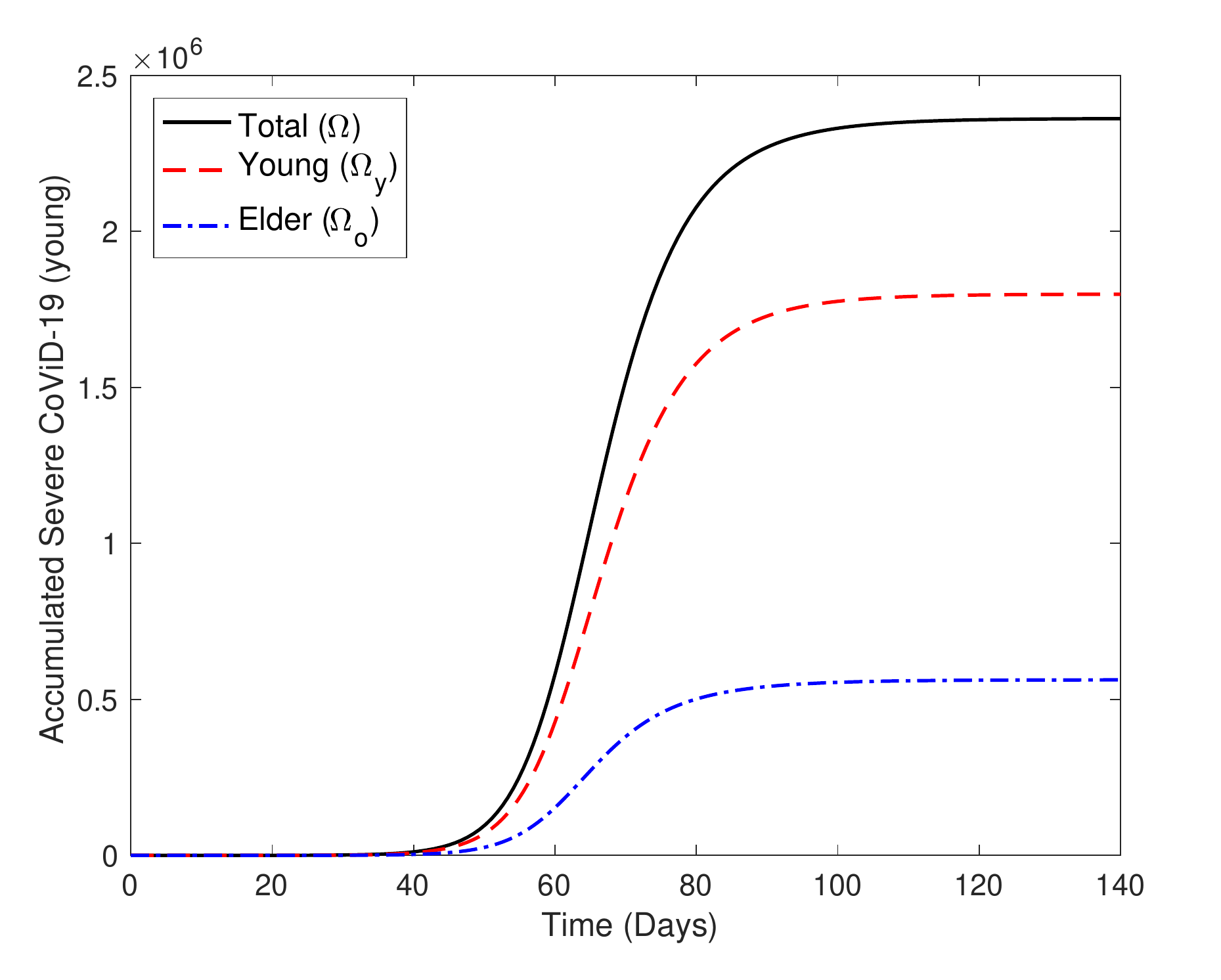}
\caption{The estimated curves of accumulated number of severe CoViD-19 ($%
\Omega _{y}$, $\Omega _{o}$ and $\Omega =\Omega _{y}+\Omega _{o}$) during
the first wave of epidemics.}%
\end{figure}

Figure 7 shows the estimated curves of accumulated number of CoViD-19 deaths
($\Pi _{y}$, $\Pi _{o}$ and $\Pi =\Pi _{y}+\Pi _{o}$), from equation (\ref%
{mort}). At $t=140$ days, $\Pi $ is approaching to plateau. The values of $%
\Pi _{y}$, $\Pi _{o}$ and $\Pi $ are at $t=140$, for the first method of
estimation, respectively, $0.6235\times 10^{5}$ ($3.47\%$), $1.883\times
10^{5}$ ($33.4\%$) and $2.507\times 10^{5}$ ($10.62\%$), and for the second
method of estimation, respectively, $1.60\times 10^{4}$ ($0.89\%$), $%
6.265\times 10^{4}$ ($11,13\%$) and $7.865\times 10^{4}$ ($3.33\%$).
Percentage between parentheses is the ratio $\Pi /\Omega $. The second
estimation method is shown in Figure 7.

\begin{figure}[!h]
\centering
\includegraphics[scale=0.55]{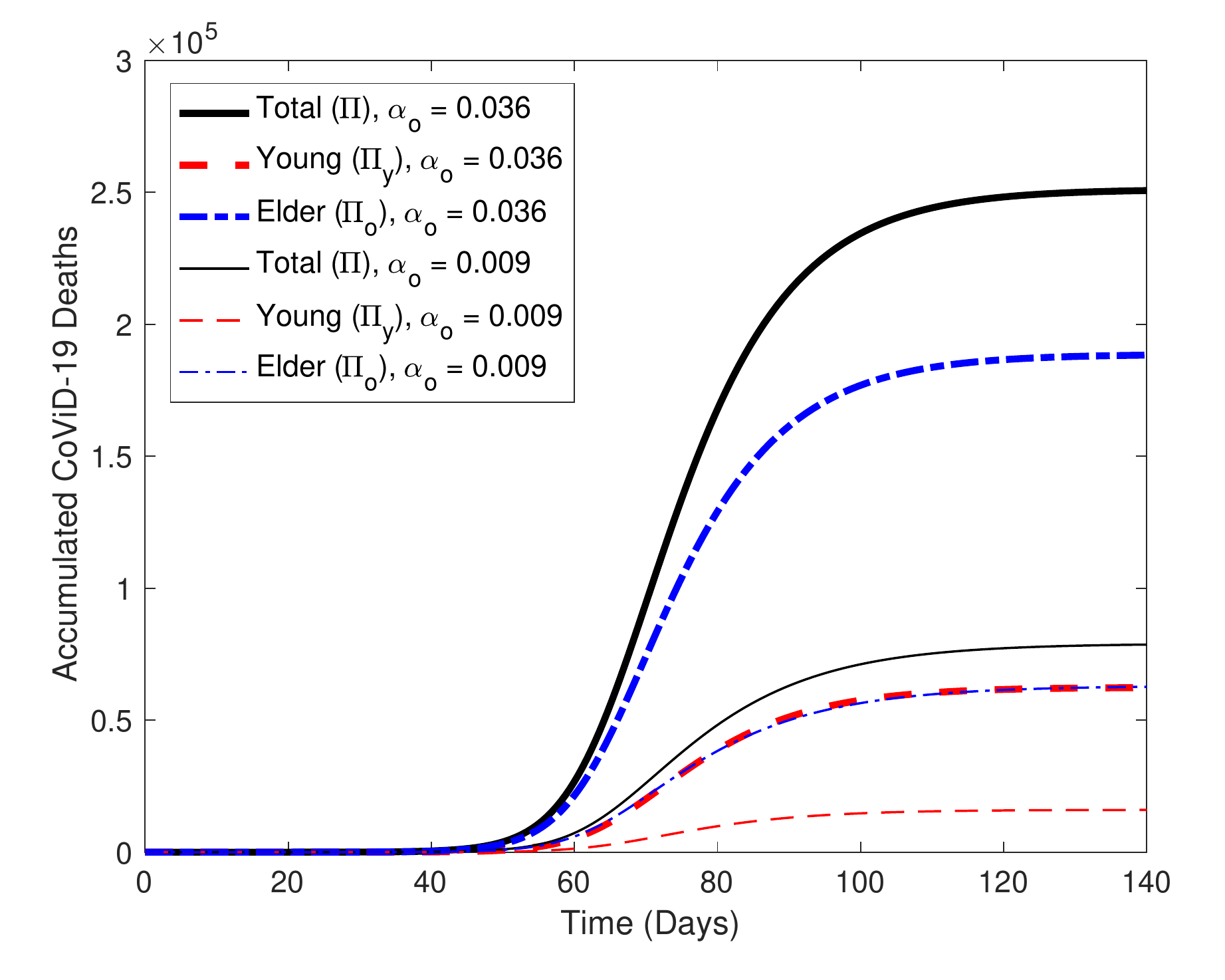}
\caption{The estimated curves of accumulated number of CoViD-19 deaths ($%
\Pi _{y}$, $\Pi _{o}$ and $\Pi =\Pi _{y}+\Pi _{o}$) during the first wave of
epidemics.}%
\end{figure}

By comparing percentages between deaths due to CoViD-19 ($\Pi $) and
accumulated severe CoViD-19 cases ($\Omega $), the first method predicts at
least $3$-times that predicted by the second method. Especially among elder
persons, second method predicts $11.13\%$, three times lower than $33.4\%$
predicted by the first method. Hence, the second estimation is more credible
than the first one. Hence, we will adopt the second estimation method for
additional mortality rates, $\alpha _{y}=0.0009$ and $\alpha _{o}=0.009$ ($%
days^{-1}$) hereafter except explicitly cited. Remember that additional
mortality rates are considered constant in all time.

Figure 8 shows the curves of the number of susceptible persons ($S_{y}$, $%
S_{o}$ and $S=S_{y}+S_{o}$). At $t=0$, the numbers of $S_{y}$, $S_{o}$ and $%
S $ are, respectively, $3.77762\times 10^{7}$, $0.68238\times 10^{7}$ and $%
4.46\times 10^{7}$, and diminish due to infection, to lower values at $t=140$
days. Notice that, after the first wave of epidemics, very few number of
susceptible persons are left behind, which are $1.23880\times 10^{5}$ ($%
0.33\%$), $0.02643\times 10^{5}$ ($0.039\%$) and $1.26523\times 10^{5}$ ($%
0.28\%$), for young, elder and total persons, respectively. Percentage
between parentheses is the ratio $S(140)/S(0)$.

\begin{figure}[!h]
\centering
\includegraphics[scale=0.55]{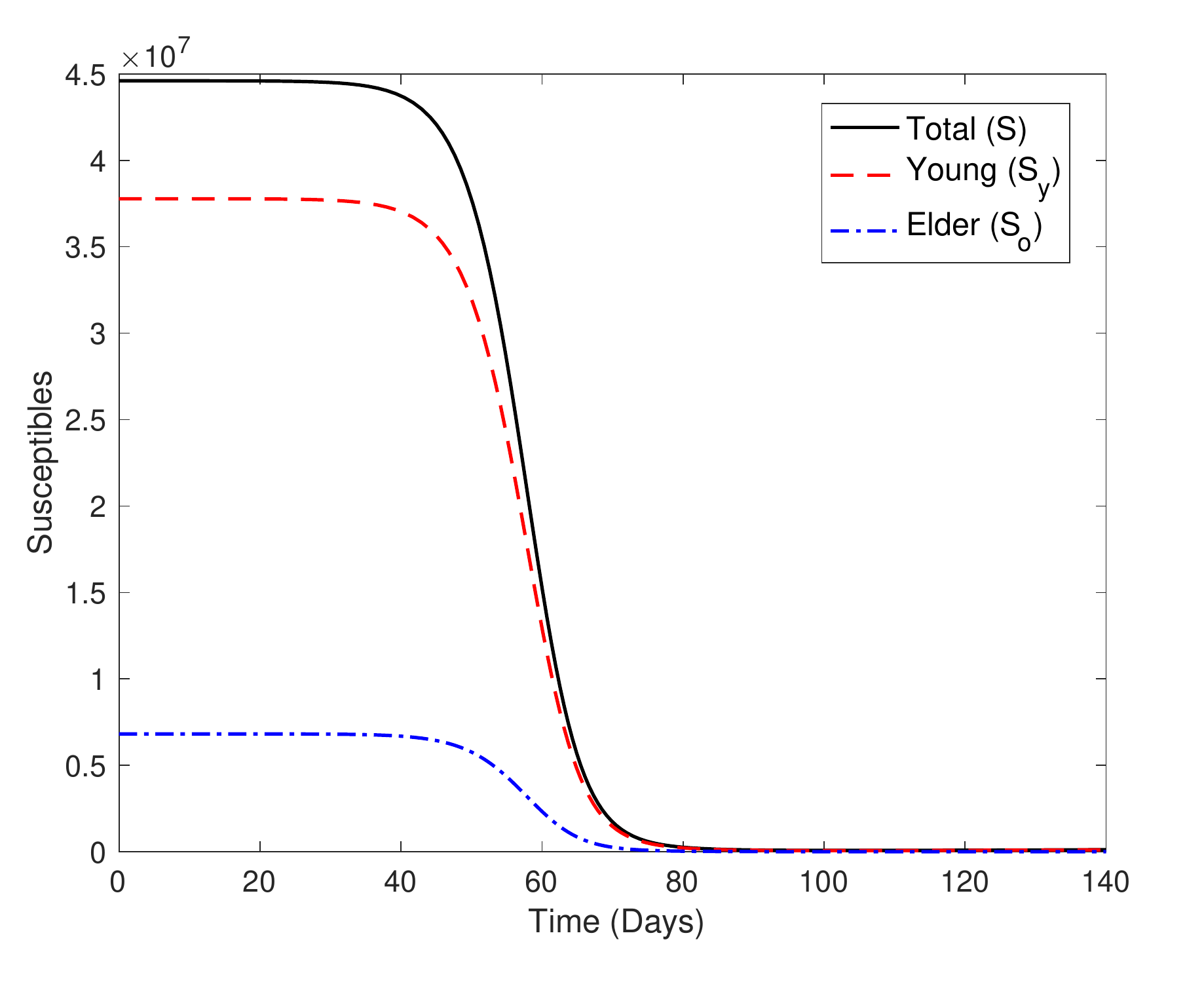}
\caption{The curves of the number of susceptible persons ($S_{y}$, $S_{o}$
and $S=S_{y}+S_{o}$) during the first wave of epidemics.
}%
\end{figure}

Figure 9 shows the curves of the number of immune persons ($I_{y}$, $I_{o}$
and $I=I_{y}+I_{o}$). At $t=0$, the number of immune persons $I_{y}$, $I_{o}$
and $I$ increase from zero to, respectively, $3.76156\times 10^{7}$ ($%
99.57\% $), $0.67234\times 10^{7}$ ($98.53\%$) and $4.43390\times 10^{7}$ ($%
99.41\%$) at $t=140$ days. Percentage between parentheses is the ratio $%
I/S(0)$.

\begin{figure}[!h]
\centering
\includegraphics[scale=0.55]{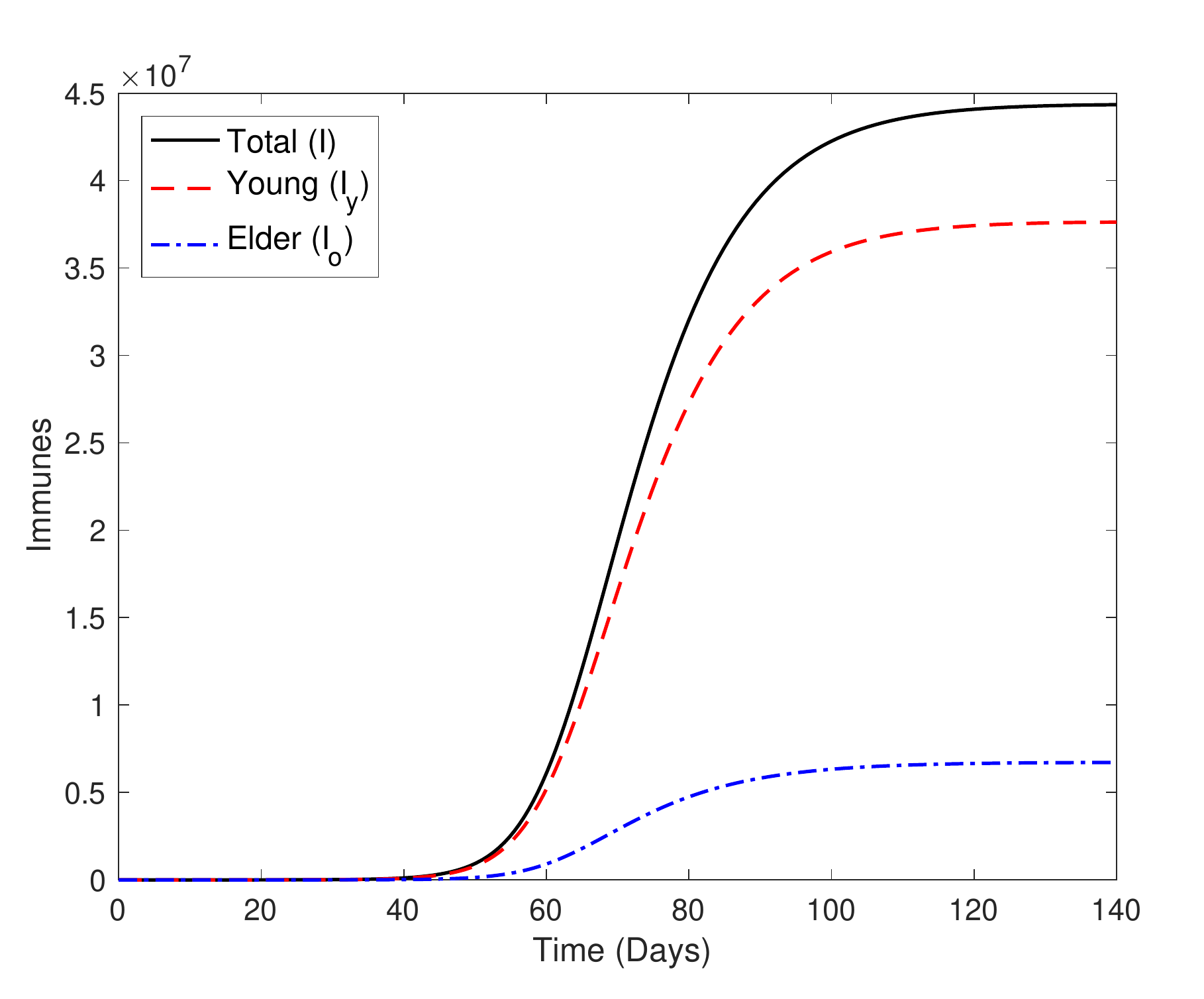}
\caption{The curves of the number of immune persons ($I_{y}$, $I_{o}$ and $%
I=I_{y}+I_{o}$) during the first wave of epidemics.}%
\end{figure}

From Figures 8 and 9, the difference between percentages of $I/S(0)$ and $%
S(140)/S(0)$ is the percentage of all persons who have had contact with new
coronavirus. Hence, the second wave of epidemics will be triggered after
elapsed very long period time waiting the accumulation of susceptible
persons to surpass its critical number \cite{yang9} \cite{yang8}. Simulating
the system of equations (\ref{system2a}), (\ref{system1c}) and (\ref%
{system2b}) for a very long time (figures not shown), the trajectories reach
the equilibrium values for susceptible persons ($s_{y}^{\ast }=S_{y}^{\ast
}/N^{\ast }=0.14660$, $s_{o}^{\ast }=S_{o}^{\ast }/N^{\ast }=0.00348$ and $%
s^{\ast }=s_{y}^{\ast }+s_{y}^{\ast }=0.15008$).

Let us estimate roughly the critical number of susceptible persons $S^{th}$
from equation (\ref{scrit}). For $R_{0}=6.915$, $S^{th}=6.450\times 10^{6}$.
Hence, for the S\~{a}o Paulo State, isolating $38.15$ million ($85.5\%$) or
above persons is necessary to avoid persistence of epidemics. The number of
young persons is $3.5$ million less than the threshold number of isolated
persons to guarantee eradication of CoViD-19. Another rough estimation is
done to isolation rate of susceptible persons $\eta _{2}$, letting $\eta
_{3}=0$ in equation (\ref{etacrit}), resulting in $\eta ^{th}=2.19\times
10^{-4}$ $years^{-1}$, for $R_{0}=6.915$. Then, for $\eta >\eta ^{th}$ the
new coronavirus epidemics fades out.

\subsection{Epidemiological scenarios considering control mechanisms}

Using estimated transmission and additional mortality rates, we solve
numerically the system of equations (\ref{system2a}), (\ref{system1c}) and (%
\ref{system2b}) considering only one control mechanism, that is, the
isolation, due to the fact that there is few number of testing kits, and
treatment and vaccine are not available yet.

In this section we fix the estimated transmission rates as $\beta _{y}=0.77$
and $\beta _{o}=\psi \beta _{y}=0.9009$ ($days^{-1}$), and the additional
mortality rates, $\alpha _{y}=0.0009$ and $\alpha _{o}=0.009$ ($days^{-1}$).

We consider two cases: Isolation without subsequent releasing of isolated
persons, and isolation followed by releasing of these persons. By varying
isolation parameters $\eta _{2y}$ and $\eta _{2o}$, and releasing parameters 
$\eta _{3y}$ and $\eta _{3o}$, we present some epidemiological scenarios. In
all scenarios, $t$ is simulation time, instead of calendar time.

\subsubsection{Scenarios -- Isolation without releasing ($\protect\eta _{3y}=%
\protect\eta _{3o}=0$)}

At $t=0$ (February 26) the first case of severe CiViD-19 was confirmed, and
at $t=27$ (March 24) isolation as mechanism of control (described by $\eta
_{2y}$ and $\eta _{2o}$) was introduced until April 22. We analyze two
cases. First, there is indiscriminated isolation for young and elder
persons, hence we assume that the same rates of isolation are applied to
young and elder persons, that is, $\eta _{2}=\eta _{2y}=\eta _{2o}$.
Further, there is discriminated (preferential) isolation of elder persons,
hence we assume that $\eta _{2o}\neq \eta _{2y}$.

\paragraph{Regime 1 -- Equal isolation of young and elder persons ($\protect%
\eta _{2}=\protect\eta _{2y}=\protect\eta _{2o}$)}

In regime 1, we call equal isolation of young and elder persons in the sense
of equal isolation rates. Recalling that $\eta _{2y}$ and $\eta _{2o}$ are
per-capita rates, both rates isolate proportionally young and elder persons,
but the actual number of isolation is higher among young persons.

We choose 7 different values for the isolation rate $\eta _{2}$ ($days^{-1}$%
) applied to young and elder persons. The values for $\eta _{2}$: $0.00021$ (%
$R_{r}=1$), $0.001$ ($R_{r}=0.23$), $0.005$ ($R_{r}=0.048$), $0.01$ ($%
R_{r}=0.024$), $0.015$ ($R_{r}=0.016$), $0.025$ ($R_{r}=0.009$) and $0.035$ (%
$R_{r}=0.007$). The value for the reduced reproduction number is $R_{r}$ is
calculated from equation (\ref{Rred}). For $\eta _{2}=0.035$, the reduced
reproduction number with respect to the basic reproduction number is reduced
in $0.1\%$. In all figures, the case $\eta _{2}=0$ ($R_{0}=6.915$) is also
shown.

Figure 10 shows curves of severe cases of CoViD-19 $D_{2j}$, $j=y,o$,
without and with isolation for different values of $\eta _{2}$. Notice that
first two curves obtained with $\eta _{2}=0$ and $0.00021$ practically
coincide, and the latter is slightly lower than the roughly estimated $\eta
^{th}=2.19\times 10^{-4}$ $years^{-1}$. We present values of peak for three
values of $\eta _{2}$. For $\eta _{2}=0$, the peak of young (first
coordinate) and elder (second coordinate) persons are ($5.532\times 10^{5}$,$%
2.061\times 10^{5}$), and for $\eta _{2}=0.01$ ($3.566\times 10^{5}$,$%
1.361\times 10^{5}$), and $0.035$ ($0.699\times 10^{5}$,$0.292\times 10^{5}$%
. The time ($days$) at which the peak occurs for young (first coordinate)
and elder (second coordinate) persons are for $\eta _{2}=0$ ($72$,$71$), $%
0.01$ ($75$,$74$) and $0.035$ ($77$,$77$). For $\eta _{2}=0.01$ in
comparison with $\eta _{2}=0$, the peaks are reduced in $64.4\%$ and $66.0\%$%
, respectively, for young and elder persons. For $\eta _{2}=0.035$, the
peaks are reduced in $12.6\%$ and $14.2\%$.

\begin{figure}[!h]
\centering
\subfloat[]{
\includegraphics[scale=0.45]{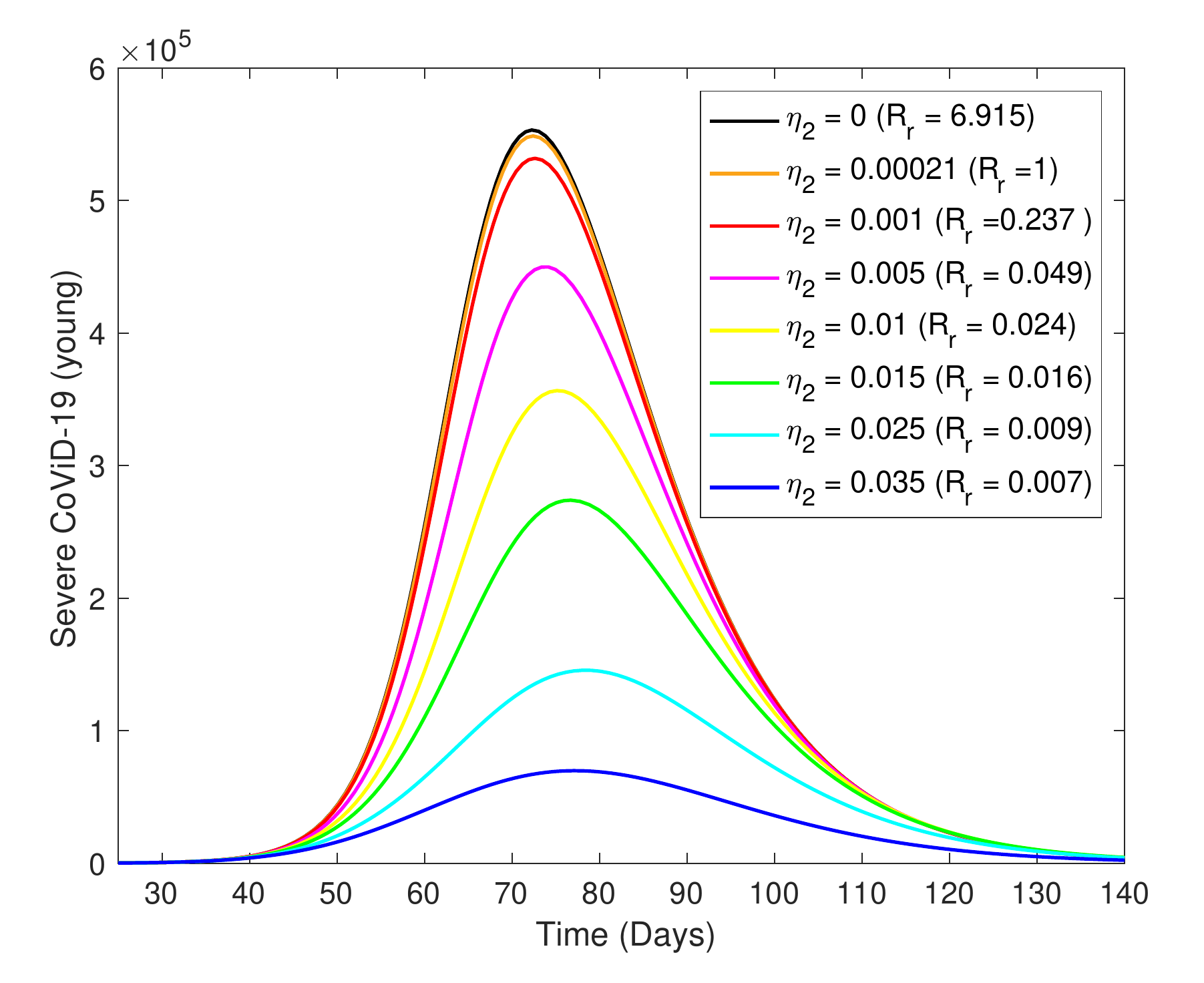}
}
\subfloat[]{
\includegraphics[scale=0.45]{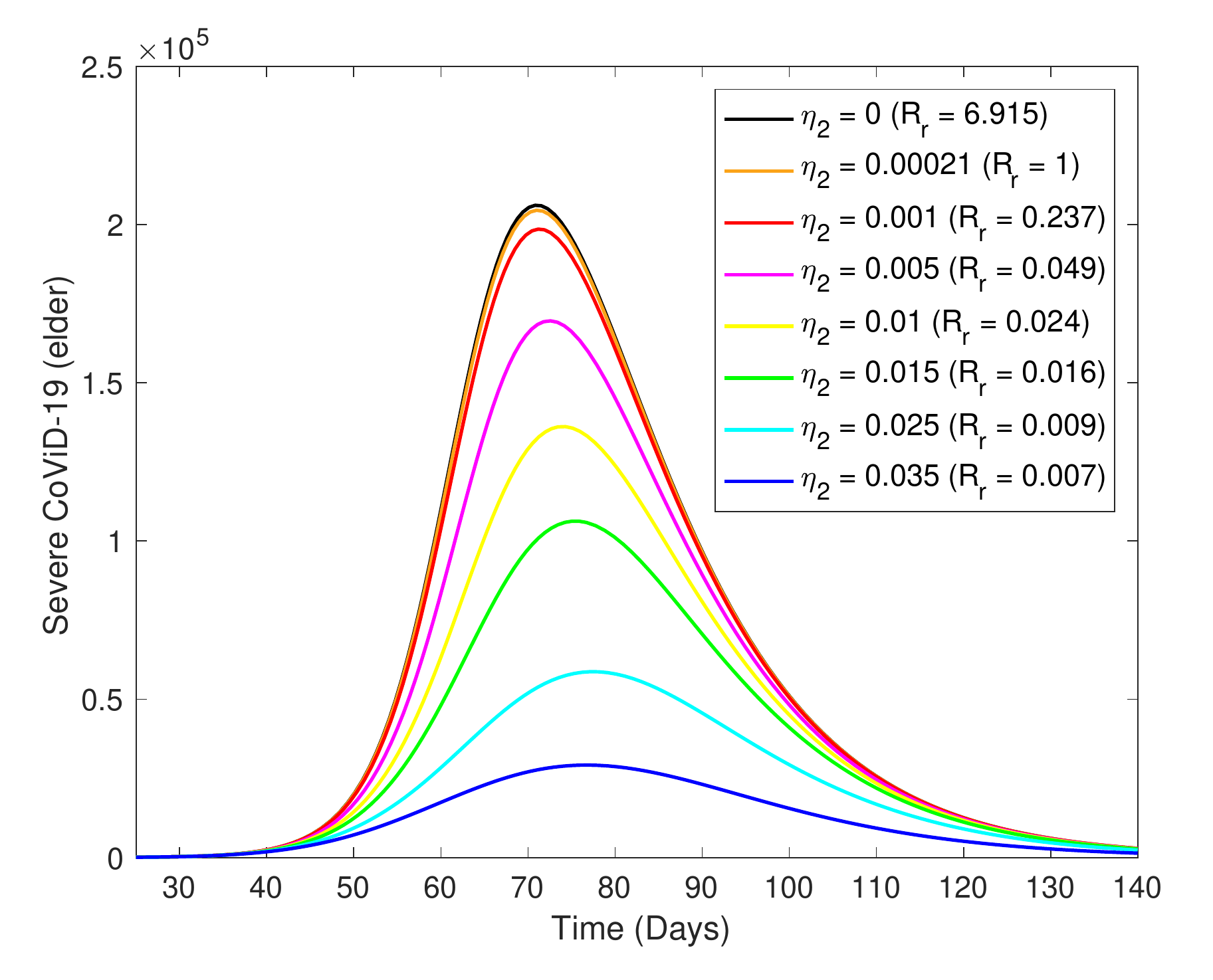}
}
\caption{The curves of severe cases of CoViD-19 $D_{2j}$, $j=y,o$, without
and with isolation for different values of $\eta _{2}$. Curves from top to
bottom corresponds to increasing $\eta _{2}$.}
\end{figure}

As isolation parameter $\eta _{2}$ increases, the diminishing peaks of
curves of $D_{2y}$ and $D_{2o}$ displace initially to right (higher times),
but at $\eta _{2}=\eta _{2}^{c}$, they change the direction and move
leftwardly. However, all curves remain inside the curve without isolation ($%
\eta _{2}=0$). The values at which the peaks change direction are $\eta
_{2y}^{c}=0.0027$ $days^{-1}$ ($t=78.35$) and $\eta _{2o}^{c}=0.0028$ $%
days^{-1}$ ($t=77.58$). In order to understand this phenomenon, we recall an
age-structured model to describe rubella infection \cite{yang5} \cite{yang6}%
. There, as vaccination rate increases, the peaks of age-depending forces of
infection initially moves to right, and, then, move leftwardly. As a
consequence, the average age at the first infection increases.

At $t=27$ isolation begun in the S\~{a}o Paulo State. For this reason, in
the system of equations (\ref{system2a}), (\ref{system1c}) and (\ref%
{system2b}), we let $\eta _{2}=0$ for $t<27$, and $\eta _{2}>0$ for $t\geq
27 $. In Figure 11 we show the estimated curves of severe CoViD-19 cases $%
D_{2}$ without ($\eta _{2}=0$ in all time) and with ($\eta _{2}=0.035$ $%
days^{-1}$) isolation, which was introduced at $t=27$. It seems that the
effects of isolation (in observed data) appears at around $t=38$ (April 5),
11 days after its introduction. Figure 11 shows an isolation scheme
described by $\eta _{2}=0.035$ $days^{-1}$ introduced at $t=27$, which
decreases the curve without isolation. The transition from without to with
isolation is under very complex dynamics, for this reason we can not assure
that $\eta _{2}=0.035$ $days^{-1}$ is a good estimation (there are so few
data). Hence, one of the curves in Figure 10 may correspond to the isolation
applied in the S\~{a}o Paulo State.

\begin{figure}[!h]
\centering
\includegraphics[scale=0.55]{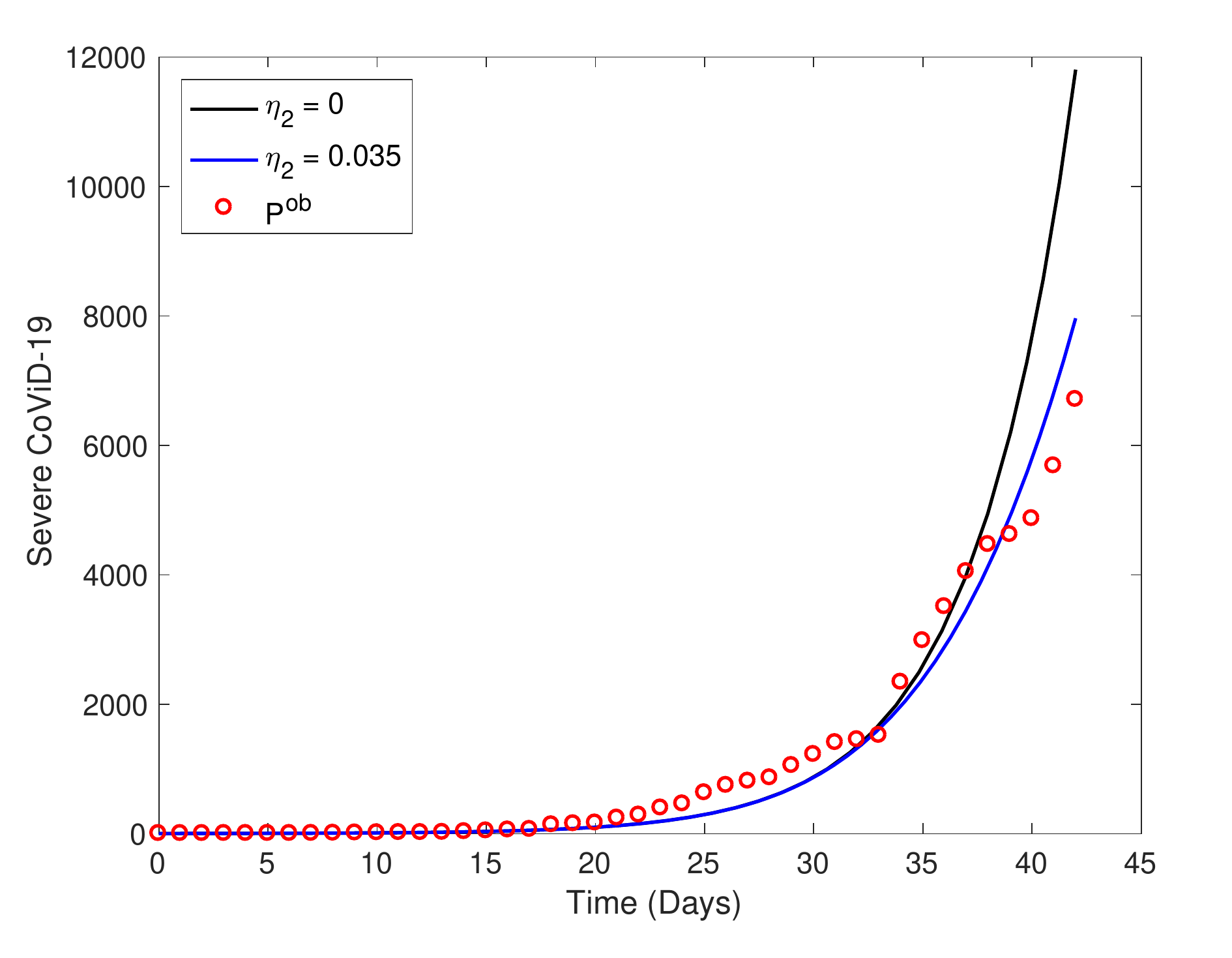}
\caption{The curves of an isolation scheme described by $\eta _{2}=0.035$ $%
days^{-1}$ introduced at $t=27$, and the curve without isolation.}
\end{figure}

The curve corresponding to $\eta _{2}=0.00021$ $days^{-1}$ in Figure 10 can
be considered as a failure isolation ($R_{r}>1$), for this reason this curve
is removed in all following figures.

Figure 12 shows curves of accumulated cases of severe CoViD-19 $\Omega _{j}$%
, $j=y,o$, without and with isolation for different values of $\eta _{2}$.
As isolation rate $\eta _{2}$ increases, the accumulated number of deaths
due to severe CoViD-19 decreases. We present at $t=140$ for three values of $%
\eta _{2}$. For $\eta _{2}=0$, the number of young (first coordinate) and
elder (second coordinate) persons are ($1.798\times 10^{6}$,$5.630\times
10^{5}$), and for $\eta _{2}=0.01$ ($1.278\times 10^{6}$,$4.063\times 10^{5}$%
), and $0.035$ ($0.372\times 10^{6}$,$1.133\times 10^{5}$). For $\eta
_{2}=0.01$ in comparison with $\eta _{2}=0$, severe CoViD-19 cases are
reduced in $71.1\%$ and $72.2\%$, respectively for young and elder persons.
For $\eta _{2}=0.035$, severe CoViD-19 cases are reduced in $20.7\%$ and $%
17.4\%$.

\begin{figure}[!h]
\centering
\subfloat[]{
\includegraphics[scale=0.45]{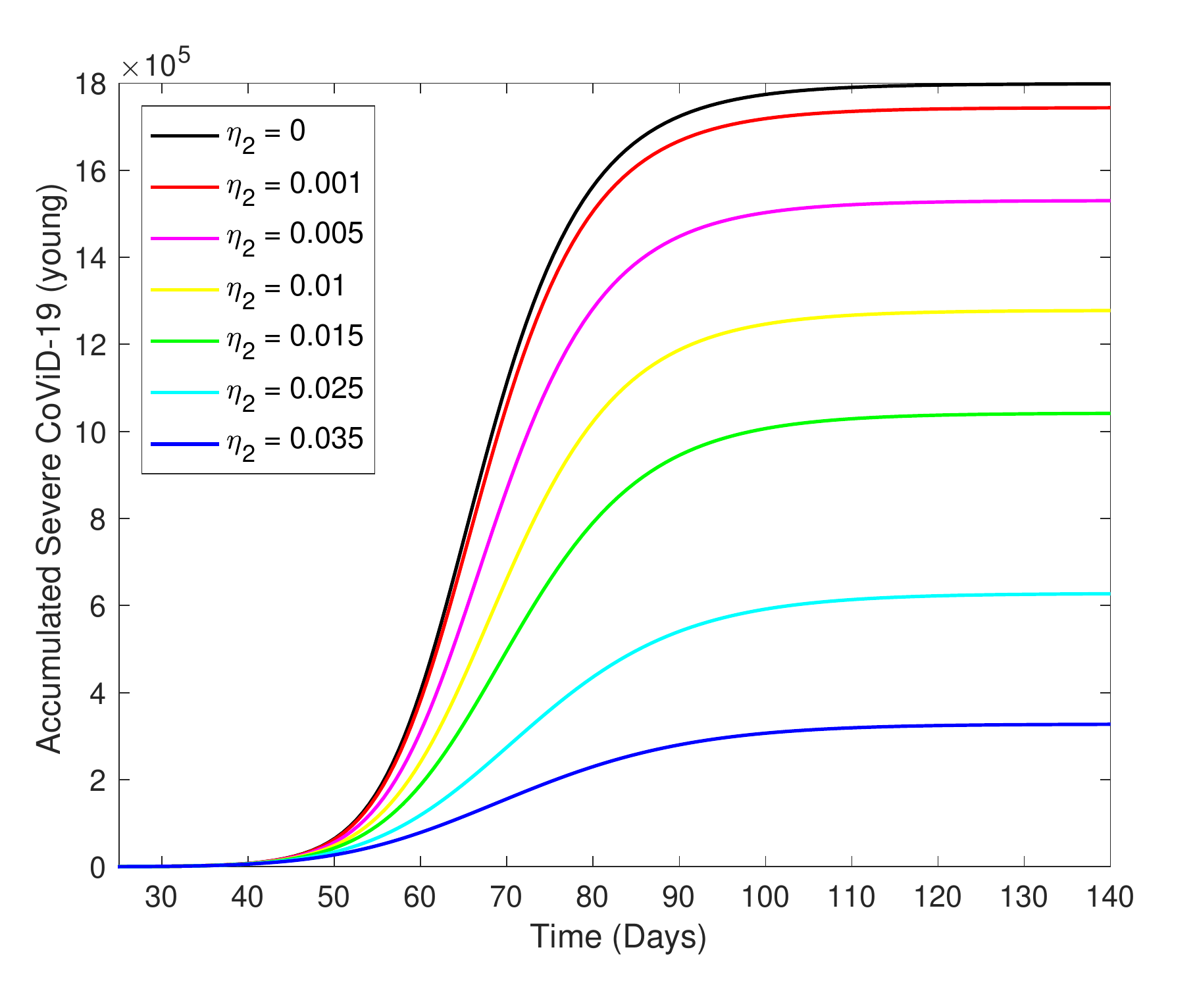}
}
\subfloat[]{
\includegraphics[scale=0.45]{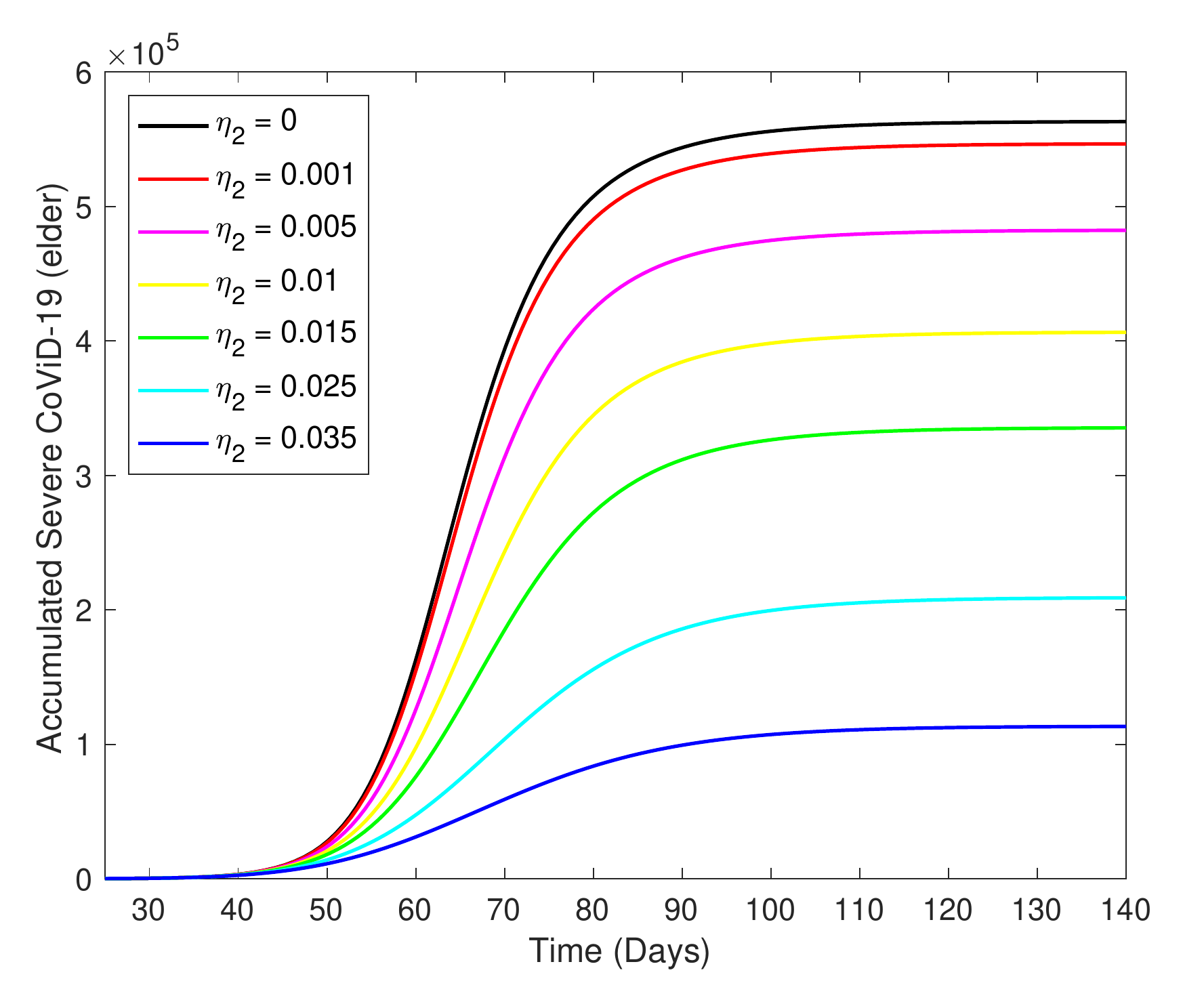}
}
\caption{The curves of accumulated cases of severe CoViD-19 $\Omega _{j}$, 
$j=y,o$, without and with isolation for different values of $\eta _{2}$.
Curves from top to bottom corresponds to increasing $\eta _{2}$. The
beginning of isolation is at $t=27$.}
\end{figure}

Figure 13 shows curves of accumulated cases of CoViD-19 deaths $\Pi _{j}$, $%
j=y,o$, without and with isolation for different values of $\eta _{2}$. We
present at $t=140$ for three values of $\eta _{2}$. For $\eta _{2}=0$, the
number of young (first coordinate) and elder (second coordinate) persons are
($1.6\times 10^{4}$,$6.265\times 10^{4}$), and for $\eta _{2}=0.01$ ($%
1.135\times 10^{4}$,$5.514\times 10^{4}$), and $0.035$ ($0.29\times 10^{4}$,$%
1.252\times 10^{4}$). For $\eta _{2}=0.01$ in comparison with $\eta _{2}=0$,
death due to CoViD-19 cases are reduced in $70.9\%$ and $88.0\%$,
respectively for young and elder persons. For $\eta _{2}=0.035$, death due
to CoViD-19 cases are reduced in $18.1\%$ and $20.0\%$.

\begin{figure}[!h]
\centering
\subfloat[]{
\includegraphics[scale=0.45]{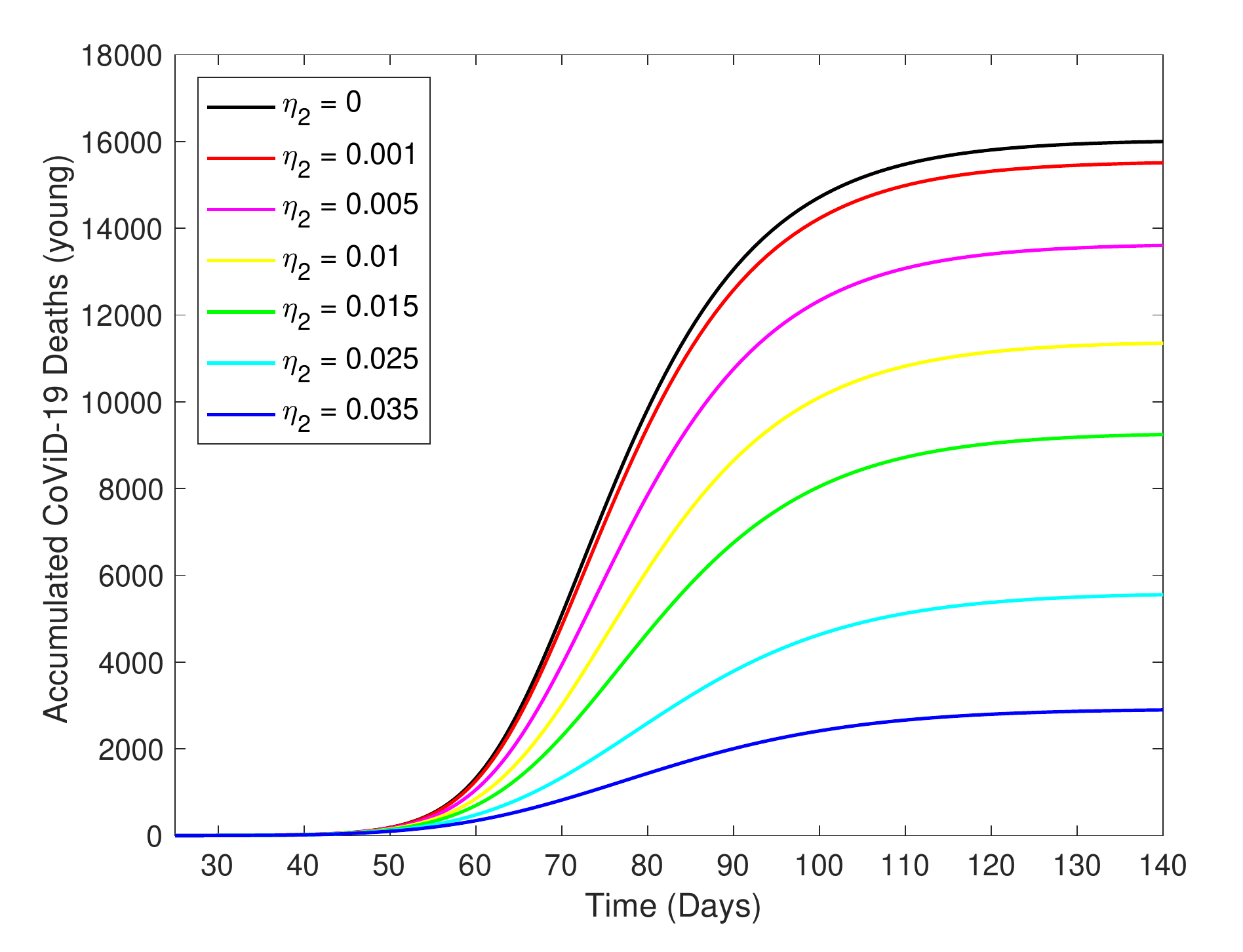}
}
\subfloat[]{
\includegraphics[scale=0.45]{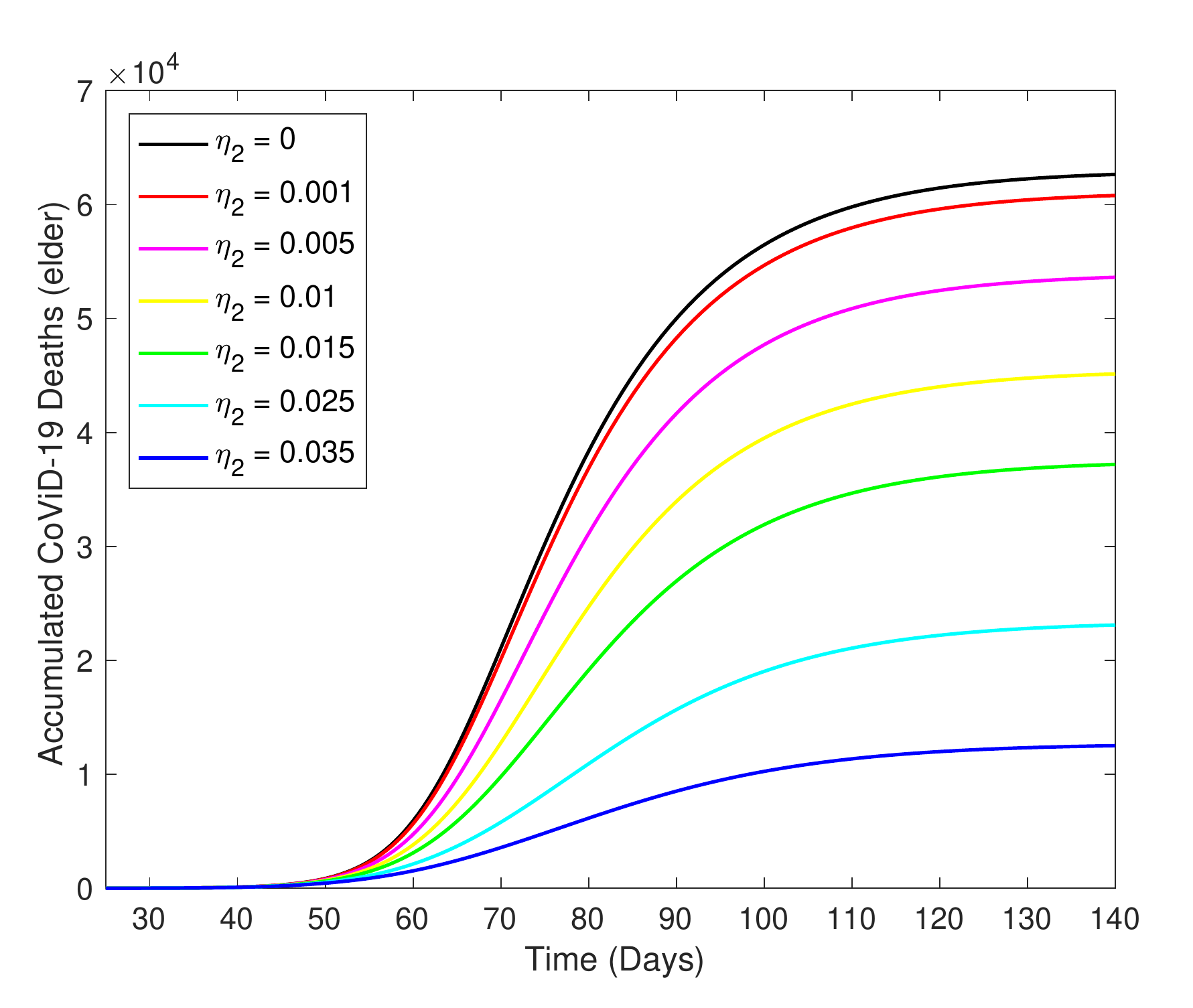}
}
\caption{The curves of accumulated cases of CoViD-19 deaths $\Pi _{j}$, $%
j=y,o$, without and with isolation for different values of $\eta _{2}$.
Curves from top to bottom corresponds to increasing $\eta _{2}$. The
beginning of isolation is at $t=27$.}
\end{figure}

Figure 14 shows curves of the number of susceptible persons $S_{j}$, $j=y,o$%
, without and with isolation for different values of $\eta _{2}$. We present
at $t=140$ for three values of $\eta _{2}$. For $\eta _{2}=0$, the number of
young (first coordinate) and elder (second coordinate) persons are ($%
1.239\times 10^{5}$,$2463$), and for $\eta _{2}=0.01$ ($1.634\times 10^{5}$,$%
8190$), and $0.035$ ($2.492\times 10^{5}$,$60620$). For $\eta _{2}=0.01$ in
comparison with $\eta _{2}=0$, susceptible persons are increased in $132\%$
and $333\%$, respectively for young and elder persons. For $\eta _{2}=0.035$%
, susceptible persons are increased in $201\%$ and $2,461\%$.

\begin{figure}[!h]
\centering
\subfloat[]{
\includegraphics[scale=0.45]{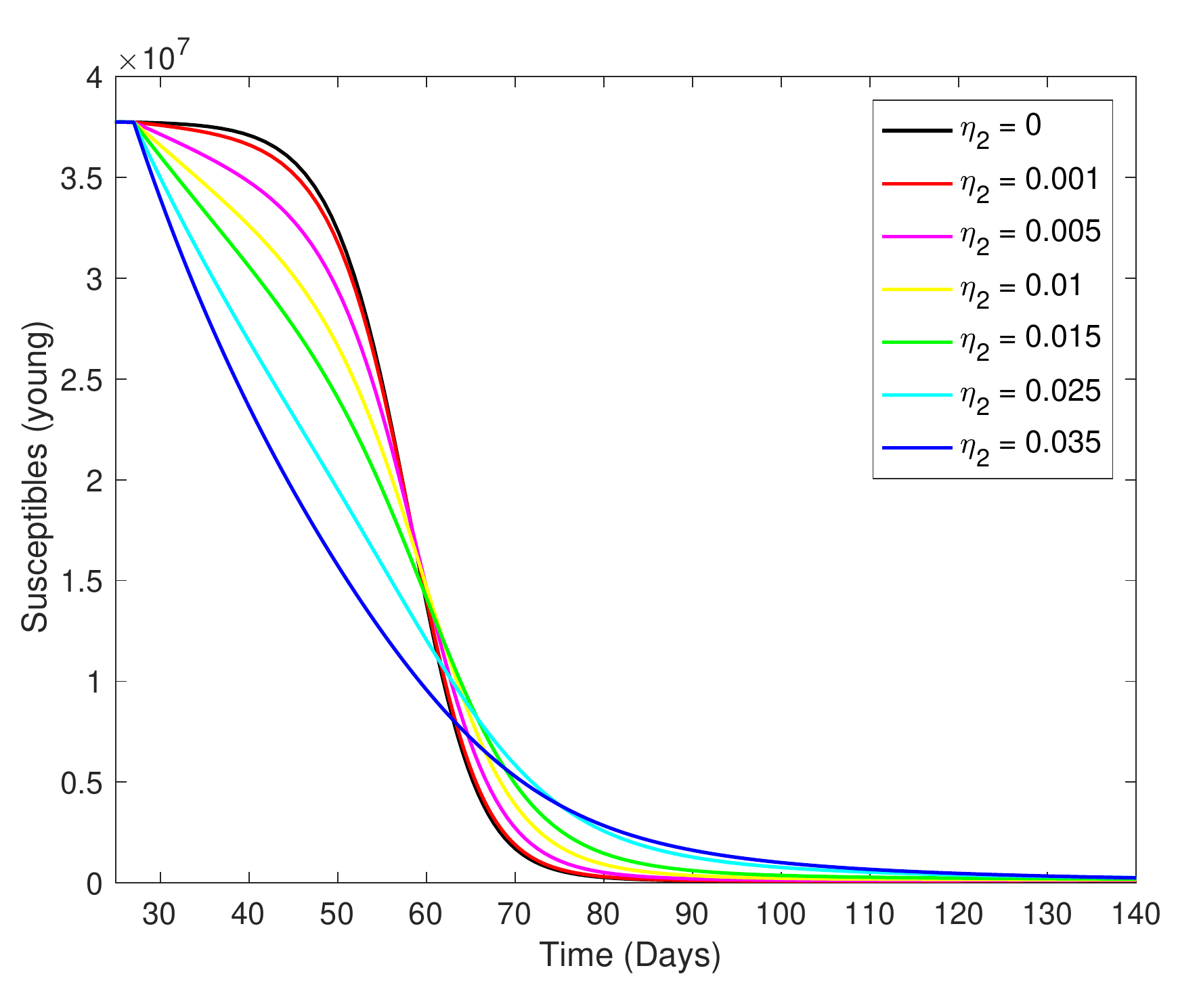}
}
\subfloat[]{
\includegraphics[scale=0.45]{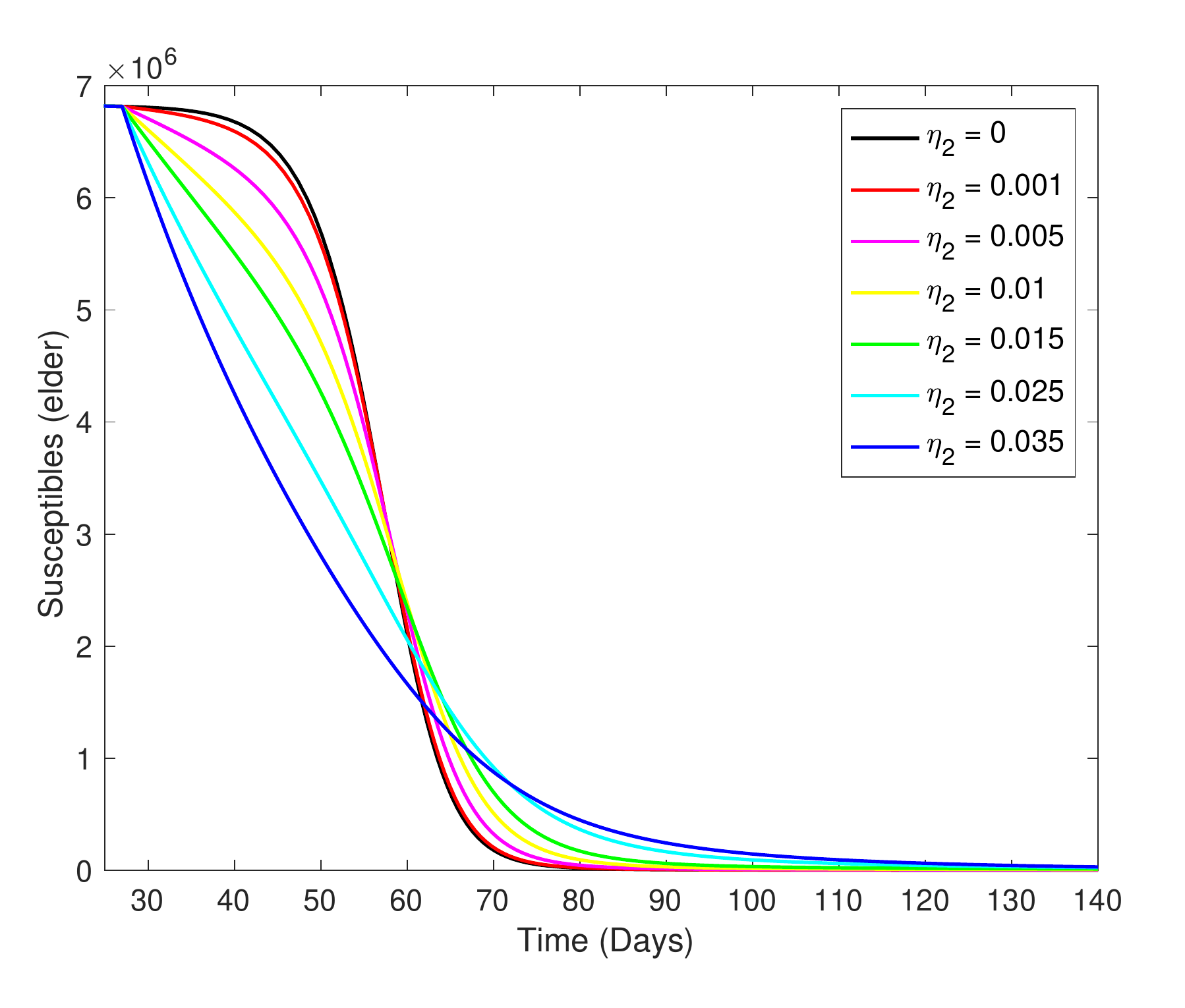}
}
\caption{The curves of the number of susceptible persons $S_{j}$, $j=y,o$,
without and with isolation for different values of $\eta _{2}$. Curves from
top to bottom corresponds to increasing $\eta _{2}$. The beginning of
isolation is at $t=27$.}
\end{figure}

As isolation parameters $\eta _{2}$ increases, the number of susceptible
persons decreases according to sigmoid shape, but, at a sufficient higher
value, follows exponential decay. Again, this phenomenon is \ understood
recalling rubella transmission model \cite{yang8}. There, as vaccination
rate increases, the fraction of susceptible persons decreases following
damped oscillations when $R_{r}>1$, attaining non-trivial equilibrium point.
However, for $R_{r}<1$, there is trivial equilibrium point and trajectories
follows two pattern: (1) if $R_{r}\ $is not so low, the fraction of
susceptible persons decreases lower than the value of trivial equilibrium
point, and must increase to attain the equilibrium value, but not surpassing
it (then there is not damped oscillations); and (2) if $R_{r}\ $is low, the
fraction of susceptible persons decreases never lower than the value of
trivial equilibrium point, for this reason attains this equilibrium value
decaying exponentially without surpassing it in any time.

Figure 15 shows curves of the number of isolated susceptible persons $%
S_{j}^{is}$, $j=y,o$, with isolation for different values of $\eta _{2}$,
from equation (\ref{propotion}). We present at $t=140$ for three values of $%
\eta _{2}$. For $\eta _{2}=0$, there is not isolated persons, and for $\eta
_{2}=0.01$ ($1.09\times 10^{7}$,$1.892\times 10^{6}$), and $0.035$ ($%
3.079\times 10^{7}$,$5.419\times 10^{6}$). For $\eta _{2}=0.01$ in
comparison with all persons $N_{0}$ (at $t=0$), isolated susceptible persons
are $2.4\%$ and $0.42\%$, respectively for young and elder persons. For $%
\eta _{2}=0.035$, isolated susceptible persons are $6.9\%$ and $1.22\%$.

\begin{figure}[!h]
\centering
\subfloat[]{
\includegraphics[scale=0.45]{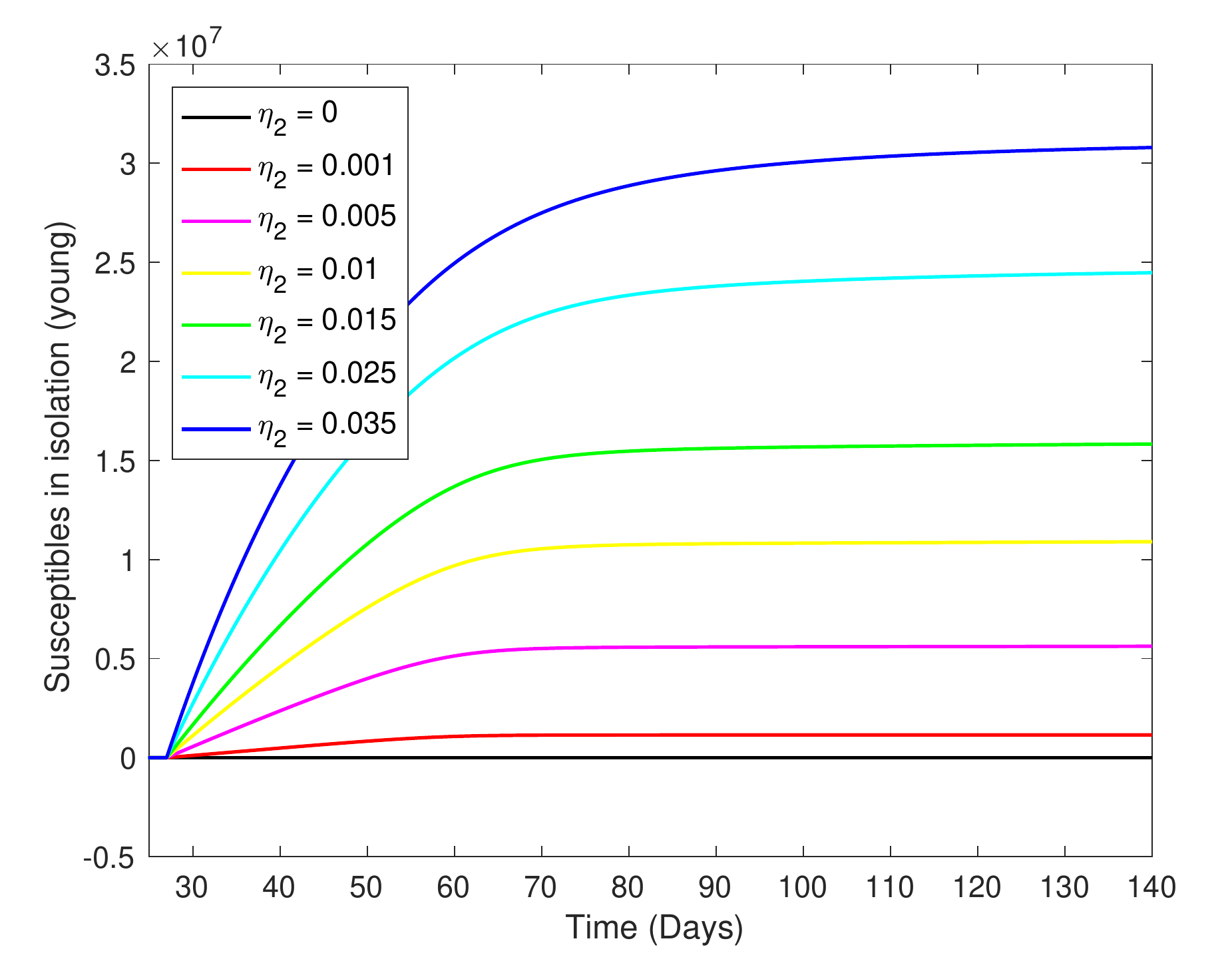}
}
\subfloat[]{
\includegraphics[scale=0.45]{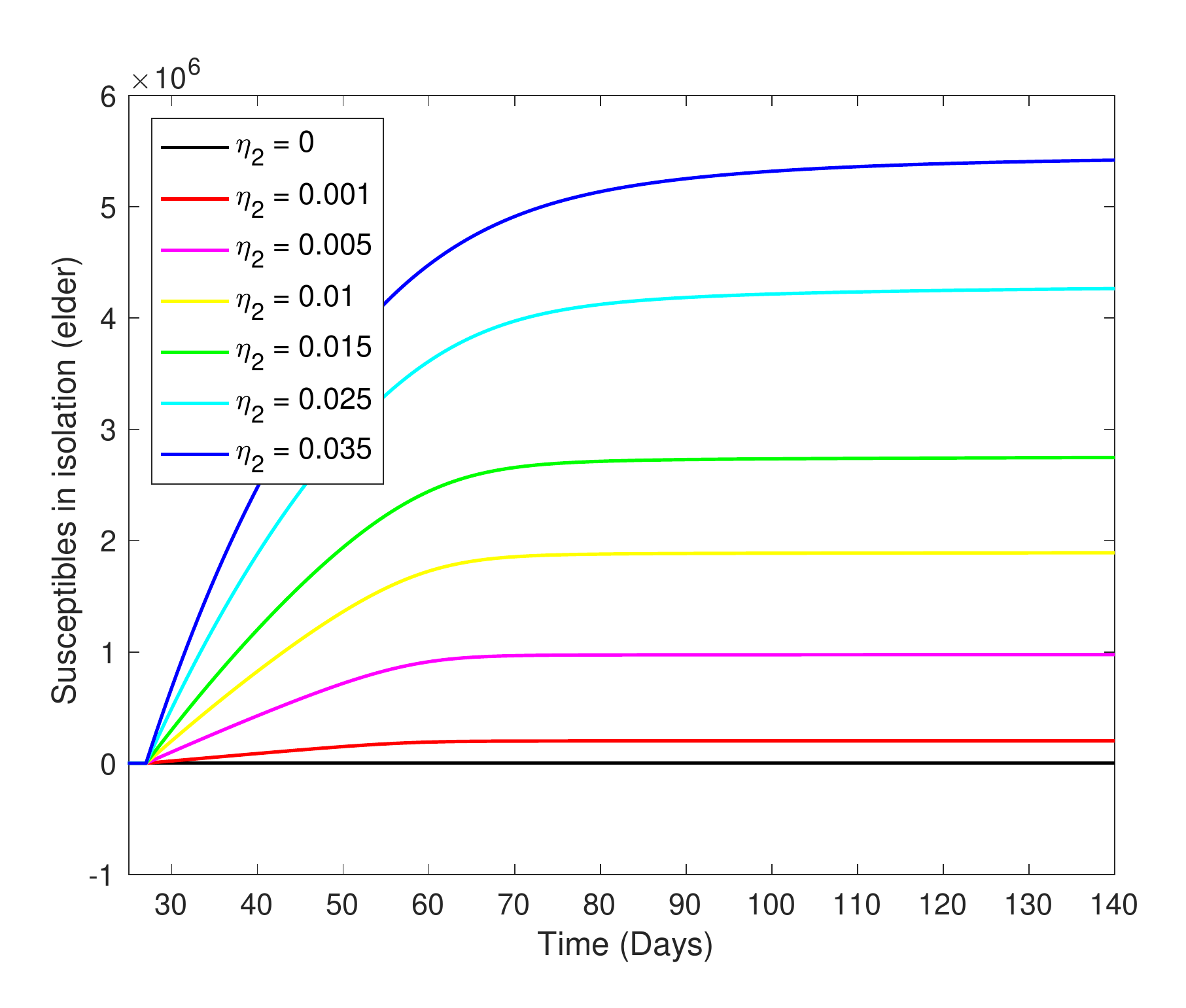}
}
\caption{The curves of the number of isolated susceptible persons $%
S_{j}^{is}$, $j=y,o$, with isolation for different values of $\eta _{2}$.
Curves from top to bottom corresponds to increasing $\eta _{2}$. The
beginning of isolation is at $t=27$.}
\end{figure}

Figure 16 shows curves of the number of immune persons $I_{j}$, $j=y,o$,
without and with isolation for different values of $\eta _{2}$. We present
at $t=140$ for three values of $\eta _{2}$. For $\eta _{2}=0$, the number of
young (first coordinate) and elder (second coordinate) persons are ($%
3.762\times 10^{7}$,$6.723\times 10^{6}$), and for $\eta _{2}=0.01$ ($%
2.671\times 10^{7}$,$4.849\times 10^{6}$), and $0.035$ ($0.683\times 10^{7}$,%
$1.349\times 10^{6}$). For $\eta _{2}=0.01$ in comparison with $\eta _{2}=0$%
, immune persons are reduced to $71.0\%$ and $72.1\%$, respectively for
young and elder persons, very close to the reductions observed in deaths due
to CoViD-19. For $\eta _{2}=0.035$, immune persons are reduced to $18.1\%$
and $20.0\%$, very close to the reductions observed in deaths due to
CoViD-19.

\begin{figure}[!h]
\centering
\subfloat[]{
\includegraphics[scale=0.45]{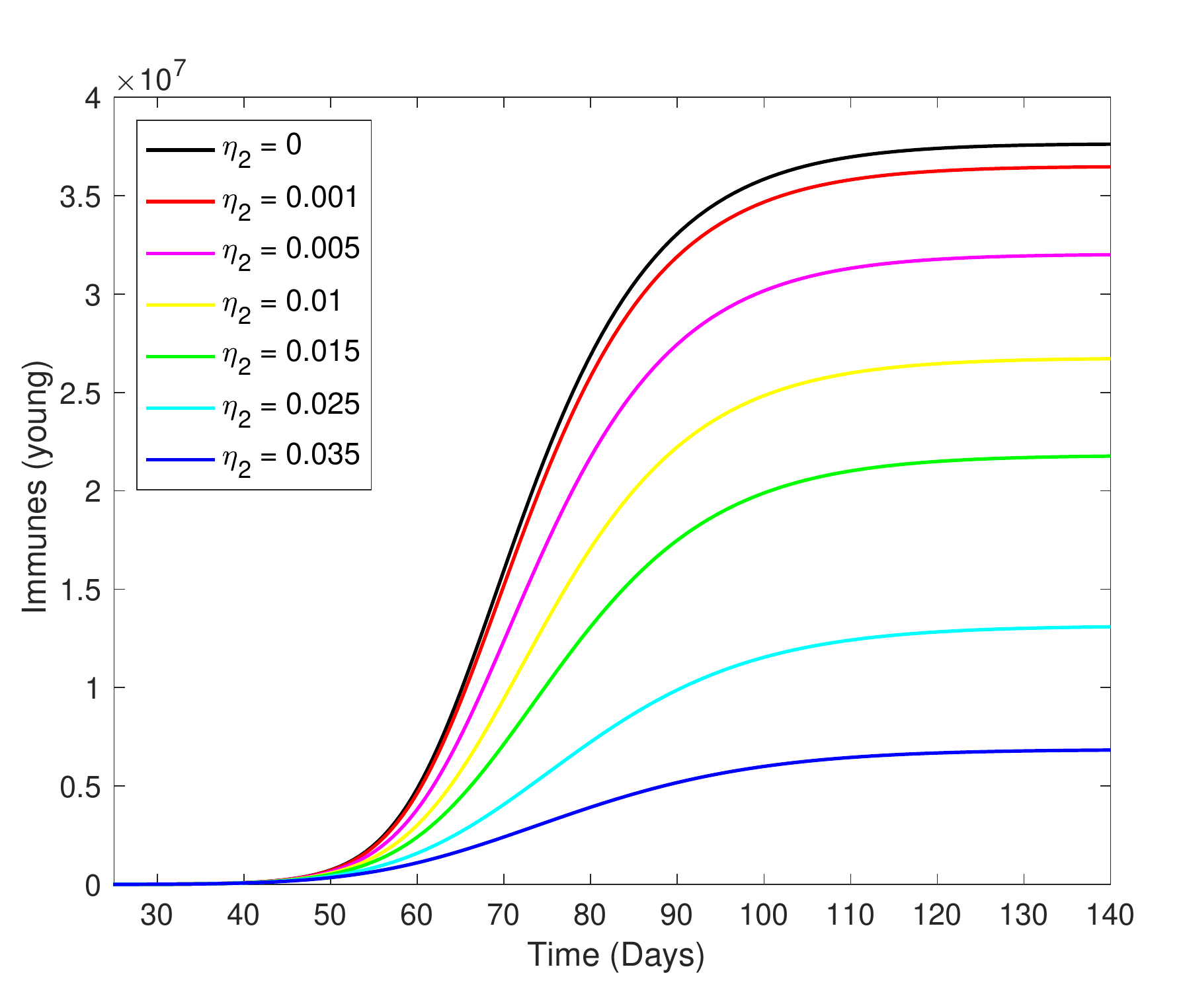}
}
\subfloat[]{
\includegraphics[scale=0.45]{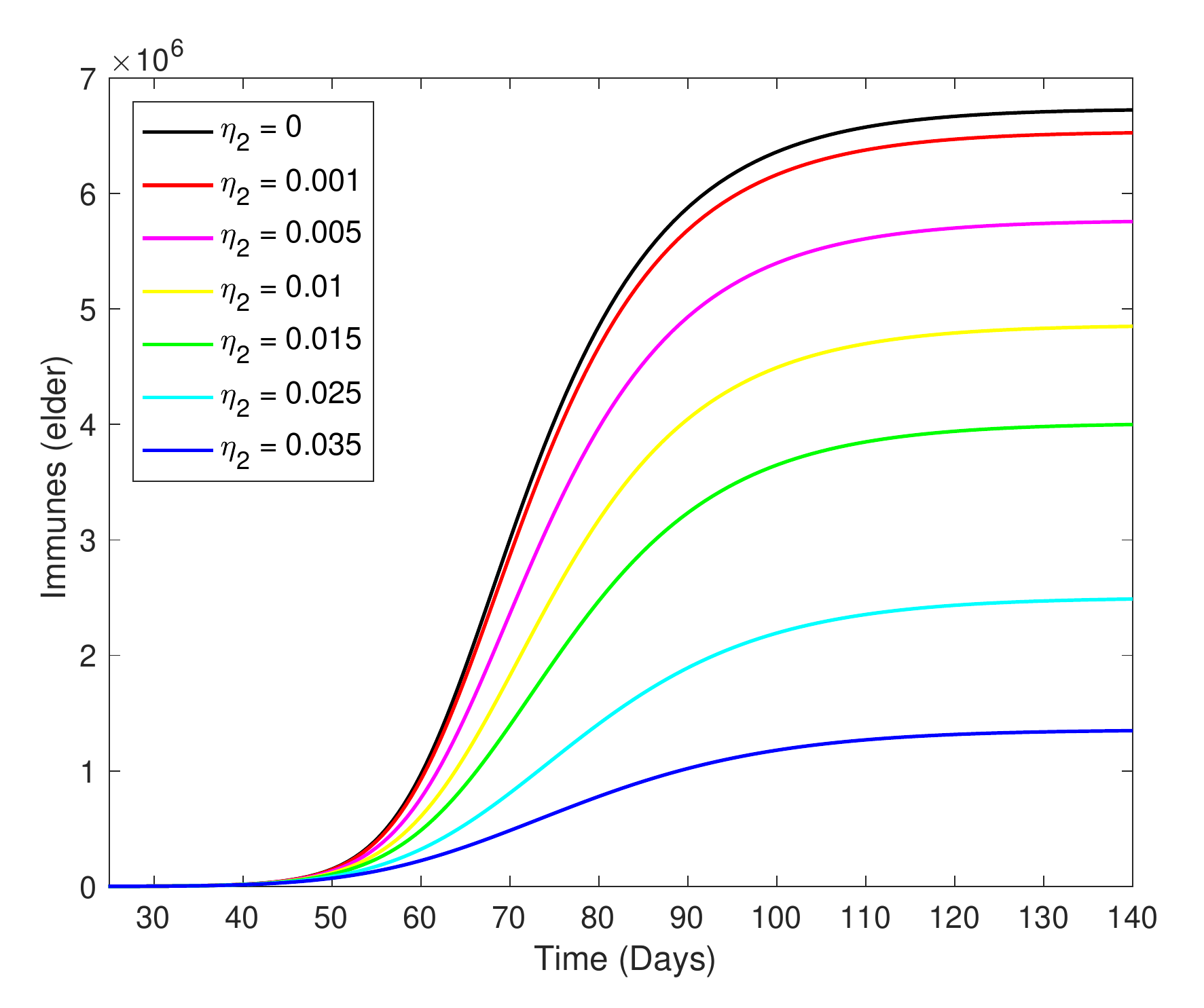}
}
\caption{The curves of the number of immune persons $I_{j}$, $j=y,o$,
without and with isolation for different values of $\eta _{2}$. Curves from
top to bottom corresponds to increasing $\eta _{2}$. The beginning of
isolation is at $t=27$.}
\end{figure}

Immunological parameters (peak of $D_{2}$, $\Omega $, ,$\Pi $ and $I$) are
reduced quite similar for $\eta _{2}=0.035$ $days^{-1}$, between $4.8$-times
($21\%$) and $8.3$-times ($12\%$), however the susceptible persons left
behind at the end of the first wave increase dramatically, $20$-times
(young) and $240$-times (elder), with $24$-times higher for elder persons.
Hence, in a second wave, there will be more infections among elder persons.

\paragraph{Regime 2 -- Different isolation of young and elder persons ($%
\protect\eta _{2o}\neq \protect\eta _{2y}$)}

In regime 2, we call different isolation of young and elder persons in the
sense that elder isolation rate is fixed, and young isolation rate is
varied, and vice-versa.

Firstly, we choose the isolation rate of elder persons $\eta _{2o}=0.01$ $%
days^{-1}$, and vary $\eta _{2y}=0.001$ ($R_{r}=0.235$), $0.005$ ($%
R_{r}=0.049$), $0.01$ ($R_{r}=0.024$), $0.015$ ($R_{r}=0.016$), $0.025$ ($%
R_{r}=0.009$), $0.035$ ($R_{r}=0.007$) and $0.1$ ($R_{r}=0.002$). The value
for the reduced reproduction number is $R_{r}$ is calculated from equation (%
\ref{Rred}).

Figure 17 shows curves of severe cases of CoViD-19 $D_{2j}$, $j=y,o$,
varying $\eta _{2y}$, fixing $\eta _{2o}=0.01$ $days^{-1}$. The decreasing
pattern of $D_{2y}$ follows that observed in regime 1, but in $D_{2o}$, as $%
\eta _{2y}$ increases, the peaks displace faster to right, and the curves
become more asymmetric (increased skewness) and spread beyond the curve
without isolation.

\begin{figure}[!h]
\centering
\subfloat[]{
\includegraphics[scale=0.45]{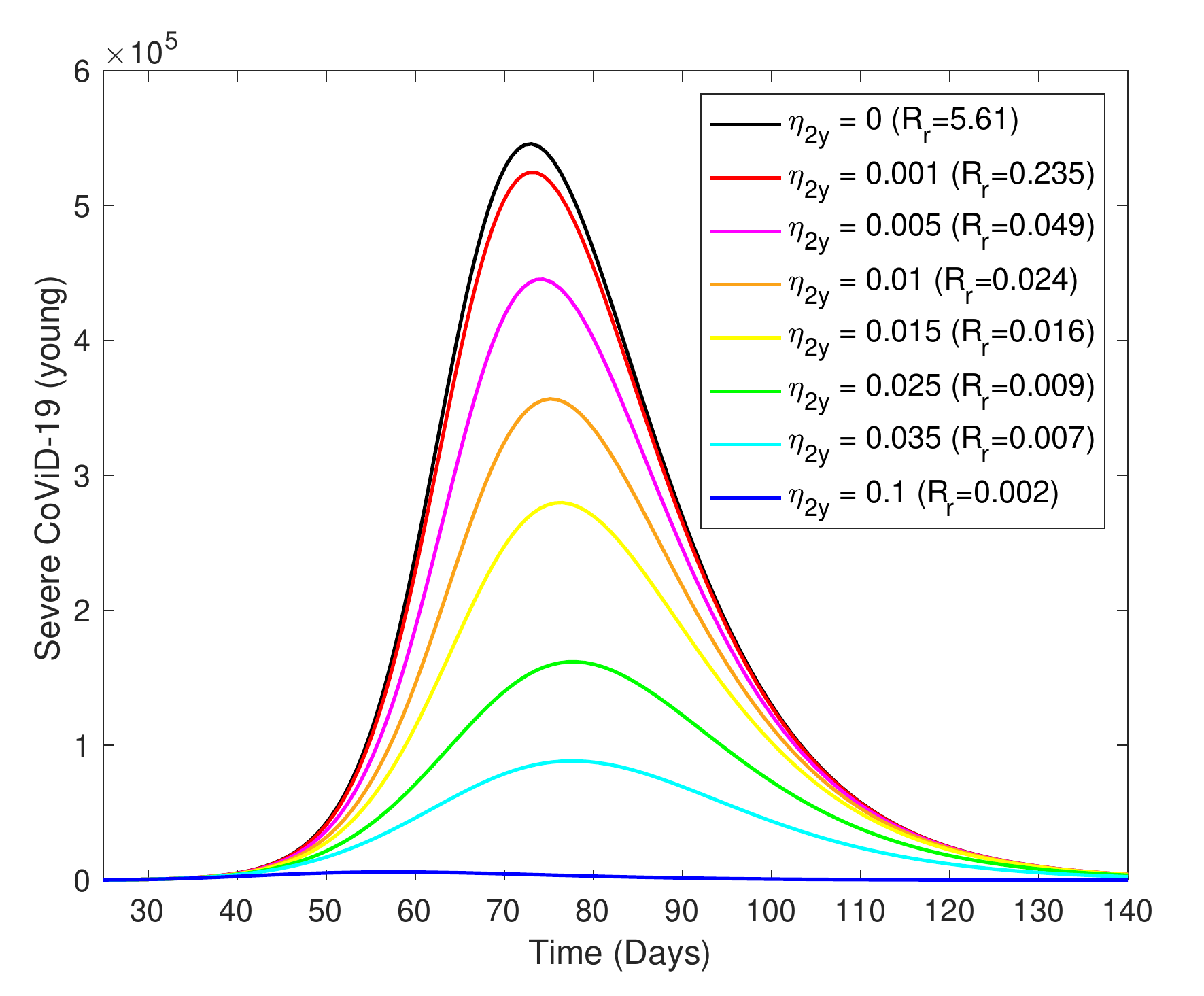}
}
\subfloat[]{
\includegraphics[scale=0.45]{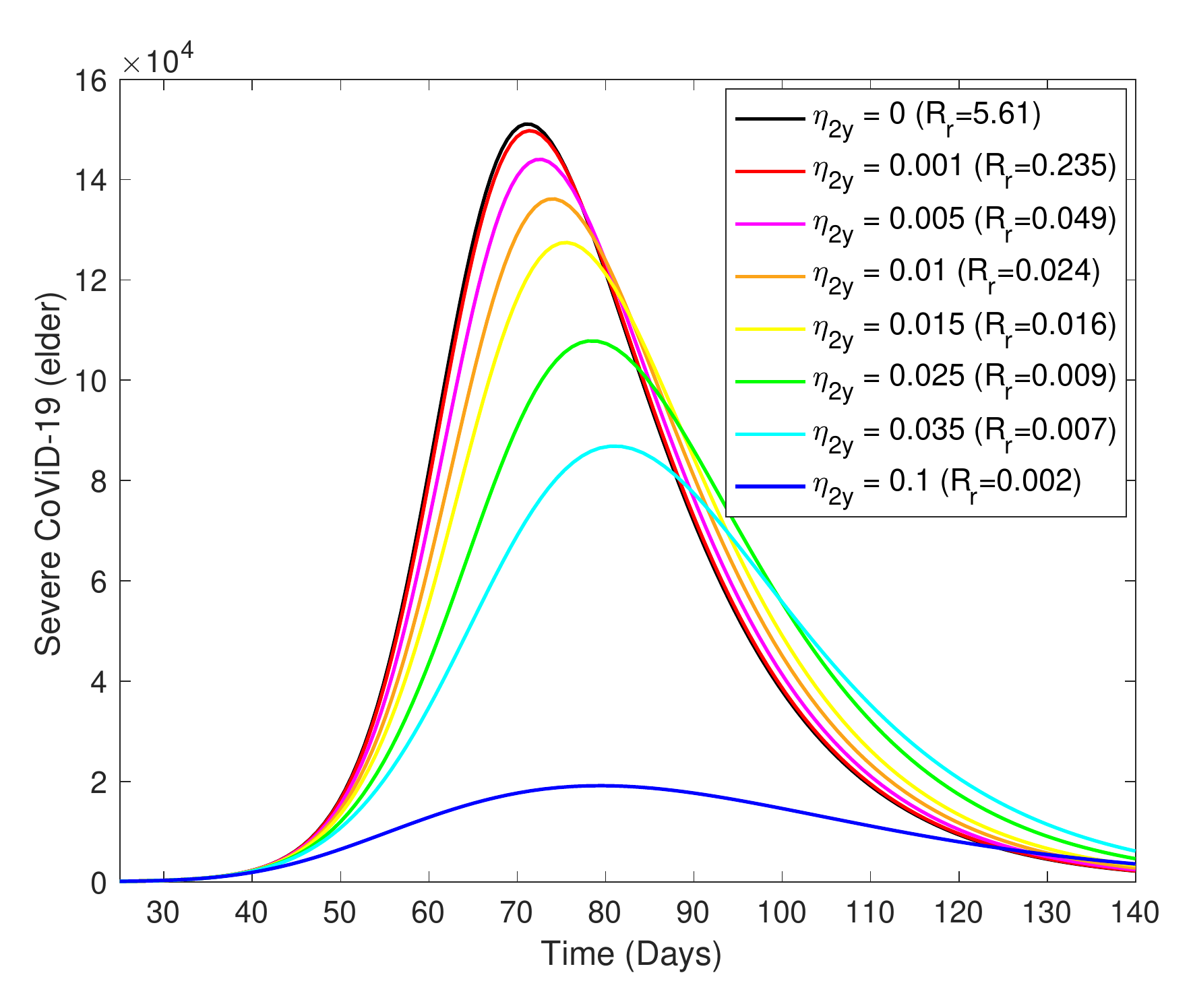}
}
\caption{The curves of severe cases of CoViD-19 $D_{2j}$, $j=y,o$, varying 
$\eta _{2y}$, fixing $\eta _{2o}=0.01$ $days^{-1}$. Curves from top to
bottom corresponds to increasing $\eta _{2y}$. The beginning of isolation is
at $t=27$.}
\end{figure}

Figure 18 shows curves of the number of susceptible persons $S_{j}$, $j=y,o$%
, varying $\eta _{2y}$, fixing $\eta _{2o}=0.01$ $days^{-1}$. The decreasing
pattern of $S_{y}$ follows that observed in regime 1 (sigmoid shape
substituted by exponential decay), but the sigmoid shaped decreasing curves
of $S_{o}$, as $\eta _{2y}$ increases, move from bottom to top, which is an
opposite pattern observed in regime 1. As isolation of young increases, the
number of susceptible young persons decreases, but the number of susceptible
elder persons increases. However, from Figure 17, severe CoViD-19 cases
decrease for both subpopulations. This can be explained by the decreasing in
immune persons: young immune persons decrease $41$-times when $\eta _{2y}$
decreases from $0.015$ to $0.1$, while elder persons decrease $4$-times (see
Table 3).

\begin{figure}[!h]
\centering
\subfloat[]{
\includegraphics[scale=0.45]{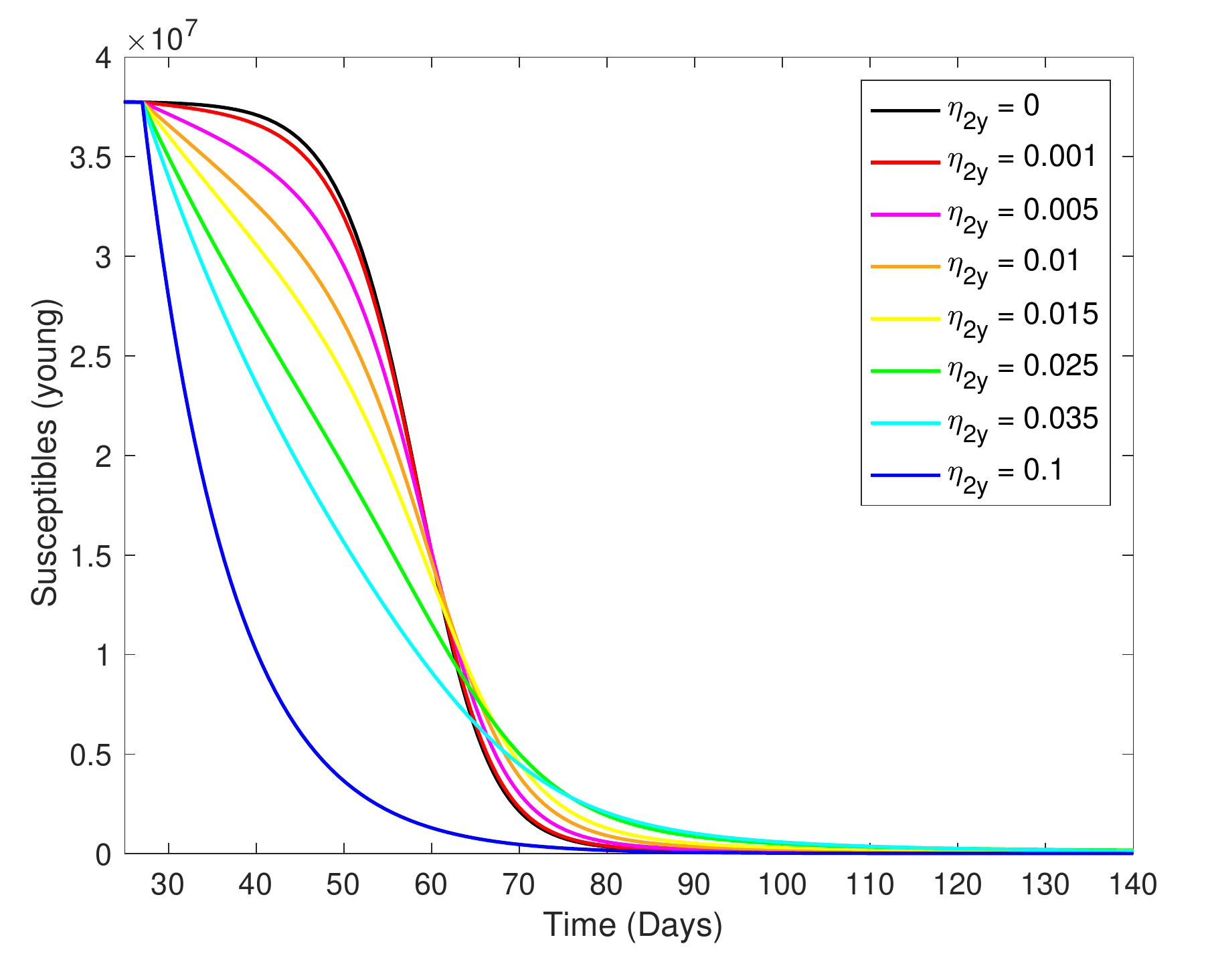}
}
\subfloat[]{
\includegraphics[scale=0.45]{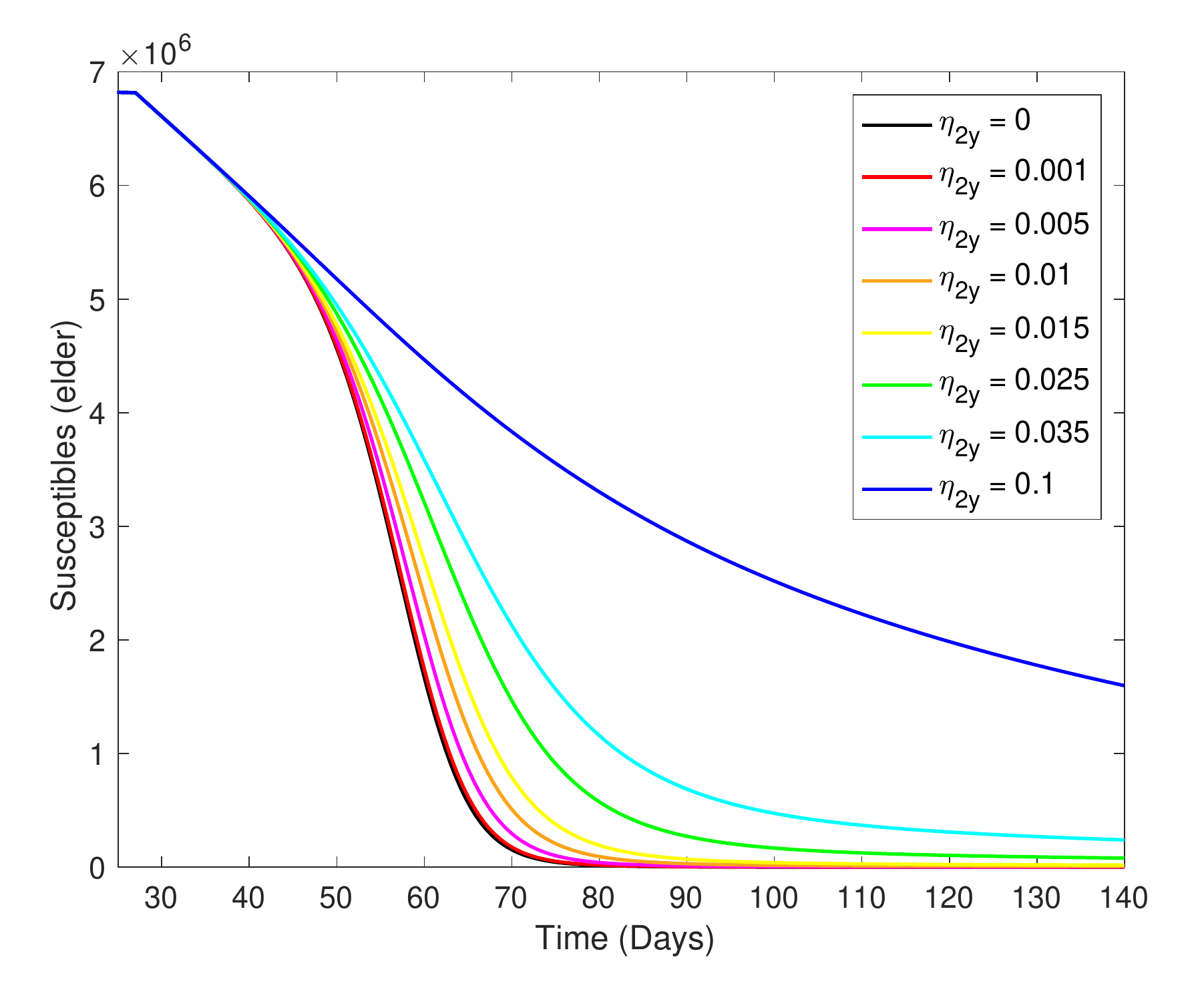}
}
\caption{The curves of the number of susceptible persons $S_{j}$, $j=y,o$,
varying $\eta _{2y}$, fixing $\eta _{2o}=0.01$ $days^{-1}$. Curves from top
to bottom corresponds to decreasing $\eta _{2y}$. The beginning of isolation
is at $t=27$.}
\end{figure}

The curves of accumulated cases of severe CoViD-19 $\Omega $, accumulated
cases of CoViD-19 deaths $\Pi $, the number of isolated susceptible person $%
S^{is}$, and the number of immune persons $I$ are similar than those shown
in foregoing section. For this reason, we present in Table 3 ($\eta
_{2o}=0.01$ $days^{-1}$ fixed) their values at $t=140$ for young, elder and
all persons, letting $\eta _{2y}=0.015$, $\eta _{2y}=0.035$ and $\eta
_{2y}=0.1$ ($days^{-1}$). For $\eta _{2o}=\eta _{2y}=0$ we have, from
foregoing section, $\Omega _{y}=1.798\times 10^{6}$, $\Omega
_{o}=5.630\times 10^{5}$ and $\Omega =2.361\times 10^{6}$; $\Pi
_{y}=1.6\times 10^{4}$, $\Pi _{o}=6.265\times 10^{4}$ and $\Pi =7.865\times
10^{4}$; $S_{y}=1.239\times 10^{5}$, $S_{o}=2463$\ and $S=1.263\times 10^{5}$%
; and $I_{y}=3.762\times 10^{7}$, $I_{o}=6.723\times 10^{6}$ and $%
I=4.434\times 10^{7}$. The percentages are calculated as the ratio between
epidemiological parameter evaluated with ($\eta _{2j}>0$) and without ($\eta
_{2y}=\eta _{2o}=0$) isolation, at $t=140$. The number of isolated
susceptible persons is $S^{is}=0$ when there is not isolation, hence the
percentage is the ratio between $S^{is}$ at $t=140$ and $N_{0}$.

\begin{table}[!h]
\centering
\caption{Values and percentages of $\Omega $, $\Pi $, $Q$ and $I$
fixing $\eta _{2o}=0.01$ $days^{-1}$ and varying $\eta _{2y}=0.015$, $\eta
_{2y}=0.035$ and $\eta _{2y}=0.1$ ($days^{-1}$). $y$, $o$ and $\Sigma $
stand for, respectively, young, elder and total persons.}
\begin{tabular}{llllllllll}
\hline
\multicolumn{1}{c}{\multirow{2}{*}{}} & \multicolumn{3}{c}{$\eta_{2y}=0.015$}
& \multicolumn{3}{c}{$\eta_{2y}=0.035$} & \multicolumn{3}{c}{$\eta_{2y}=0.1$}
\\ 
\multicolumn{1}{c}{} & \multicolumn{1}{c}{$y$} & \multicolumn{1}{c}{$o$} & 
\multicolumn{1}{c}{$\Sigma$} & \multicolumn{1}{c}{$y$} & \multicolumn{1}{c}{$%
o$} & \multicolumn{1}{c}{$\Sigma$} & \multicolumn{1}{c}{$y$} & 
\multicolumn{1}{c}{$o$} & \multicolumn{1}{c}{$\Sigma$} \\ \hline
$\Omega \;(10^6)$ & $1.051$ & $0.3982$ & $1.4492$ & $0.396$ & $0.3342$ & $%
0.7302$ & $0.026$ & $0.098$ & $0.124$ \\ 
$\Pi$ & $9330$ & $44170$ & $53500$ & $3500$ & $36700$ & $40200$ & $230$ & $%
10540$ & $10770$ \\ 
$S \;(10^5)$ & $1.748$ & $0.1907$ & $1.9387$ & $1.476$ & $2.4$ & $3.876$ & $%
0.167$ & $15.98$ & $16.15$ \\ 
$Q \;(10^7)$ & $1.566$ & $0.1978$ & $1.7638$ & $2.945$ & $0.252$ & $%
3.197$ & $3.739$ & $0.4013$ & $4.14$ \\ 
$I \;(10^7)$ & $2.196$ & $0.4749$ & $2.6709$ & $0.827$ & $0.3966$ & $1.224$
& $0.054$ & $0.1148$ & $0.1688$ \\ \hline
&  &  &  &  &  &  &  &  &  \\ 
$\Omega \;(\%)$ & $58.45$ & $70.73$ & $61.38$ & $22$ & $59.36$ & $30.93$ & $%
1.45$ & $17.41$ & $5.25$ \\ 
$\Pi\; (\%)$ & $58.31$ & $70.50$ & $68.02$ & $21.88$ & $58.58$ & $51.11$ & $%
1.44$ & $16.82$ & $13.69$ \\ 
$S \;(\%)$ & $141.08$ & $774.26$ & $153.43$ & $119.13$ & $9744$ & $306.74$ & 
$13.48$ & $64880$ & $1278$ \\ 
$Q \;(\%)$ & $41.45$ & $29.00$ & $39.55$ & $77.95$ & $36.95$ & $71.68$
& $98.97$ & $58.84$ & $92.82$ \\ 
$I \;(\%)$ & $58.37$ & $7.06$ & $60.23$ & $21.98$ & $5.899$ & $27.60$ & $%
1.44 $ & $1.708$ & $3.81$ \\ \hline
\end{tabular}
\end{table}

Figures 17 and 18 and Table 3 portrait preferential isolation of young
persons, but maintaining elder persons isolated at a fixed level. Hence, the
increasing in $\eta _{2y}$ of course protects young persons, but elder
persons are also benefitted .

Now, we choose the isolation rate of young persons $\eta _{2y}=0.01$ $%
days^{-1}$, and vary the isolation rate of elder persons $\eta _{2o}$ ($%
days^{-1}$) for 7 different values: $\eta _{2o}=0.001$ ($R_{r}=0.025$), $%
0.005$ ($R_{r}=0.02444$), $0.01$ ($R_{r}=0.02442$), $0.015$ ($R_{r}=0.024416$%
), $0.025$ ($R_{r}=0.024413$), $0.035$ ($R_{r}=0.024411$) and $0.1$ ($%
R_{r}=0.02440$).

Figure 19 shows curves of severe cases of CoViD-19 $D_{2j}$, $j=y,o$,
varying $\eta _{2o}$, fixing $\eta _{2y}=0.01$ $days^{-1}$. The same pattern
observed in Figure 17, changing $D_{2y}$ by $D_{2o}$, but more smooth.

\begin{figure}[!h]
\centering
\subfloat[]{
\includegraphics[scale=0.45]{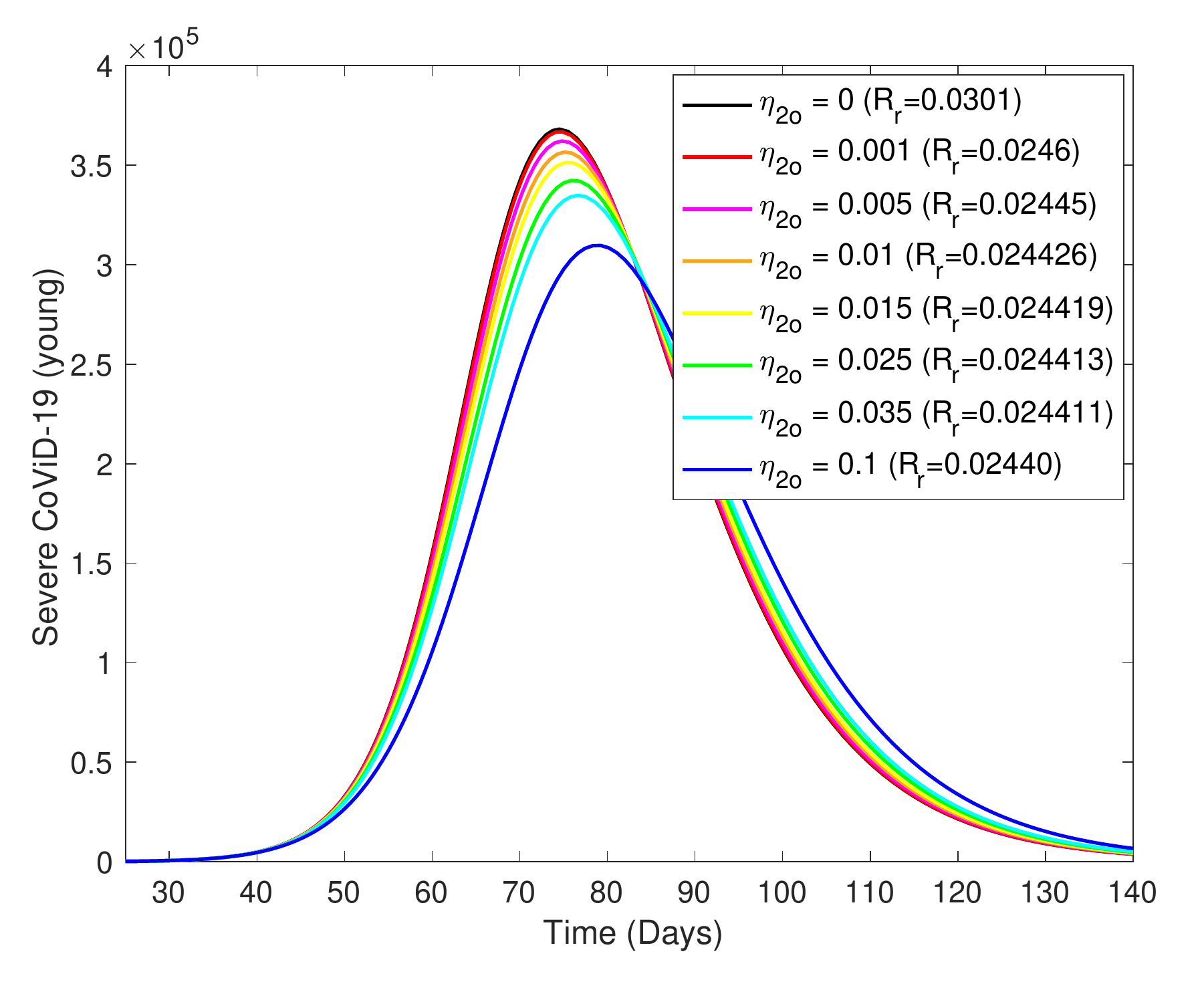}
}
\subfloat[]{
\includegraphics[scale=0.45]{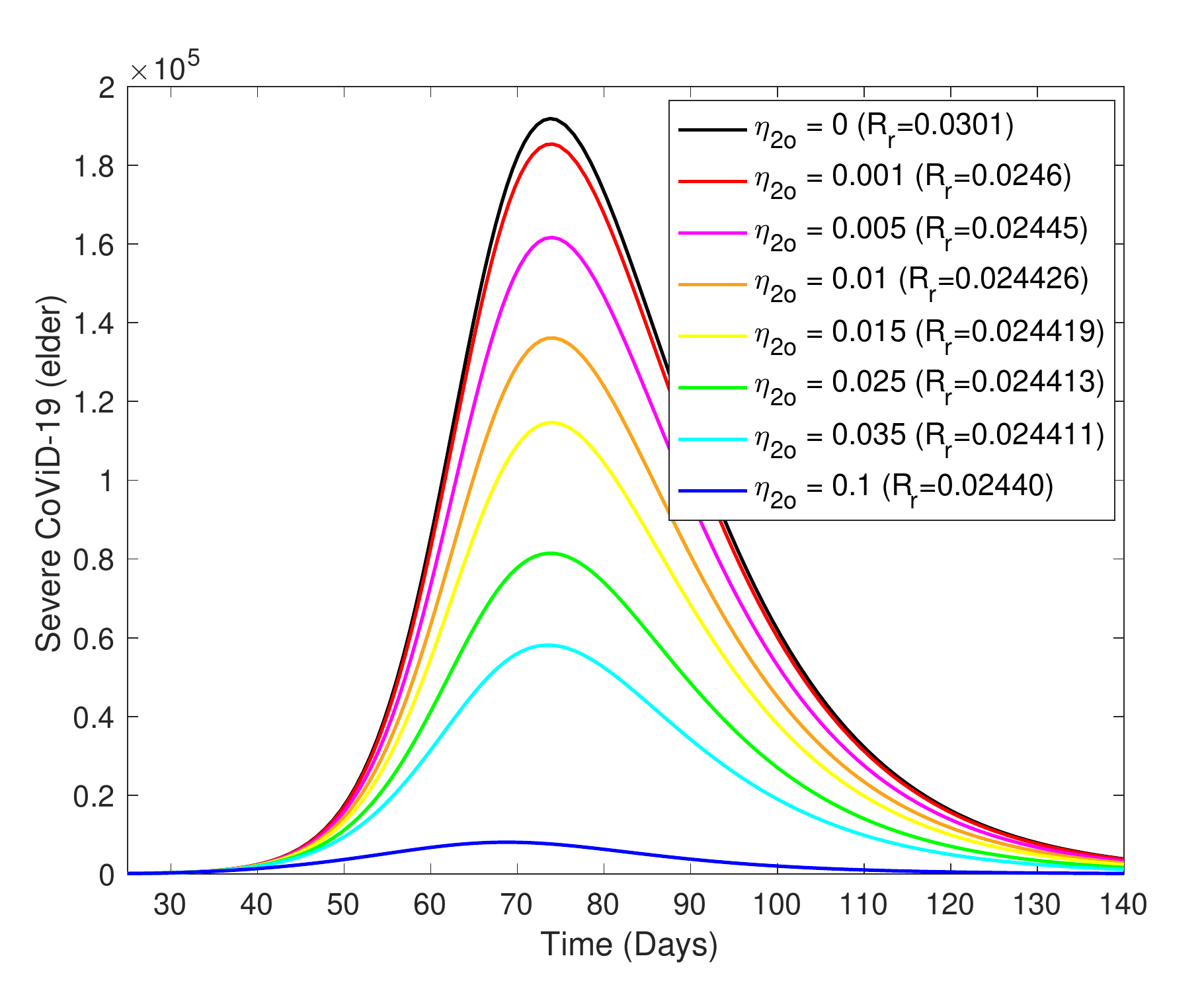}
}
\caption{The curves of severe cases of CoViD-19 $D_{2j}$, $j=y,o$, varying 
$\eta _{2o}$, fixing $\eta _{2y}=0.01$ $days^{-1}$. Curves from top to
bottom corresponds to increasing $\eta _{2}$. The beginning of isolation is
at $t=27$.}
\end{figure}

Figure 20 shows curves of the number of susceptible persons $S_{j}$, $j=y,o$%
, varying $\eta _{2o}$, fixing $\eta _{2y}=0.01$ $days^{-1}$. The same
pattern observed in Figure 18, changing $S_{y}$ by $S_{o}$.

\begin{figure}[!h]
\centering
\subfloat[]{
\includegraphics[scale=0.45]{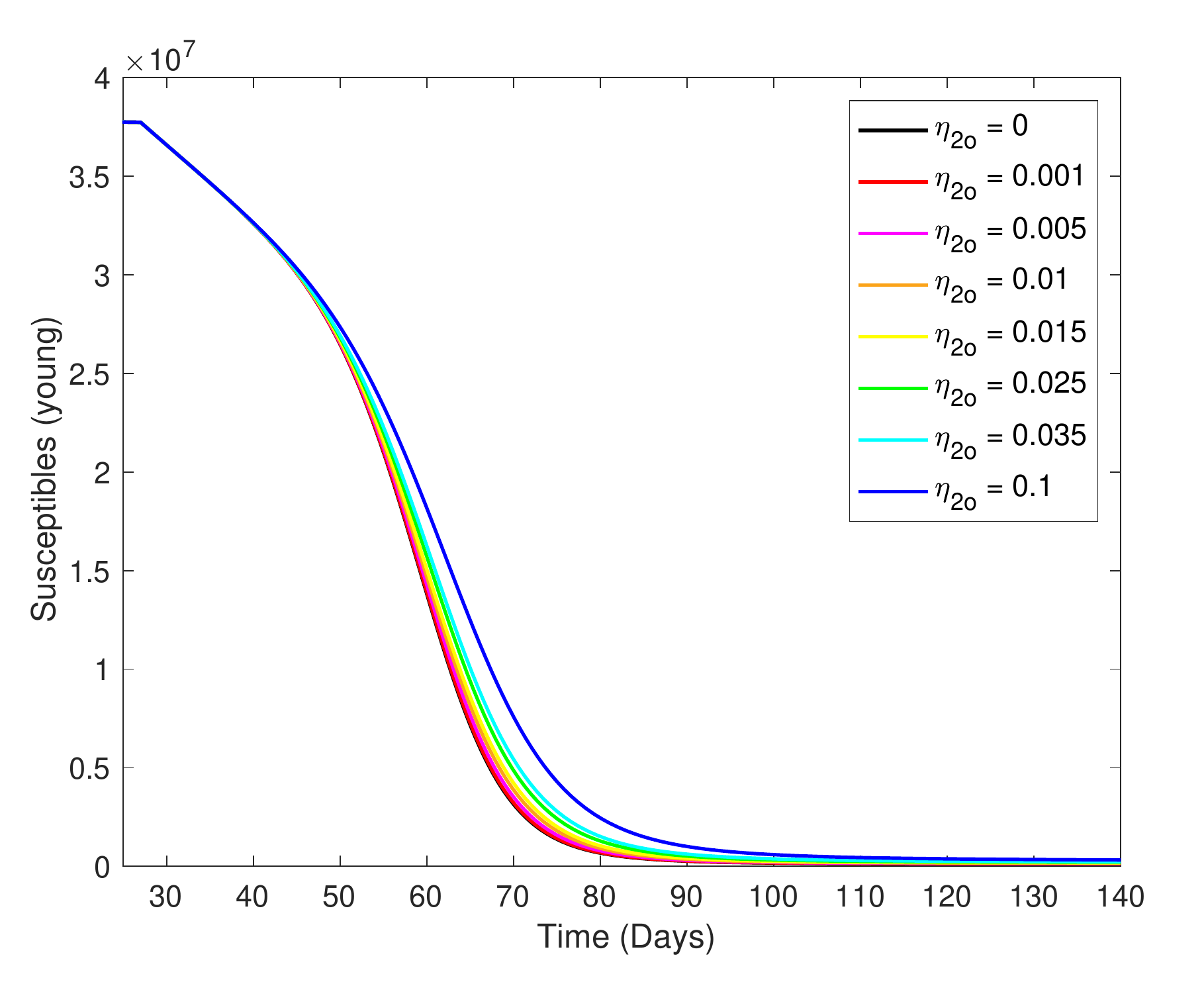}
}
\subfloat[]{
\includegraphics[scale=0.45]{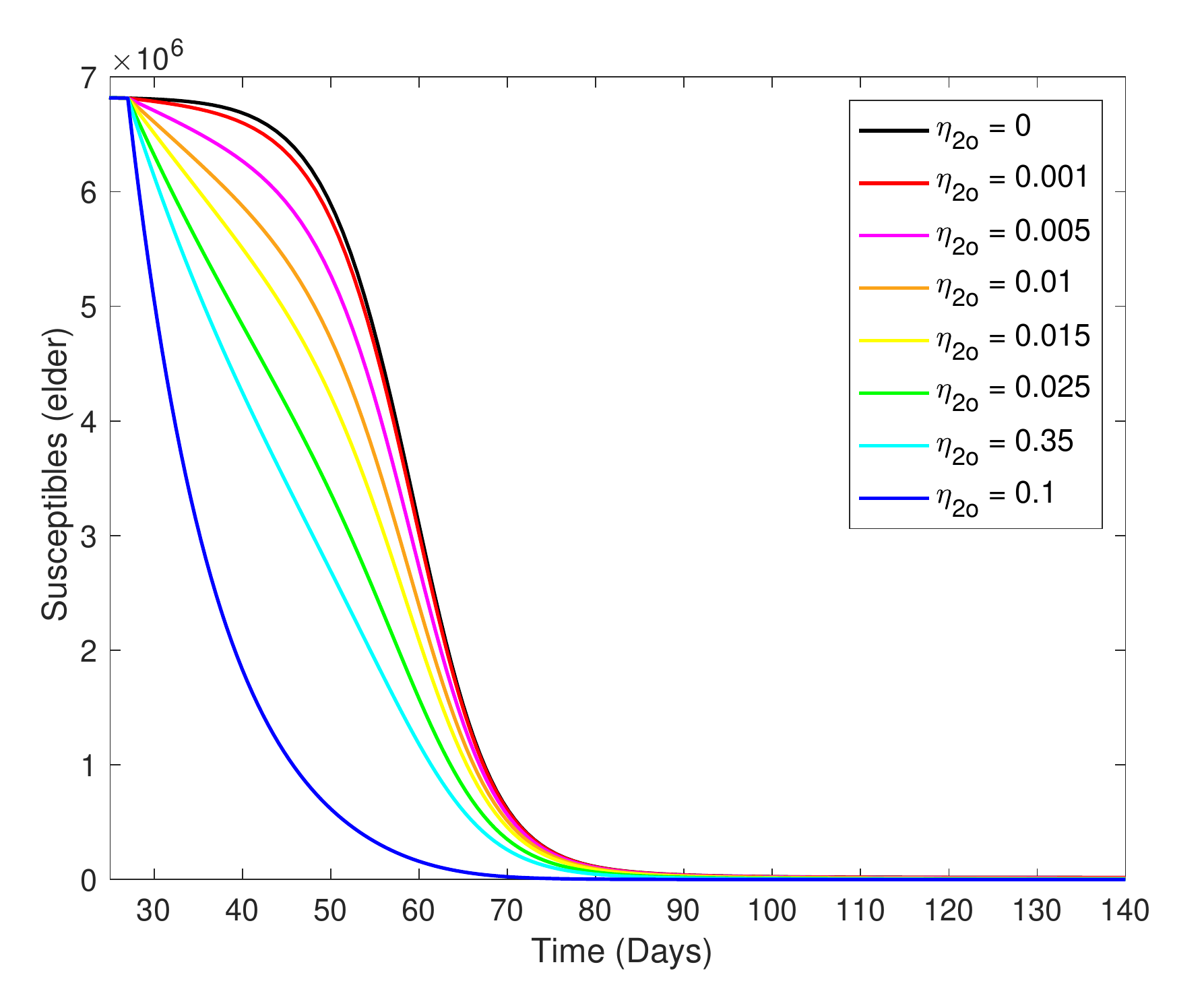}
}
\caption{The curves of the number of susceptible persons $S_{j}$, $j=y,o$,
varying $\eta _{2o}$, fixing $\eta _{2y}=0.01$ $days^{-1}$. Curves from top
to bottom corresponds to decreasing $\eta _{2o}$. The beginning of isolation
is at $t=27$.}
\end{figure}

The curves of accumulated cases of severe CoViD-19 $\Omega $, accumulated
cases of CoViD-19 deaths $\Pi $, the number of isolated susceptible person $%
S^{is}$, and the number of immune persons $I$ are similar than those shown
in foregoing section. For this reason, we present in Table 4 ($\eta
_{2y}=0.01$ $days^{-1}$ fixed) their values at $t=140$ for young, elder and
all persons, letting $\eta _{2o}=0.015$, $\eta _{2o}=0.035$ and $\eta
_{2o}=0.1$ ($days^{-1}$). Values for $\Omega $, $\Pi $, $S^{is}$ and $I$,
for $\eta _{2o}=\eta _{2y}=0$, are those used in Table 3, as well as the
definitions of the percentages.

\begin{table}[!h]
\centering
\caption{Values and percentages of $\Omega $, $\Pi $, $Q$ and $I$
fixing $\eta _{2y}=0.01$ $days^{-1}$ and varying $\eta _{2o}=0.015$, $\eta
_{2o}=0.035$ and $\eta _{2o}=0.1$ ($days^{-1}$). $y$, $o$ and $\Sigma $
stand for, respectively, young, elder and total persons.}
\begin{tabular}{llllllllll}
\hline
\multicolumn{1}{c}{\multirow{2}{*}{}} & \multicolumn{3}{c}{$\eta_{2o}=0.015$}
& \multicolumn{3}{c}{$\eta_{2o}=0.035$} & \multicolumn{3}{c}{$\eta_{2o}=0.1$}
\\ 
\multicolumn{1}{c}{} & \multicolumn{1}{c}{$y$} & \multicolumn{1}{c}{$o$} & 
\multicolumn{1}{c}{$\Sigma$} & \multicolumn{1}{c}{$y$} & \multicolumn{1}{c}{$%
o$} & \multicolumn{1}{c}{$\Sigma$} & \multicolumn{1}{c}{$y$} & 
\multicolumn{1}{c}{$o$} & \multicolumn{1}{c}{$\Sigma$} \\ \hline
$\Omega \;(10^6)$ & $1.272$ & $0.3452$ & $1.6172$ & $1.251$ & $0.1802$ & $%
1.4312$ & $1.216$ & $0.0274$ & $1.2434$ \\ 
$\Pi$ & $11300$ & $38350$ & $49650$ & $11110$ & $20030$ & $31140$ & $10840$
& $3050$ & $13890$ \\ 
$S \;(10^5)$ & $1.787$ & $0.05501$ & $1.84201$ & $2.347$ & $0.00943$ & $%
2.35643$ & $3.288$ & $0.0002$ & $3.2882$ \\ 
$Q \;(10^7)$ & $1.101$ & $0.2636$ & $1.3646$ & $1.138$ & $0.4642$ & $%
1.6022$ & $1.202$ & $0.6498$ & $1.8518$ \\ 
$I \;(10^7)$ & $2.659$ & $0.412$ & $3.071$ & $2.615$ & $0.2151$ & $2.8301$ & 
$2.539$ & $0.0327$ & $2.5717$ \\ \hline
&  &  &  &  &  &  &  &  &  \\ 
$\Omega \;(\%)$ & $70.75$ & $61.31$ & $68.50$ & $69.58$ & $32.01$ & $60.62$
& $67.63$ & $4.87$ & $52.66$ \\ 
$\Pi\; (\%)$ & $70.63$ & $61.21$ & $63.13$ & $69.44$ & $31.97$ & $39.59$ & $%
67.75$ & $4.87$ & $17.66$ \\ 
$S \;(\%)$ & $144.2$ & $223.4$ & $145.8$ & $189.4$ & $38.3$ & $186.5$
& $265.4$ & $0.81$ & $260.22$ \\ 
$Q \;(\%)$ & $29.14$ & $38.65$ & $30.60$ & $30.12$ & $68.06$ & $35.92$
& $31.82$ & $95.28$ & $41.52$ \\ 
$I \;(\%)$ & $70.68$ & $6.13$ & $69.26$ & $69.51$ & $3.20$ & $63.82$ & $%
67.49 $ & $0.49$ & $58.00$ \\ \hline
\end{tabular}
\end{table}

Figures 19 and 20 and Table 4 portrait preferential isolation of elder
persons, but maintaining young persons isolated at a fixed level. Hence, the
increasing in $\eta _{2o}$ of course protects elder persons, but young
persons are also benefitted.

Tables 3 and 4 show two kinds isolation for two different goals. If the
objective is diminishing the total number of severe CoViD-19 cases $\Omega $%
, the better strategy is isolating more young than elder persons. However,
if the goal is the reduction of fatality cases $\Pi $, the better strategy
is the isolating more elder than young persons, but if the isolation is very
intense ($\eta _{2y}=\eta _{2o}=0.1$), then isolating more young persons is
recommended. Notice that only strategy $\eta _{2o}=0.01$ and $\eta _{2y}=0.1$
attains the number of isolated susceptible persons above the threshold $%
3.815\times 10^{7}$.

\subsubsection{Scenarios -- Isolation and releasing}

When releasing is introduced, then equation (\ref{propotion}) is not anymore
valid to evaluated the accumulated number of isolated susceptible persons.
Hence, we use $Q_{y}$, $Q_{o}$ and $Q=Q_{y}+Q_{o}$ for the numbers of
isolated susceptible, respectively, young, elder and total persons. $Q_{y}$
and $Q_{o}$ are solutions of the system of equations (\ref{system2a}), (\ref%
{system1c}) and (\ref{system2b}).

At $t=0$ (February 26) the first case of severe CiViD-19 was confirmed, and
at $t=27$ (March 24) isolation as mechanism of control (described by $\eta
_{2y}$ and $\eta _{2o}$) was introduced until April 22.\footnote{%
In April 6 the isolation was extended until April 22.} Hence, the beginning
of releasing of isolated persons will occur at the simulation time $t=56$.%
\footnote{%
Simulations were done in April 10.} We assume that same rates of releasing
are applied to young and elder persons, that is, $\eta _{3}=\eta _{3y}=\eta
_{3o}$, and consider regime 1-type isolation, that is, $\eta _{2}=\eta
_{2o}=\eta _{2y}$. Hence, from time $0$ to $27$ we have $R_{0}=6.915$ (no
isolation), we have regime 1-type isolation from $27$ to $56$ with $%
R_{r}=0.007$, and sinceafter $56$, we have isolation and releasing with
value of $R_{r}$ depending on $\eta _{3}$.

In order to assess epidemiological scenarios when isolated persons are
released, we fix $\eta _{2}=0.035$ ($days^{-1}$), and vary $\eta _{3}=0$ ($%
R_{r}=0.007$), $0.0055$ ($R_{r}=0.84$), $0.01$ ($R_{r}=1.49$), $0.015$ ($%
R_{r}=2.02$), $0.25$ ($R_{r}=2.82$), $0.035$ ($R_{r}=3.39$) and $0.1$ ($%
R_{r}=5.07$). The value for the reduced reproduction number is $R_{r}$ is
calculated from equation (\ref{Rred}).

Figure 21 shows curves of severe cases of CoViD-19 $D_{2j}$, $j=y,o$, fixing 
$\eta _{2}=0.035$ $days^{-1}$, and varying $\eta _{3}$. The beginning of
release is at $t=56$, date proposed by the S\~{a}o Paulo State authorities.
For instance, when $\eta _{3}=0.035$ $days^{-1}$, the peaks are for young
and elder persons, respectively, $2.31\times 10^{5}$ and $9.06\times 10^{4}$%
, which occur at $t=99$ and $t=98$.

\begin{figure}[!h]
\centering
\subfloat[]{
\includegraphics[scale=0.45]{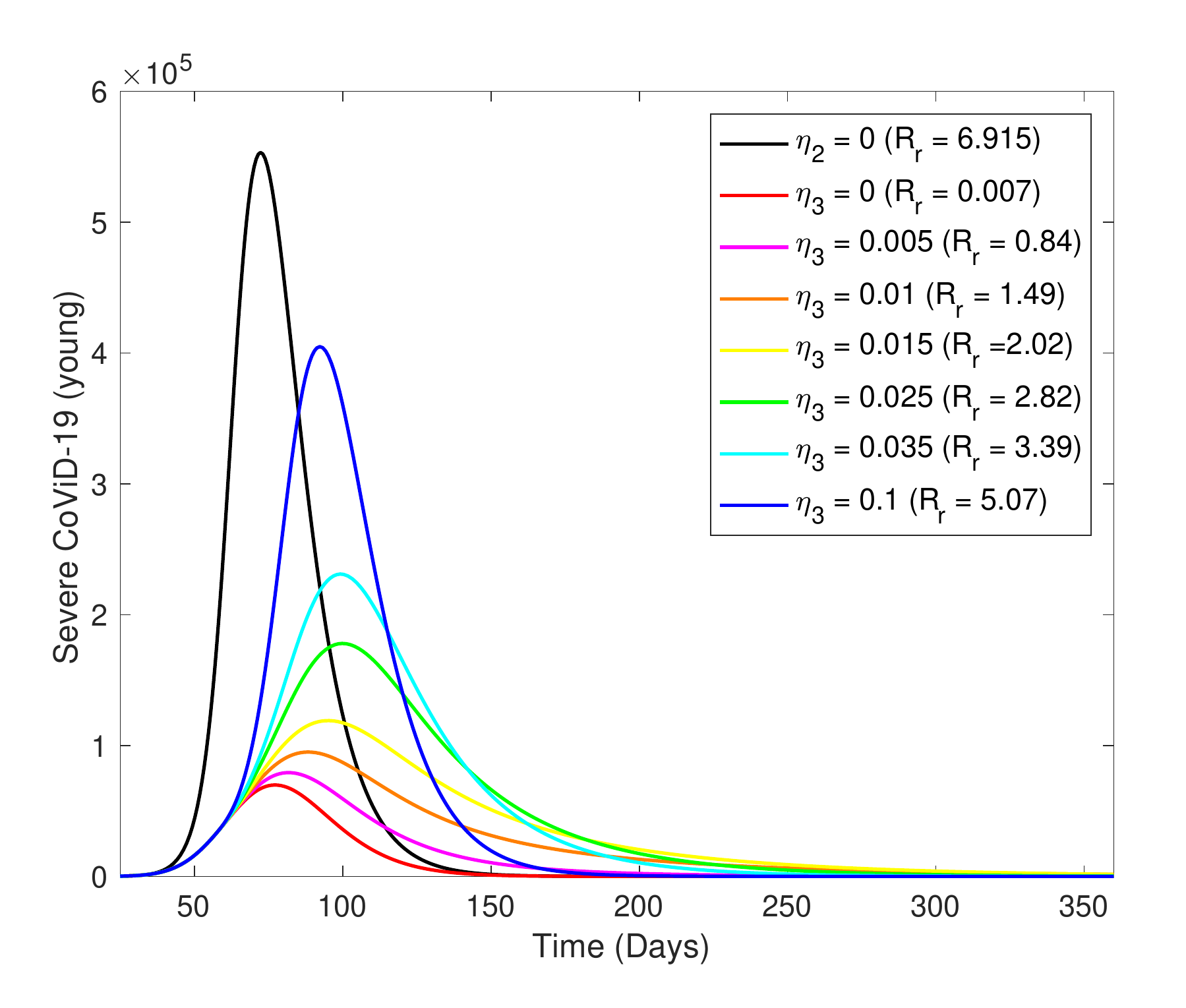}
}
\subfloat[]{
\includegraphics[scale=0.45]{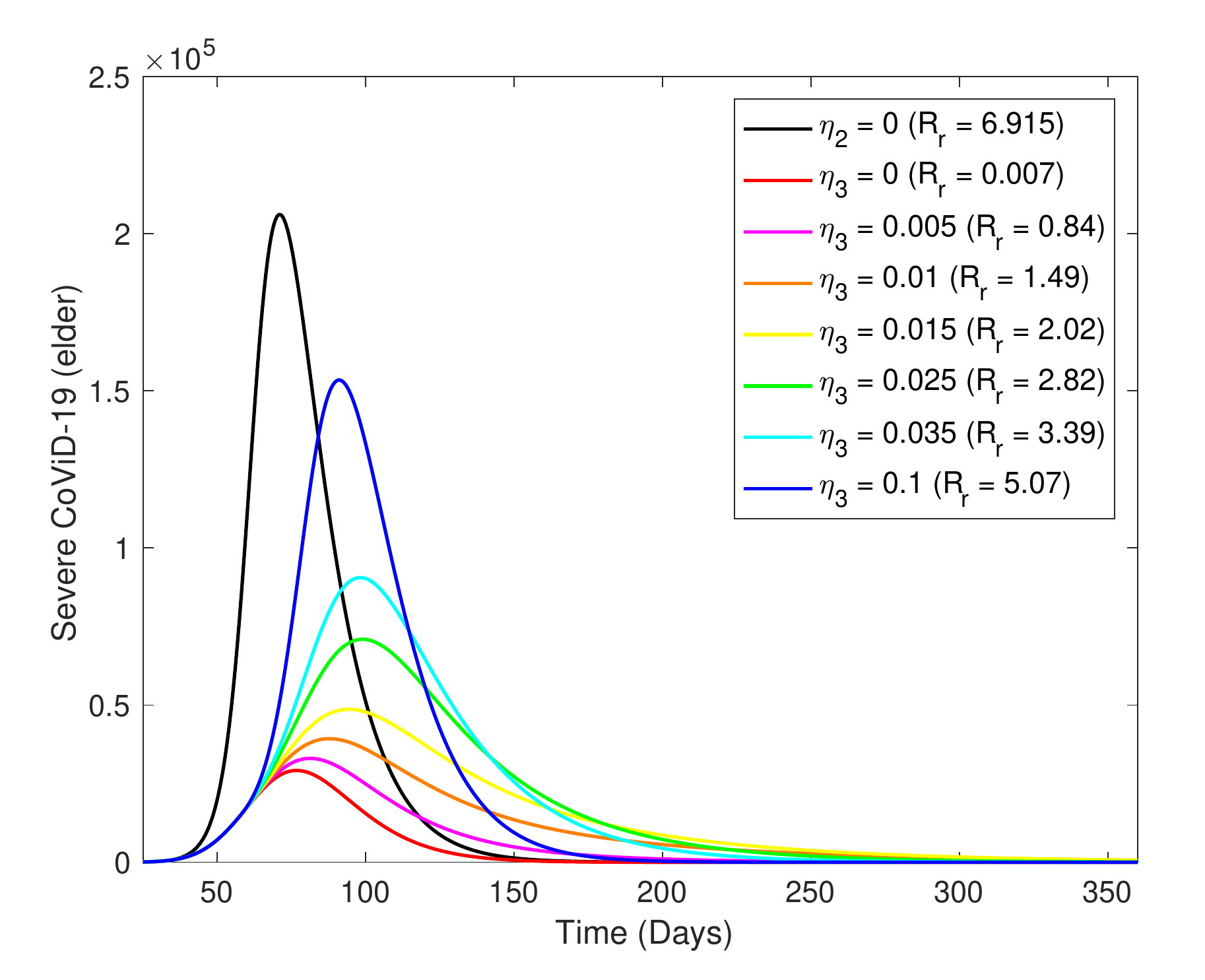}
}
\caption{The curves of severe cases of CoViD-19 $D_{2j}$, $j=y,o$, fixing $%
\eta _{2}=0.035$ $days^{-1}$, and varying $\eta _{3}$. Curves from top to
bottom corresponds to decreasing $\eta _{3}$. The beginning of release is at 
$t=56$.}
\end{figure}

The curves of accumulated cases of severe CoViD-19 $\Omega $, accumulated
cases of CoViD-19 deaths $\Pi $, the number of isolated susceptible person $%
S^{is}$, and the number of immune persons $I$ are similar than those shown
in foregoing section. For this reason, we present in Table 5 ($\eta
_{2}=0.035$ $days^{-1}$ fixed) their values at $t=360$ for young, elder and
all persons, letting $\eta _{3o}=0.015$, $\eta _{3o}=0.035$ and $\eta
_{3o}=0.1$ ($days^{-1}$). Values for $\Omega $, $\Pi $, $S^{is}$ and $I$,
for $\eta _{2}=0$, are those used in Table 3, as well as the definitions of
the percentages.

\begin{table}[!h]
\centering
\caption{Values and percentages of $\Omega $, $\Pi $, $Q$ and $I$
fixing $\eta _{2y}=\eta _{2o}=0.035$ $days^{-1}$ and varying $\eta _{3}=0.015
$, $\eta _{3}=0.035$ and $\eta _{3}=0.1$ ($days^{-1}$). $y$, $o$ and $\Sigma 
$ stand for, respectively, young, elder and total persons. Releasing
initiates at $t=56$.}
\begin{tabular}{llllllllll}
\hline
\multirow{2}{*}{} & \multicolumn{3}{c}{$\eta_3 = 0.015$}                                             & \multicolumn{3}{c}{$\eta_3 = 0.035$}                                             & \multicolumn{3}{c}{$\eta_3 = 0.1$}                                               \\
                  & \multicolumn{1}{c}{$y$} & \multicolumn{1}{c}{$o$} & \multicolumn{1}{c}{$\Sigma$} & \multicolumn{1}{c}{$y$} & \multicolumn{1}{c}{$o$} & \multicolumn{1}{c}{$\Sigma$} & \multicolumn{1}{c}{$y$} & \multicolumn{1}{c}{$o$} & \multicolumn{1}{c}{$\Sigma$} \\ \hline
$\Omega \;(10^6)$  & $1.121$                 & $0.3723$                & $1.4933$                     & $1.531$                 & $0.4927$                & $2.0237$                     & $1.75$                  & $0.552$                 & $2.302$                      \\
$\Pi$             & $9985$                  & $41570$                 & $51555$                      & $13650$                 & $55110$                 & $68760$                      & $15600$                 & $61720$                 & $77320$                      \\
$S \;(10^6)$      & $4.317$                 & $0.676$                 & $4.993$                      & $3.012$                 & $0.422$                 & $3.434$                      & $1.098$                 & $0.1004$                & $1.198$                      \\
$Q \;(10^7)$ & $1.016$                 & $0.161$                 & $1.177$                      & $0.299$                 & $0.0422$                 & $0.3412$                      & $0.0381$                 & $0.0035$                 & $0.0416$                      \\
$I \;(10^7)$      & $2.331$                 & $0.442$                 & $2.773$                      & $3.183$                 & $0.585$                 & $3.768$                      & $3.636$                 & $0.655$                 & $4.291$                      \\ \hline
                  &                         &                         &                              &                         &                         &                              &                         &                         &                              \\
$\Omega \;(\%)$    & $62.35$                 & $66.13$                 & $63.25$                      & $85.15$                 & $87.51$                 & $85.71$                      & $97.33$                 & $98.01$                 & $97.49$                      \\
$\Pi \;(\%)$      & $62.41$                 & $66.35$                 & $65.55$                      & $85.31$                 & $87.96$                 & $87.43$                      & $97.50$                 & $98.52$                 & $98.31$                      \\
$S \;(\%)$        & $3484$                  & $27446$                 & $3951$                       & $2431$                  & $17133$                 & $2717$                       & $886$                   & $4076$                  & $948$                        \\
$Q \;(\%)$   & $26.89$                & $23.61$                & $26.39$                     & $7.91$                & $6.19$                & $7.65$                     & $1.01$                 & $0.51$                & $0.93$                      \\
$I \;(\%)$        & $61.96$                 & $6.57$                  & $62.54$                      & $84.61$                 & $8.70$                  & $84.97$                      & $96.65$                 & $9.74$                  & $96.77$                      \\ \hline
\end{tabular}
\end{table}

Figure 22 shows curves of severe cases of CoViD-19 $D_{2j}$, $j=y,o$, fixing 
$\eta _{2}=0.035$ $days^{-1}$, and varying $\eta _{3}$. The beginning of
release is at $t=49$, a week \ earlier. For instance, when $\eta _{3}=0.035$ 
$days^{-1}$, the peaks are for young and elder persons, respectively, $%
2.515\times 10^{5}$ and $9.827\times 10^{4}$, which occur at $t=93$ and $92$%
. In comparison with Figure 21, the peaks are increased for young and elder
persons in, respectively, $8.9\%$ and $8.5\%$, which are both anticipated in 
$6$ days.

\begin{figure}[!h]
\centering
\subfloat[]{
\includegraphics[scale=0.45]{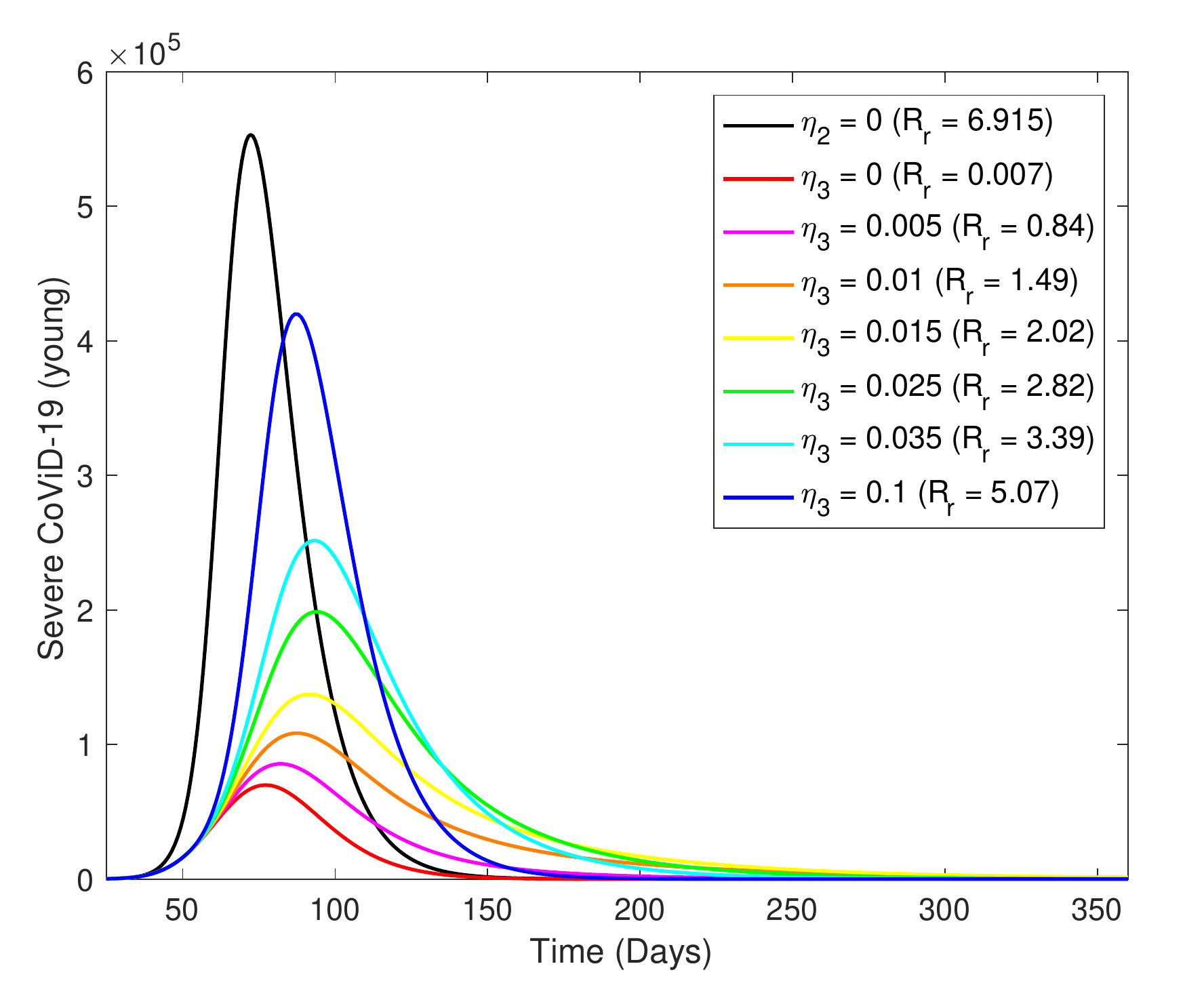}
}
\subfloat[]{
\includegraphics[scale=0.45]{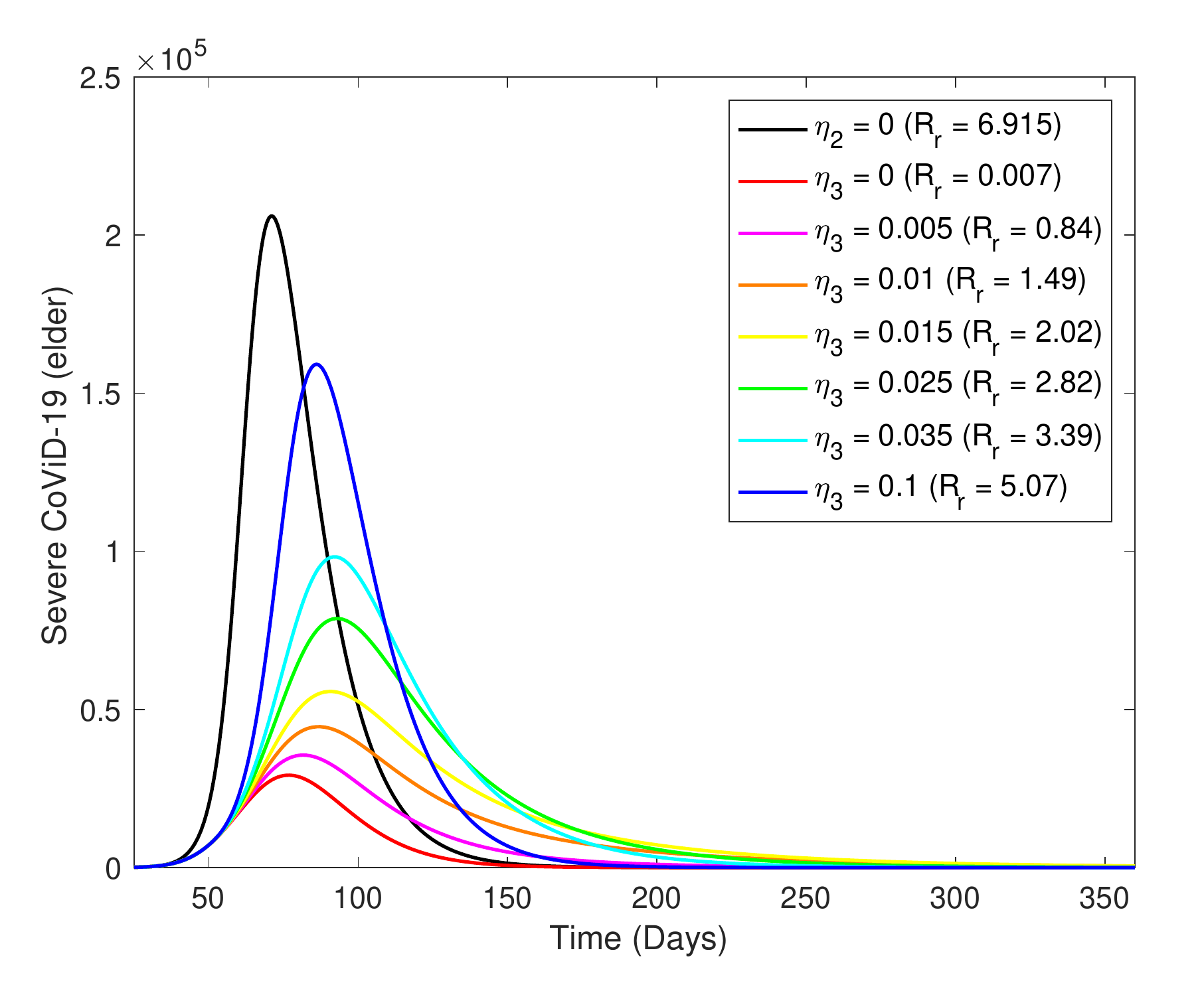}
}
\caption{The curves of severe cases of CoViD-19 $D_{2j}$, $j=y,o$, fixing $%
\eta _{2}=0.035$ $days^{-1}$, and varying $\eta _{3}$. Curves from top to
bottom corresponds to decreasing $\eta _{3}$. The beginning of release is at 
$t=49$.}
\end{figure}

The curves of accumulated cases of severe CoViD-19 $\Omega $, accumulated
cases of CoViD-19 deaths $\Pi $, the number of isolated susceptible person $%
S^{is}$, and the number of immune persons $I$ are similar than those shown
in foregoing section. For this reason, we present in Table 6 ($\eta
_{2}=0.035$ $days^{-1}$ fixed) their values at $t=360$ for young, elder and
all persons, letting $\eta _{3o}=0.015$, $\eta _{3o}=0.035$ and $\eta
_{3o}=0.1$ ($days^{-1}$). Values for $\Omega $, $\Pi $, $S^{is}$ and $I$,
for $\eta _{2}=0$, are those used in Table 3, as well as the definitions of
the percentages.

\begin{table}[!h]
\centering
\caption{Values and percentages of $\Omega $, $\Pi $, $Q$ and $I$
fixing $\eta _{2y}=0.01$ $days^{-1}$ and varying $\eta _{2o}=0.015$, $\eta
_{2o}=0.035$ and $\eta _{2o}=0.1$ ($days^{-1}$). $y$, $o$ and $\Sigma $
stand for, respectively, young, elder and total persons. Releasing initiates
at $t=49$.}
\begin{tabular}{llllllllll}
\hline
\multirow{2}{*}{} & \multicolumn{3}{c}{$\eta_3 = 0.015$}                                             & \multicolumn{3}{c}{$\eta_3 = 0.035$}                                             & \multicolumn{3}{c}{$\eta_3 = 0.1$}                                               \\
                  & \multicolumn{1}{c}{$y$} & \multicolumn{1}{c}{$o$} & \multicolumn{1}{c}{$\Sigma$} & \multicolumn{1}{c}{$y$} & \multicolumn{1}{c}{$o$} & \multicolumn{1}{c}{$\Sigma$} & \multicolumn{1}{c}{$y$} & \multicolumn{1}{c}{$o$} & \multicolumn{1}{c}{$\Sigma$} \\ \hline
$\Omega \;(10^6)$  & $1.131$                 & $0.3748$                & $1.5058$                     & $1.535$                 & $0.4937$                & $2.0287$                     & $1.751$                 & $0.5523$                & $2.3$                     \\
$\Pi$             & $10080$                 & $41870$                 & $51950$                      & $13690$                 & $55220$                 & $68910$                      & $15620$                 & $61770$                 & $77390$                      \\
$S \;(10^6)$      & $4.271$                 & $0.67$                  & $4.941$                      & $2.971$                 & $0.4166$                & $3.3876$                     & $1.074$                 & $0.0967$                & $1.17$                     \\
$Q \;(10^7)$      & $1.001$                 & $0.158$                 & $1.159$                      & $0.2949$                & $0.042$                 & $0.3369$                     & $0.037$                 & $0.0034$                & $0.04$                     \\
$I \;(10^7)$      & $2.352$                 & $0.445$                 & $2.797$                      & $3.191$                 & $0.586$                 & $3.777$                      & $3.639$                 & $0.656$                 & $4.295$                      \\ \hline
                  &                         &                         &                              &                         &                         &                              &                         &                         &                              \\
$\Omega \;(\%)$    & $62.90$                 & $66.57$                 & $63.78$                      & $85.37$                 & $87.69$                 & $85.93$                      & $97.39$                 & $98.10$                 & $97.56$                      \\
$\Pi \;(\%)$      & $63.00$                 & $66.83$                 & $66.05$                      & $85.56$                 & $88.14$                 & $87.62$                      & $97.63$                 & $98.60$                 & $98.40$                      \\
$S \;(\%)$        & $3447$               & $27203$               & $3910$                    & $2398$               & $16914$               & $2681$                    & $867$                & $3926$                & $926$                     \\
$Q \;(\%)$        & $26.50$                 & $23.17$                 & $25.99$                      & $7.81$                  & $6.16$                  & $7.55$                       & $0.98$                  & $0.50$                  & $0.91$                       \\
$I \;(\%)$        & $62.52$                 & $6.62$                  & $63.08$                      & $84.82$                 & $8.72$                  & $85.18$                      & $96.73$                 & $9.76$                  & $96.86$                      \\ \hline
\end{tabular}
\end{table}

Figure 23 shows curves of severe cases of CoViD-19 $D_{2j}$, $j=y,o$, fixing 
$\eta _{2}=0.035$ $days^{-1}$, and varying $\eta _{3}$. The beginning of
release is at $t=63$, a week later. For instance, when $\eta _{3}=0.035$ $%
days^{-1}$, the peaks are for young and elder persons, respectively, $%
2.084\times 10^{5}$ and $8.197\times 10^{4}$, which occur at $t=108$ and $%
107 $. In comparison with Figure 21, the peaks are decreased for young and
elder persons in, respectively, $9.8\%$ and $9.5\%$, which are both delayed
in $9$ days.

\begin{figure}[!h]
\centering
\subfloat[]{
\includegraphics[scale=0.45]{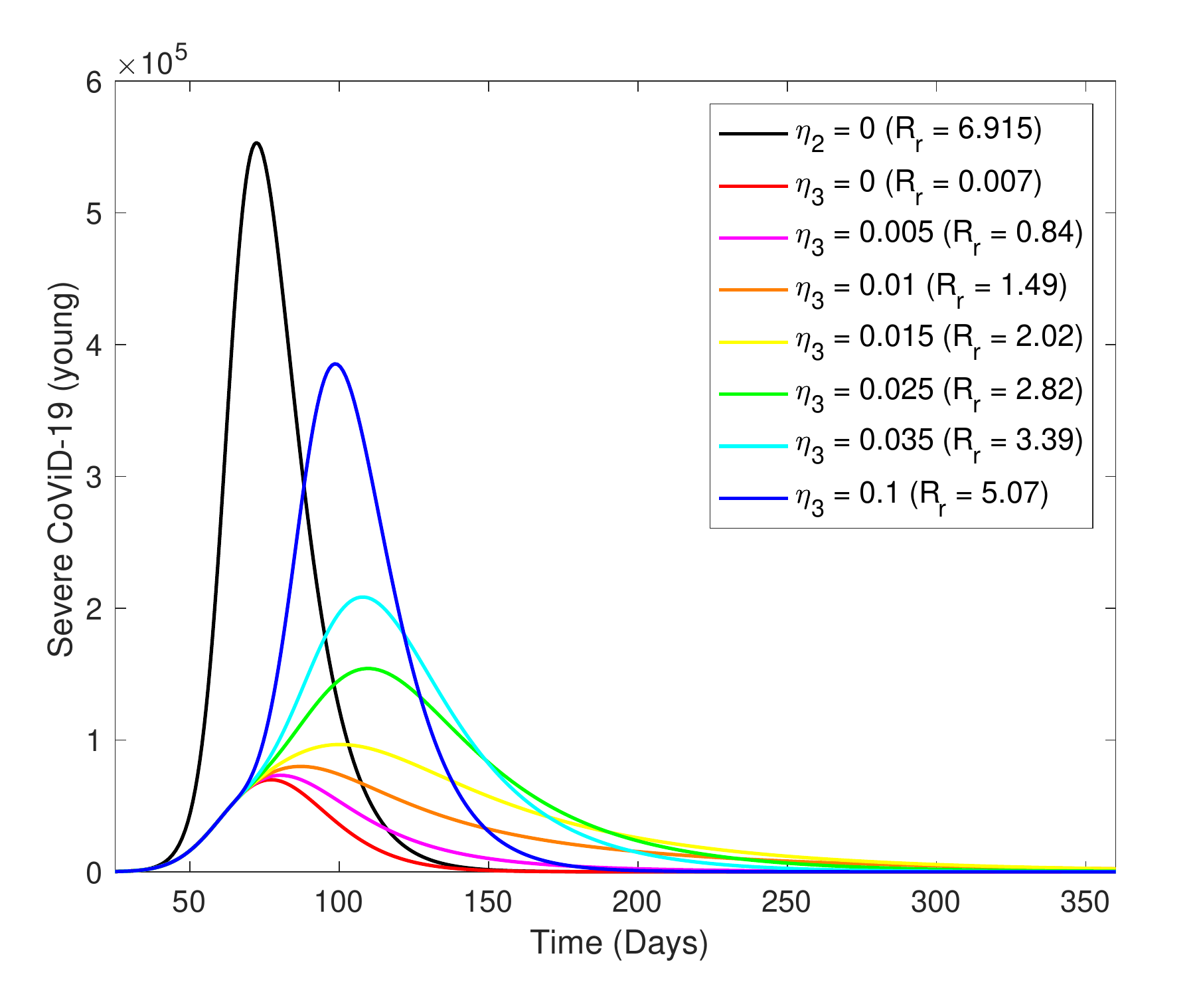}
}
\subfloat[]{
\includegraphics[scale=0.45]{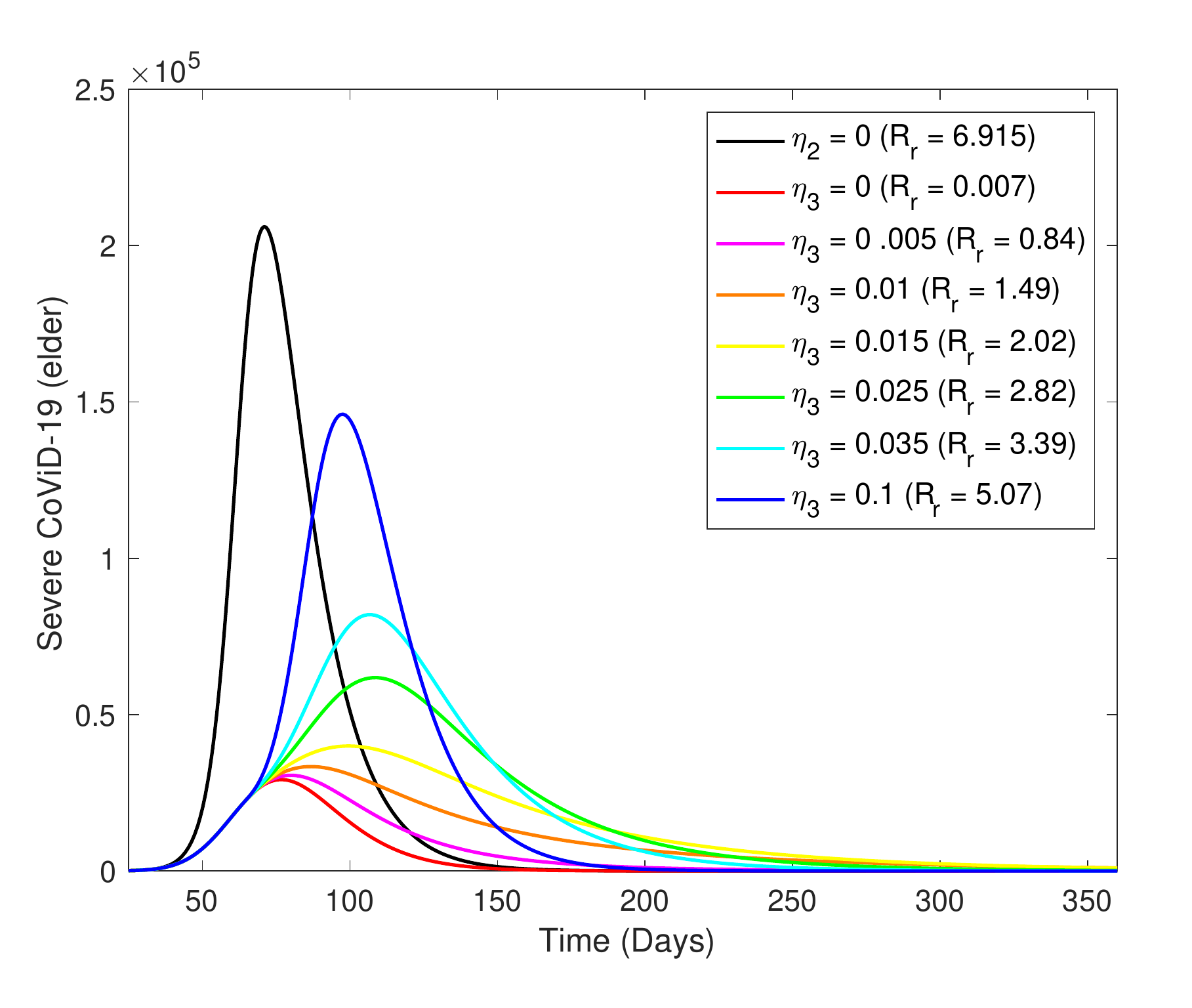}
}
\caption{The curves of severe cases of CoViD-19 $D_{2j}$, $j=y,o$, fixing $%
\eta _{2}=0.035$ $days^{-1}$, and varying $\eta _{3}$. Curves from top to
bottom corresponds to decreasing $\eta _{3}$. The beginning of release is at 
$t=63$.}
\end{figure}

The curves of accumulated cases of severe CoViD-19 $\Omega $, accumulated
cases of CoViD-19 deaths $\Pi $, the number of isolated susceptible person $%
S^{is}$, and the number of immune persons $I$ are similar than those shown
in foregoing section. For this reason, we present in Table 7 ($\eta
_{2}=0.035$ $days^{-1}$ fixed) their values at $t=360$ for young, elder and
all persons, letting $\eta _{3o}=0.015$, $\eta _{3o}=0.035$ and $\eta
_{3o}=0.1$ ($days^{-1}$). Values for $\Omega $, $\Pi $, $S^{is}$ and $I$,
for $\eta _{2}=0$, are those used in Table 3, as well as the definitions of
the percentages.

\begin{table}[!h]
\centering
\caption{Values and percentages of $\Omega $, $\Pi $, $Q$ and $I$
fixing $\eta _{2y}=0.01$ $days^{-1}$ and varying $\eta _{2o}=0.015$, $\eta
_{2o}=0.035$ and $\eta _{2o}=0.1$ ($days^{-1}$). $y$, $o$ and $\Sigma $
stand for, respectively, young, elder and total persons. Releasing initiates
at $t=63$.}
\begin{tabular}{llllllllll}
\hline
\multicolumn{1}{c}{\multirow{2}{*}{}} & \multicolumn{3}{c}{$\eta_3 = 0.015$}                                             & \multicolumn{3}{c}{$\eta_3 = 0.035$}                                             & \multicolumn{3}{c}{$\eta_3 = 0.1$}                                               \\
\multicolumn{1}{c}{}                  & \multicolumn{1}{c}{$y$} & \multicolumn{1}{c}{$o$} & \multicolumn{1}{c}{$\Sigma$} & \multicolumn{1}{c}{$y$} & \multicolumn{1}{c}{$o$} & \multicolumn{1}{c}{$\Sigma$} & \multicolumn{1}{c}{$y$} & \multicolumn{1}{c}{$o$} & \multicolumn{1}{c}{$\Sigma$} \\ \hline
$\Omega \;(10^6)$                      & $1.111$                 & $0.37$                  & $1.481$                      & $1.527$                 & $0.49$                  & $2.017$                      & $1.747$                 & $0.55$                  & $2.297$                      \\
$\Pi$                                 & $9885$                  & $41260$                 & $51145$                      & $13620$                 & $55020$                 & $68640$                      & $15580$                 & $61660$                 & $77240$                      \\
$S \;(10^6)$                          & $4.358$                 & $0.68$                  & $5.038$                      & $3.045$                 & $0.43$                  & $3.475$                      & $1.126$                 & $0.105$                 & $1.231$                      \\
$Q \;(10^7)$                          & $1.032$                 & $0.162$                 & $1.194$                      & $0.3024$                & $0.043$                 & $0.3454$                     & $0.039$                 & $0.0037$                & $0.0427$                     \\
$I \;(10^7)$                          & $2.309$                 & $0.439$                 & $2.748$                      & $3.176$                 & $0.585$                 & $3.761$                      & $3.632$                 & $0.655$                 & $4.287$                      \\ \hline
                                      &                         &                         &                              &                         &                         &                              &                         &                         &                              \\
$\Omega \;(\%)$                        & $61.79$                 & $65.72$                 & $62.73$                      & $84.93$                 & $87.03$                 & $85.43$                      & $97.16$                 & $97.69$                 & $97.29$                      \\
$\Pi \;(\%)$                          & $61.78$                 & $65.86$                 & $65.03$                      & $85.13$                 & $87.82$                 & $87.27$                      & $97.38$                 & $98.42$                 & $98.21$                      \\
$S \;(\%)$                            & $3517$               & $27609$              & $3987$                    & $2458$               & $17458$              & $2750$                    & $909$                & $4263$               & $974$                     \\
$Q \;(\%)$                            & $27.32$                 & $23.75$                 & $26.77$                      & $8.00$                  & $6.30$                  & $7.74$                       & $1.03$                  & $0.54$                  & $0.96$                       \\
$I \;(\%)$                            & $61.38$                 & $6.53$                  & $61.97$                      & $84.42$                 & $8.70$                  & $84.82$                      & $96.54$                 & $9.74$                  & $96.68$                      \\ \hline
\end{tabular}
\end{table}

Comparing Figures 21, 22 and 23, the peaks are increased in $9\%$ and
anticipated in $6$ days if isolation is released $7$ days earlier, while the
peaks are decreased in $10\%$ and delayed in $9$ days if isolation is
released $7$ days later. From Tables 5, 6 and 7, the increase in severe
coViD-19 cases and deaths due to this disease by anticipating isolation in $%
7 $ days are $0.9\%$, $0.3\%$ and $0.06\%$ for, respectively, $\eta
_{3}=0.015$, $0.035$ and $0.1$ ($days^{-1}$); while by delaying in $7$ days,
both are decreased in $0.9\%$, $0.6\%$ and $0.2\%$ for, respectively, $\eta
_{3}=0.015 $, $0.035$ and $0.1$ ($days^{-1}$). However, $0.9\%$ represents $%
95$ deaths.

In Figure 24 we show releasing occurring without isolation, that is, from
time $0$ to $27$, we have $R_{0}=6.915$ (no isolation), we have regime
1-type isolation from $27$ to $56$, $R_{r}=0.007$, and sinceafter $56$, we
have only releasing with $R_{0}=6.915$ ($\eta _{2}=0$ and $\eta _{3}=0.035$ $%
days^{-1}$).

\begin{figure}[!h]
\centering
\subfloat[]{
\includegraphics[scale=0.45]{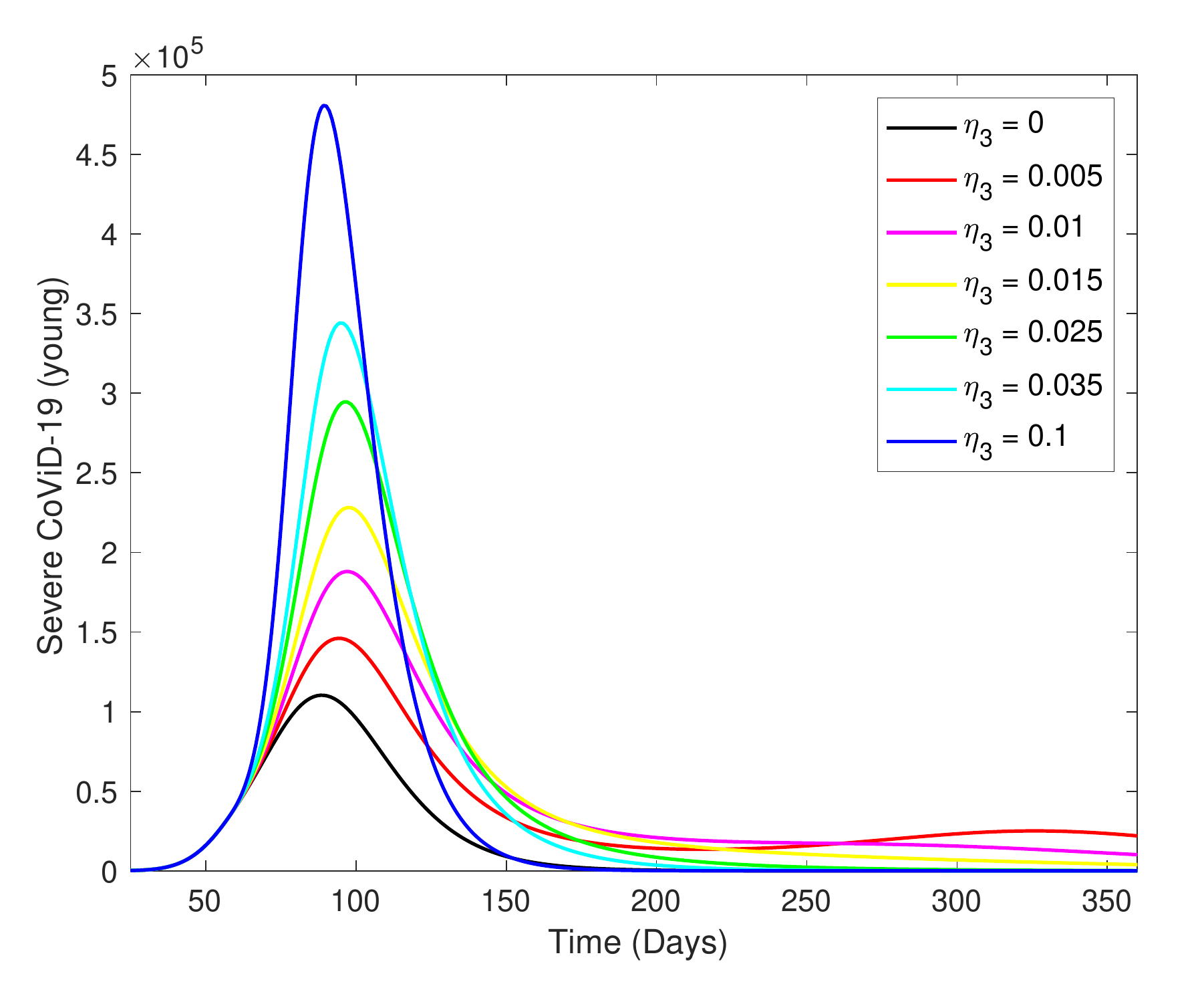}
}
\subfloat[]{
\includegraphics[scale=0.45]{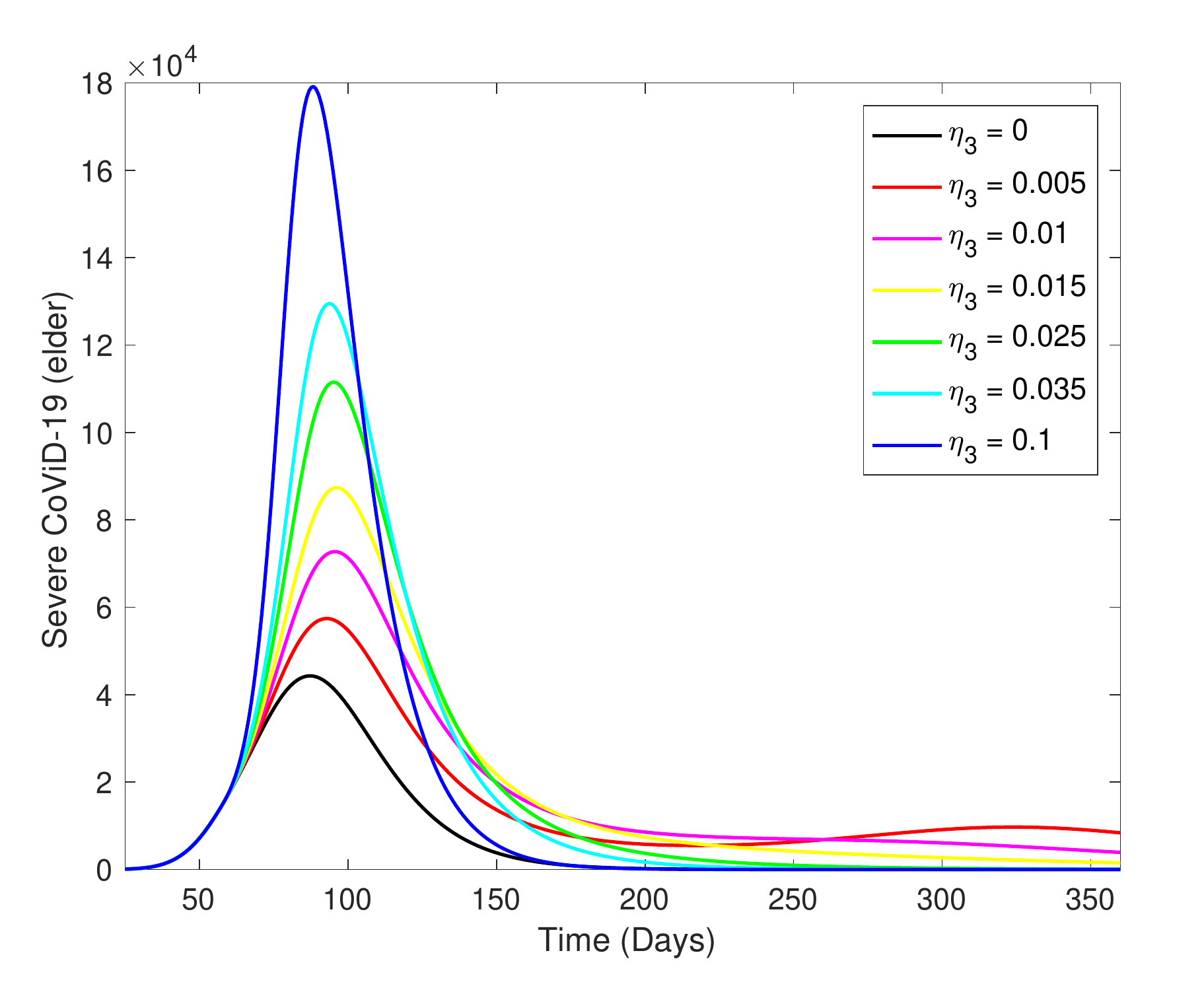}
}
\caption{The curves of severe cases of CoViD-19 $D_{2j}$, $j=y,o$, letting 
$\eta _{2}=0.035$ and\ $\eta _{3}=0$ ($days^{-1}$) during isolation, and $%
\eta _{2}=0$ and\ $\eta _{3}=0.035$ ($days^{-1}$) when releasing begins at $%
t=56$. Curves from top to bottom corresponds to decreasing $\eta _{3}$.}
\end{figure}

When releasing is done without new isolations, there appears small epidemics
(see curve for $\eta _{3}=0.005$), which is delayed as $\eta _{3}$ increases
(for other values, the second small epidemics does not appear until $360$
days). If releasing strategy is done, the first wave does not vanish
completely, except for huge releasing scheme (higher $\eta _{3}$). This is a
good epidemiological scenario due to not only in the diminishing in the
pressure for hospitalization (consequently, decreases deaths), but also in
the increasing in immune persons, hence decreasing the effective
reproduction number (known as herd immunity).

\section{Discussion}

System of equations (\ref{system2a}), (\ref{system1c}) and (\ref{system2b})
were simulated to providing epidemiological scenarios. These scenarios are
more reliable if based on credible values assigned to model parameters. We
used ratio $4:1$ for the ratios of asymptomatic:symptomatic and mild:severe
(non-hospitalized:hospitalized) CoViD-19 \cite{bepi8}. Also, we let $\alpha
_{y}=0.1\alpha _{o}$, and $\alpha _{o}$ must be such that deaths will occur
in $10\%$ of hospitalized elder persons, hence, $1\%$ of hospitalized young
persons will die \cite{who}. We used overvalued parameters, except maybe the
ratio between asymptomatic:symptomatic, which is completely unknown. In many
viruses, the ratio is higher than $4:1$, but for new coronavirus is unknown.
When mass testing against new coronavirus could be done, then this ratio can
be estimated.

The least square estimation method was approximated by the sum of the square
of the distance between parametrized curve and observed data. When
estimation of epidemic curves are based on few available data, in general
parameters are overestimated. Hence, both transmission and mortality rates
were overestimated. Fortunately, there was another information to use: $10\%$
of fatality among elder hospitalized persons. Taking into account this
information, we estimated lower mortality rates, but estimated transmission
rates were those based on few available data. Hence, the basic reproduction
number $R_{0}=6.915$ seems overestimated.

Let us consider estimation of transmission and mortality rates based on few
data. From Figures 7 and 8, it is expected at the end of the fist wave of
epidemics, $2.36$ million of severe (hospitalized) CoViD-19 cases, and $250$
thousand of deaths due to this disease in the S\~{a}o Paulo State. If we
consider a $5$-times higher inhabitants than the S\~{a}o Paulo State, it is
expected $11.8$ million of severe (hospitalized) CoViD-19 cases, and $1,250$
thousand of deaths. Approximately these numbers of cases and deaths were
projected to Brazil by Ferguson \textit{et al}. \cite{ferguson}. However,
the second method of estimation for fatality rates resulted in $78.7$
thousand of deaths in the S\~{a}o Paulo State, but the number of severe
cases is the same. Hence, extrapolating to Brazil, the number is $383$
thousand of deaths.

We address the question of the discrepancy in providing number of deaths
during the first wave of epidemics. Mathematical and computational
(especially agent based models) models that are based on data to estimate
model parameters, these models must be fed continuously with new data and
reestimated model parameters. As the number of data increases, their
estimations become more and more reliable. Hence, initial estimations and
forecasting are extremely bad, and, moreover, they become dangerous when
predicting catastrophic scenarios, which can lead to formulate mistaken
public health policies.

With respect to isolation of susceptible persons, depending on the target we
have two strategies. If the goal is decreasing the number of CoViD-19 cases
in order to adequate capacity of Hospital and ICU, the better strategy is
isolating mor young than elder persons. However, if death due to CoViD-19 is
the main goal, better strategy is isolating more elder than young persons.

We also studied releasing strategies. We compare the releasing that will be
initiated in April 22, with releasing one week earlier (April 19) and one
week later (April 29).

The estimated basic reproduction number and its partial values were $%
R_{0}=6.915$ (partials $R_{0y}=5.606$ and $R_{0o}=1.309$), and the
asymptotic fraction of\ susceptible persons and its partial fractions
provided by Runge-Kutta method were $s^{\ast }=0.15008$, $s_{y}^{\ast
}=0.14660$ and $s_{o}^{\ast }=0.00348$. Using equation (\ref{rinv}), we
obtain $1/R_{0}=0.1446$. Clearly, $s^{\ast }$ is not the inverse of the
basic reproduction number $R_{0}$, and $f(s^{\ast },s_{y}^{\ast
},s_{o}^{\ast })$ in equation (\ref{rinv}) is not $s^{\ast }=s_{y}^{\ast
}+s_{o}^{\ast }$, neither $R_{0y}s_{y}^{\ast }+R_{0o}s_{o}^{\ast }$. The
analysis of the non-trivial equilibrium point to find $f(s^{\ast
},s_{y}^{\ast },s_{o}^{\ast })$ is left to a further work. In order to
understand this question, we suppose that new coronavirus is circulating in
non-communicating young and elder sub-populations, then each population
approach to $s_{y}^{\ast }=1/R_{0y}=0.178$ or $s_{o}^{\ast }=1/R_{0o}=0.764$
at steady state (non-trivial equilibrium point $P^{\ast }$). But, new
coronavirus is circulating in a homogeneously mixed populations of young and
elder persons (this is a strong assumption of the modeling). Using equation (%
\ref{force}), let us calculate the forces of infection $\lambda _{1}=\beta
_{1y}A_{y}+\beta _{2y}D_{1y}$ (contribution due to infectious young
persons), $\lambda _{2}=\beta _{1o}A_{o}+\beta _{2o}D_{1o}$ (elder persons)
and $\lambda =\lambda _{1}+\lambda _{2}$ (both classes), which are shown in
Figure 25 ($\lambda $ is the force of infection acting on young persons, and
for elder persons, it is enough multiplying by the factor $\psi $).

\begin{figure}[!h]
\centering
\includegraphics[scale=0.55]{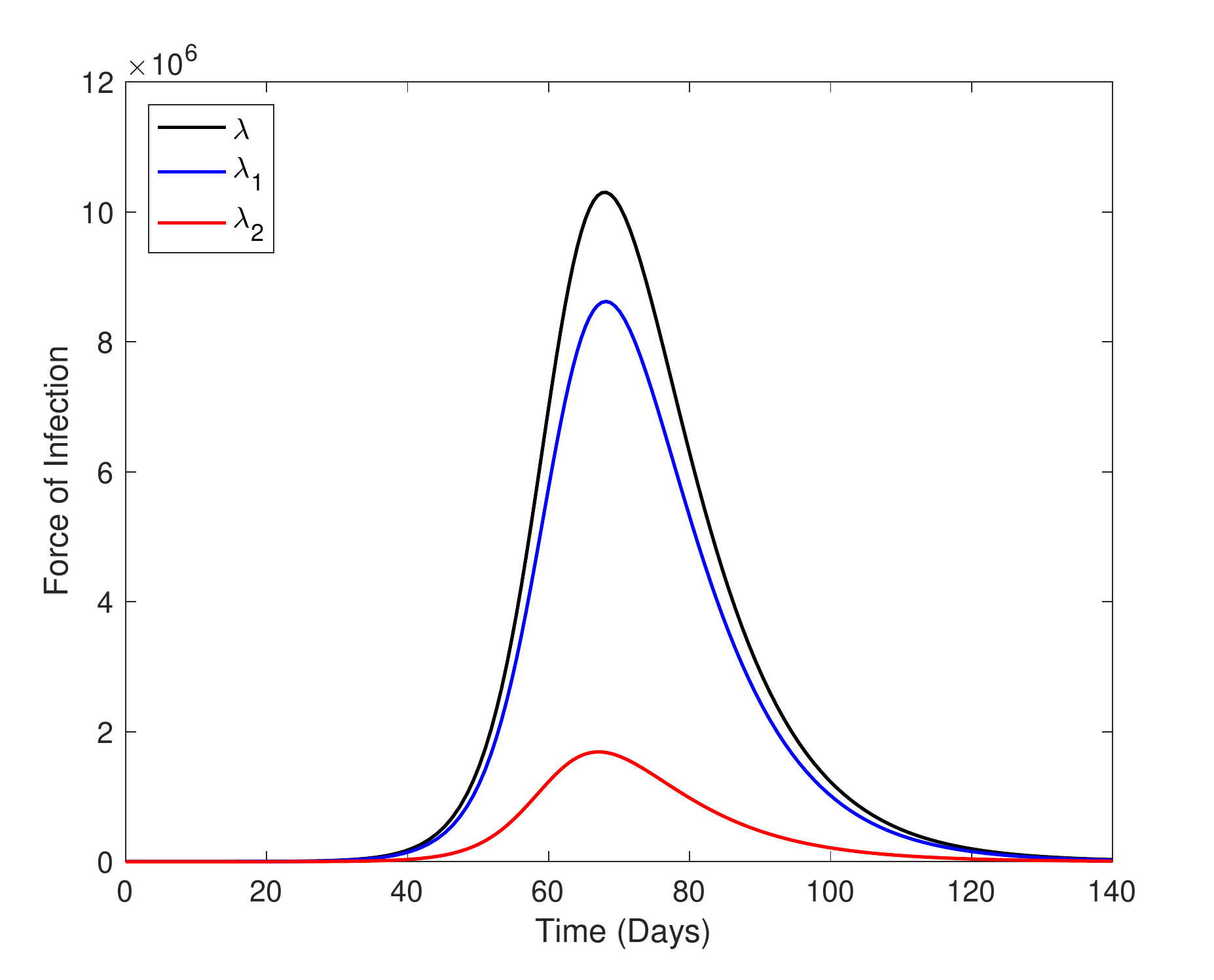}
\caption{The forces of infection $\lambda _{1}=\beta _{1y}A_{y}+\beta
_{2y}D_{1y}$ (young persons), $\lambda _{2}=\beta _{1o}A_{o}+\beta
_{2o}D_{1o}$ (elder persons) and $\lambda =\lambda _{1}+\lambda _{2}$ (both
classes).}
\end{figure}

The peaks of the force of infection for $\lambda _{1}$, $\lambda _{2}$ and $%
\lambda $ are, respectively, $8.62\times 10^{6}$, $1.69\times 10^{6}$ and $%
10.31\times 10^{6}$, which occur at $68.18$, $66.94$\ and $68.18$ (days),
and contributions at peak of $\lambda _{1}$ and $\lambda _{2}$ with respect
to $\lambda $ are $83.6\%$ and $16.4\%$. The ratio between peaks $\lambda
_{1}$:$\lambda _{2}$ is $5.1$:$1$, which is close to the ratio between
numbers of young:elder $5.5$:$1$. When virus circulates in mixed
populations, young and elder persons are infected additionally by,
respectively, elder ($\lambda _{2}$) and young ($\lambda _{1}$) persons.
This is the reason for the actual equilibrium values are bigger ($%
s_{y}^{\ast }>1/R_{0y}$ and $s_{o}^{\ast }>R_{0o}$), but among elder persons
the increase ($220$-times) is huge ($\lambda _{1}$, very big, acting in
relatively small population $S_{o}$). For this reason contacts between elder
and young persons must be avoided.

\section{Conclusion}

We formulated a mathematical model considering two subpopulations comprised
by young and elder persons to study CoViD-19 in the S\~{a}o Paulo State,
Brazil. The model considered continuos but constant rates of isolation and
releasing. In a future work, we change rates to describe isolation and
releasing by proportions of susceptible persons being isolated or released.
The reason behind this is the absence of translation of rates to proportions.

Our model estimated quite same number of severe CoViD-19 cases predicted by
Ferguson \textit{et al}. \cite{ferguson} for Brazil, but $3.3$-times lower
for deaths due to CoViD-19. The difference is mainly done by estimation of
the additional mortality rates. It is also expected that $R_{0}$ must be
lower if additional information may exist, or more data will be available.
As consequence, maybe severe CoViD019 cases should be much lower
(consequently, deaths also). If currently adopted lockdown is indeed based
on the goal of decreasing hospitalized CoViD-19 cases, then our model
agrees, since it predicts that higher number of young and elder persons must
be isolated in order to achieve this objective. However, if the goal is
reduction in the number of deaths due to CoViD-19, as much elder persons
must be isolated, but not so much young persons. Remember that in a mixing
of young and elder persons, the infection is much harmfull in elder than
young persons, which is reason to avoid contact between them. An optimal
rates of isolation of young and elder persons to reduce both CoViD-19 cases
and deaths can be obtained by optimal control theory \cite{thome}.

If vaccine and efficient treatments are available, the pandemic of new
coronavirus should not be considered a threaten to public health. However,
currently, there is not vaccine, neither efficient treatment. For this
reason the adoption of isolation or lockdown is a recommended strategy,
which can be less hardly implemented if there is enough kit to test against
new coronavirus. Remember that all isolation strategies considered in our
model assume the identification of susceptible persons. Hence isolation as
control mechanism allows an additional time to seek for cure (medicine)
and/or develop vaccine.

\appendix

\renewcommand{\theequation}{\Alph{section}.\arabic{equation}} %
\setcounter{equation}{0}

\section{Trivial equilibrium and its stability}

By the fact that $N$ is varying, the system is non-autonomous non-linear
differential equations. To obtain autonomous system of equations, we use
fractions of individuals in each compartment, defined by, with $j=y$ and $o$,%
\[
\begin{array}{ccc}
x_{j}=\frac{X_{j}}{N}, & \mathrm{where} & 
X=S_{j},Q_{j},E_{j},A_{j},Q_{1_{j}},D_{1_{j}},Q_{2_{j}},D_{2_{j}},I,%
\end{array}%
\]%
resulting in%
\[
\frac{d}{dt}x_{j}\equiv \frac{d}{dt}\frac{X_{j}}{N}=\frac{1}{N}\frac{d}{dt}%
X_{j}-x_{j}\frac{1}{N}\frac{d}{dt}N=\frac{1}{N}\frac{d}{dt}X_{j}-x\left(
\phi -\mu \right) +x_{j}\left( \alpha _{y}d_{2y}+\alpha _{o}d_{2o}\right) , 
\]%
using equation (\ref{nvar}) for $N$. Hence, equations (\ref{system2a}), (\ref%
{system1c}) and (\ref{system2b}) in terms of fractions become, for
susceptible persons,%
\begin{equation}
\left\{ 
\begin{array}{rll}
\displaystyle\frac{d}{dt}s_{y} & = & \phi -\left( \eta _{2y}+\varphi +\phi
\right) s_{y}-\lambda s_{y}+\eta _{3y}q_{y}+s_{y}\left( \alpha
_{y}d_{2y}+\alpha _{o}d_{2o}\right) \\ 
\displaystyle\frac{d}{dt}s_{o} & = & \varphi s_{y}-\left( \eta _{2o}+\phi
\right) s_{o}-\lambda \psi s_{o}+\eta _{3o}q_{o}+s_{o}\left( \alpha
_{y}d_{2y}+\alpha _{o}d_{2o}\right) ,%
\end{array}%
\right.  \label{sis1}
\end{equation}%
for infected persons,%
\begin{equation}
\left\{ 
\begin{array}{rll}
\displaystyle\frac{d}{dt}q_{j} & = & \eta _{2j}s_{j}-\left( \eta _{3j}+\phi
\right) q_{j}+q_{j}\left( \alpha _{y}d_{2y}+\alpha _{o}d_{2o}\right) \\ 
\displaystyle\frac{d}{dt}e_{j} & = & \lambda \left( \delta _{jy}+\psi \delta
_{jo}\right) s_{j}-\left( \sigma _{j}+\phi \right) e_{j}+e_{j}\left( \alpha
_{y}d_{2y}+\alpha _{o}d_{2o}\right) \\ 
\displaystyle\frac{d}{dt}a_{j} & = & p_{j}\sigma _{j}e_{j}-\left( \gamma
_{j}+\eta _{j}+\chi _{j}+\phi \right) a_{j}+a_{j}\left( \alpha
_{y}d_{2y}+\alpha _{o}d_{2o}\right) \\ 
\displaystyle\frac{d}{dt}q_{1j} & = & \left( \eta _{j}+\chi _{j}\right)
a_{j}-\left( \gamma _{j}+\phi \right) q_{1j}+q_{1j}\left( \alpha
_{y}d_{2y}+\alpha _{o}d_{2o}\right) \\ 
\displaystyle\frac{d}{dt}d_{1j} & = & \left( 1-p_{j}\right) \sigma
_{j}e_{j}-\left( \gamma _{1j}+\eta _{1j}+\phi \right) d_{1j}+d_{1j}\left(
\alpha _{y}d_{2y}+\alpha _{o}d_{2o}\right) \\ 
\displaystyle\frac{d}{dt}q_{2j} & = & \left( \eta _{1j}+m_{j}\gamma
_{1j}\right) d_{1j}-\left( \gamma _{j}+\xi _{j}+\phi \right)
q_{2j}+q_{2j}\left( \alpha _{y}d_{2y}+\alpha _{o}d_{2o}\right) \\ 
\displaystyle\frac{d}{dt}d_{2j} & = & \left( 1-m_{j}\right) \gamma
_{1j}d_{1j}+\xi _{j}q_{2j}-\left( \gamma _{2j}+\theta _{j}+\phi +\alpha
_{j}\right) d_{2j}+d_{2j}\left( \alpha _{y}d_{2y}+\alpha _{o}d_{2o}\right) ,%
\end{array}%
\right.  \label{sis2}
\end{equation}%
and for immune persons%
\begin{equation}
\begin{array}{rll}
\displaystyle\frac{d}{dt}i & = & \gamma _{y}a_{y}+\gamma _{y}q_{1y}+\gamma
_{y}q_{2y}+\left( \gamma _{2y}+\theta _{y}\right) d_{2y}+\gamma
_{o}a_{o}+\gamma _{o}q_{1o}+\gamma _{o}q_{2o}+\left( \gamma _{2o}+\theta
_{o}\right) d_{2o}-\phi i+ \\ 
&  & i\left( \alpha _{y}d_{2y}+\alpha _{o}d_{2o}\right) ,%
\end{array}
\label{sis3}
\end{equation}%
where $\lambda $ is the force of infection given by equation (\ref{force}),
and%
\[
\sum_{j=y,o}\left(
s_{j}+q_{j}+e_{j}+a_{j}+q_{1j}+d_{1j}+q_{2j}+d_{2j}\right) +i=1, 
\]%
which is autonomous system of equations. We remember that all classes vary
with time, however their fractions attain steady state (the sum of
derivatives of all classes is zero). This system of equations is not easy to
determine non-trivial (endemic) equilibrium point $P^{\ast }$. Hence, we
restrict our analysis with respect to trivial (disease free) equilibrium
point.

The trivial or disease free equilibrium $P^{0}$ is given by%
\[
P^{0}=\left(
s_{j}^{0},q_{j}^{0},e_{j}^{0}=0,a_{j}^{0}=0,q_{1j}^{0}=0,d_{1j}^{0}=0,q_{2j}^{0}=0,d_{2j}^{0}=0,i^{0}=0\right) , 
\]%
for $j=y$ and $o$, where%
\begin{equation}
\left\{ 
\begin{array}{l}
s_{y}^{0}=\displaystyle\frac{\phi \left( \eta _{3y}+\phi \right) }{\phi
\left( \eta _{2y}+\eta _{3y}+\phi \right) +\varphi \left( \eta _{3y}+\phi
\right) } \\ 
q_{y}^{0}=\displaystyle\frac{\phi \eta _{2y}}{\phi \left( \eta _{2y}+\eta
_{3y}+\phi \right) +\varphi \left( \eta _{3y}+\phi \right) } \\ 
s_{o}^{0}=\displaystyle\frac{\varphi \left( \eta _{3y}+\phi \right) \left(
\eta _{3o}+\phi \right) }{\left[ \phi \left( \eta _{2y}+\eta _{3y}+\phi
\right) +\varphi \left( \eta _{3y}+\phi \right) \right] \left( \eta
_{2o}+\eta _{3o}+\phi \right) } \\ 
q_{o}^{0}=\displaystyle\frac{\varphi \eta _{2o}\left( \eta _{3y}+\phi
\right) }{\left[ \phi \left( \eta _{2y}+\eta _{3y}+\phi \right) +\varphi
\left( \eta _{3y}+\phi \right) \right] \left( \eta _{2o}+\eta _{3o}+\phi
\right) },%
\end{array}%
\right.  \label{trivial}
\end{equation}%
with $s_{y}^{0}+q_{y}^{0}+s_{o}^{0}+q_{o}^{0}=1$.

Due to $17$ equations, we do not deal with characteristic equation
corresponding to Jacobian matrix evaluated at $P^{0}$, but we apply the next
generation matrix theory \cite{diekman}.

The next generation matrix, evaluated at the trivial equilibrium $P^{0}$, is
obtained considering the vector of variables $x=\left(
e_{y},a_{y},d_{1y},e_{o},a_{o},d_{1o}\right) $. We apply method proposed in 
\cite{yang1} and proved in \cite{yang2}. There are control mechanisms
(isolation), hence we obtain the reduced reproduction number $R_{r}$ by
isolation.

In order to obtain the reduced reproduction number, diagonal matrix $V$ is
considered. Hence, the vectors $f$ and $v$ are%
\begin{equation}
f^{T}=\left( 
\begin{array}{c}
\lambda s_{y}+e_{y}\left( \alpha _{y}d_{2y}+\alpha _{o}d_{2o}\right) \\ 
p_{y}\sigma _{y}e_{y}+a_{y}\left( \alpha _{y}d_{2y}+\alpha _{o}d_{2o}\right)
\\ 
\left( 1-p_{y}\right) \sigma _{y}e_{y}+d_{1y}\left( \alpha _{y}d_{2y}+\alpha
_{o}d_{2o}\right) \\ 
\lambda \psi s_{o}+e_{o}\left( \alpha _{y}d_{2y}+\alpha _{o}d_{2o}\right) \\ 
p_{o}\sigma _{o}e_{o}+a_{o}\left( \alpha _{y}d_{2y}+\alpha _{o}d_{2o}\right)
\\ 
\left( 1-p_{o}\right) \sigma _{o}e_{o}+d_{1o}\left( \alpha _{y}d_{2y}+\alpha
_{o}d_{2o}\right)%
\end{array}%
\right)  \label{fv1}
\end{equation}%
and%
\begin{equation}
\begin{array}{l}
v^{T}=\left( 
\begin{array}{c}
\left( \sigma _{y}+\phi \right) e_{y} \\ 
\left( \gamma _{y}+\eta _{y}+\chi _{y}+\phi \right) a_{y} \\ 
\left( \gamma _{1y}+\eta _{1y}+\phi \right) d_{1y} \\ 
\left( \sigma _{o}+\phi \right) e_{o} \\ 
\left( \gamma _{o}+\eta _{o}+\chi _{o}+\phi \right) a_{o} \\ 
\left( \gamma _{1o}+\eta _{1o}+\phi \right) d_{1o}%
\end{array}%
\right) ,%
\end{array}
\label{fv2}
\end{equation}%
where the superscript $T$ stands for the transposition of a matrix, from
which we obtain the matrices $F$ and $V$ (see \cite{diekman}) evaluated at
the trivial equilibrium $P^{0}$, which were omitted. The next generation
matrix $FV^{-1}$ is%
\[
FV^{-1}=\left[ 
\begin{array}{cccccc}
0 & \frac{\beta _{1y}s_{y}^{0}}{\gamma _{y}+\eta _{y}+\chi _{y}+\phi } & 
\frac{\beta _{2y}s_{y}^{0}}{\gamma _{1y}+\eta _{1y}+\phi } & 0 & \frac{\beta
_{1o}s_{y}^{0}}{\gamma _{o}+\eta _{o}+\chi _{o}+\phi } & \frac{\beta
_{2o}s_{y}^{0}}{\gamma _{1o}+\eta _{1o}+\phi } \\ 
\displaystyle\frac{p_{y}\sigma _{y}}{\sigma _{y}+\phi } & 0 & 0 & 0 & 0 & 0
\\ 
\frac{\left( 1-p_{y}\right) \sigma _{y}}{\sigma _{y}+\phi } & 0 & 0 & 0 & 0
& 0 \\ 
0 & \frac{\beta _{1y}\psi s_{o}^{0}}{\gamma _{y}+\eta _{y}+\chi _{y}+\phi }
& \frac{\beta _{2y}\psi s_{o}^{0}}{\gamma _{1y}+\eta _{1y}+\phi } & 0 & 
\frac{\beta _{1o}\psi s_{o}^{0}}{\gamma _{o}+\eta _{o}+\chi _{o}+\phi } & 
\frac{\beta _{2o}\psi s_{o}^{0}}{\gamma _{1o}+\eta _{1o}+\phi } \\ 
0 & 0 & 0 & \frac{p_{o}\sigma _{o}}{\sigma _{o}+\phi } & 0 & 0 \\ 
0 & 0 & 0 & \frac{\left( 1-p_{o}\right) \sigma _{o}}{\sigma _{o}+\phi } & 0
& 0%
\end{array}%
\right] , 
\]%
and the characteristic equation corresponding to $FV^{-1}$ is%
\begin{equation}
\lambda ^{4}\left( \lambda ^{2}-R_{r}\right) =0,  \label{charact_gross}
\end{equation}%
where the reduced reproduction number $R_{r}$ and its partial reduced
reproduction numbers $R_{ry}$ and $R_{ro}$ are%
\begin{equation}
\begin{array}{ccccc}
R_{r}=R_{ry}+R_{ro}, & \mathrm{where} & \left\{ 
\begin{array}{l}
R_{ry}=R_{0y}s_{y}^{0} \\ 
R_{ro}=R_{0o}\psi s_{o}^{0},%
\end{array}%
\right. & \mathrm{with} & \left\{ 
\begin{array}{l}
R_{0y}=p_{y}R_{0y}^{1}+\left( 1-p_{y}\right) R_{0y}^{2} \\ 
R_{0o}=p_{o}R_{0o}^{1}+\left( 1-p_{o}\right) R_{0o}^{2},%
\end{array}%
\right.%
\end{array}
\label{Rr}
\end{equation}%
and $R_{0y}$ and $R_{0o}$ are the basic partial reproduction numbers defined
by%
\begin{equation}
\left\{ 
\begin{array}{lll}
R_{0y}^{1}=\displaystyle\frac{\sigma _{y}}{\sigma _{y}+\phi }\frac{\beta
_{1y}}{\gamma _{y}+\eta _{y}+\chi _{y}+\phi }, & \mathrm{and} & R_{0y}^{2}=%
\displaystyle\frac{\sigma _{y}}{\sigma _{y}+\phi }\frac{\beta _{2y}}{\gamma
_{1y}+\eta _{1y}+\phi } \\ 
R_{0o}^{1}=\displaystyle\frac{\sigma _{o}}{\sigma _{o}+\phi }\frac{\beta
_{1o}}{\gamma _{o}+\eta _{o}+\chi _{o}+\phi }, & \mathrm{and} & R_{0o}^{2}=%
\displaystyle\frac{\sigma _{o}}{\sigma _{o}+\phi }\frac{\beta _{2o}}{\gamma
_{1o}+\eta _{1o}+\phi }.%
\end{array}%
\right.  \label{Rr1}
\end{equation}%
Actually, we must have $\eta _{j}=\chi _{j}=\eta _{1j}=\chi _{1j}=0$, with $%
j=i,o$, to be fit in the definition of the basic reproduction number.

Instead of calculating the spectral radius ($\rho \left( FV^{-1}\right) =%
\sqrt{R_{r}}$), we apply procedure in \cite{yang1} (the sum of coefficients
of characteristic equation), resulting in a threshold $R_{r}$. Hence, the
trivial equilibrium point $P^{0}$ is locally asymptotically stable (LAS) if $%
R_{r}<1$.

In order to obtain the fraction of susceptible individuals, $M$ must be the
simplest (matrix with least number of non-zeros). Hence, the vectors $f$ and 
$v$ are%
\[
\begin{array}{lll}
f^{T}=\left( 
\begin{array}{c}
\lambda s_{y} \\ 
0 \\ 
0 \\ 
\lambda \psi s_{o} \\ 
0 \\ 
0%
\end{array}%
\right) & \mathrm{and} & v^{T}=\left( 
\begin{array}{c}
\left( \sigma _{y}+\phi \right) e_{y}-e_{y}\left( \alpha _{y}d_{2y}+\alpha
_{o}d_{2o}\right) \\ 
-p_{y}\sigma _{y}e_{y}+\left( \gamma _{y}+\eta _{y}+\chi _{y}+\phi \right)
a_{y}-a_{y}\left( \alpha _{y}d_{2y}+\alpha _{o}d_{2o}\right) \\ 
-\left( 1-p_{y}\right) \sigma _{y}e_{y}+\left( \gamma _{1y}+\eta _{1y}+\phi
\right) d_{1y}-d_{1y}\left( \alpha _{y}d_{2y}+\alpha _{o}d_{2o}\right) \\ 
\left( \sigma _{o}+\phi \right) e_{o}-e_{o}\left( \alpha _{y}d_{2y}+\alpha
_{o}d_{2o}\right) \\ 
-p_{o}\sigma _{o}e_{o}+\left( \gamma _{o}+\eta _{o}+\chi _{o}+\phi \right)
a_{o}-a_{o}\left( \alpha _{y}d_{2y}+\alpha _{o}d_{2o}\right) \\ 
-\left( 1-p_{o}\right) \sigma _{o}e_{o}+\left( \gamma _{1o}+\gamma
_{3o}+\eta _{1o}+\phi \right) d_{1o}-d_{1o}\left( \alpha _{y}d_{2y}+\alpha
_{o}d_{2o}\right)%
\end{array}%
\right) ,%
\end{array}%
\]%
where superscript $T$ stands for the transposition of a matrix, from which
we obtain the matrices $F$ and $V$ evaluated at the trivial equilibrium $%
P^{0}$, which were omitted. The next generation matrix $FV^{-1}$ is%
\[
FV^{-1}=\left[ 
\begin{array}{cccccc}
R_{0y}s_{y}^{0} & \frac{\beta _{1y}s_{y}^{0}}{\gamma _{y}+\eta _{y}+\chi
_{y}+\phi } & \frac{\beta _{2y}s_{y}^{0}}{\gamma _{1y}+\eta _{1y}+\phi } & 
R_{0o}s_{y}^{0} & \frac{\beta _{1o}s_{y}^{0}}{\gamma _{o}+\eta _{o}+\chi
_{o}+\phi } & \frac{\beta _{2o}s_{y}^{0}}{\gamma _{1o}+\eta _{1o}+\phi } \\ 
0 & 0 & 0 & 0 & 0 & 0 \\ 
0 & 0 & 0 & 0 & 0 & 0 \\ 
R_{0y}\psi s_{o}^{0} & \frac{\beta _{1y}\psi s_{o}^{0}}{\gamma _{y}+\eta
_{y}+\chi _{y}+\phi } & \frac{\beta _{2y}\psi s_{o}^{0}}{\gamma _{1y}+\eta
_{1y}+\phi } & R_{0o}\psi s_{o}^{0} & \frac{\beta _{1o}\psi s_{o}^{0}}{%
\gamma _{o}+\eta _{o}+\chi _{o}+\phi } & \frac{\beta _{2o}\psi s_{o}^{0}}{%
\gamma _{1o}+\eta _{1o}+\phi } \\ 
0 & 0 & 0 & 0 & 0 & 0 \\ 
0 & 0 & 0 & 0 & 0 & 0%
\end{array}%
\right] , 
\]%
and the characteristic equation corresponding to $FV^{-1}$ is%
\[
\lambda ^{5}\left( \lambda -R_{r}\right) =0. 
\]%
The spectral radius is $\rho \left( FV^{-1}\right) =R_{r}=R_{ry}+R_{ro}$
given by equation (\ref{Rr}). Hence, the trivial equilibrium point $P^{0}$
is LAS if $\rho <1$.

Both procedures resulted in the same threshold, hence, according to \cite%
{yang3}, the inverse of the reduced reproduction number $R_{r}$ given by
equation (\ref{Rr}) is a function of the fraction of susceptible individuals
at endemic equilibrium $s^{\ast }$ through%
\begin{equation}
\displaystyle f\left( s^{\ast },s_{y}^{\ast },s_{o}^{\ast }\right) =\frac{1}{%
R_{r}}=\frac{1}{R_{ry}+R_{ro}}=\frac{1}{R_{0y}s_{y}^{0}+R_{0o}\psi s_{o}^{0}}%
,  \label{rinv}
\end{equation}%
where $s^{\ast }=s_{y}^{\ast }+s_{o}^{\ast }$ (see \cite{yang10} \cite{yang3}%
). For this reason, the effective reproduction number $R_{e}$ \cite{yang4},
which varies with time, can not be defined by $R_{e}=R_{0}\left( s_{y}+\psi
s_{o}\right) $, or $R_{e}=R_{0y}s_{y}+R_{0o}\psi s_{o}$. The function $%
f\left( \varkappa \right) $ is determined by calculating the coordinates of
the non-trivial equilibrium point $P^{\ast }$. For instance, for dengue
transmission model, $f\left( s_{1}^{\ast },s_{2}^{\ast }\right) =s_{1}^{\ast
}\times s_{2}^{\ast }$, where $s_{1}^{\ast }$ and $s_{2}^{\ast }$ are the
fractions at equilibrium of, respectively, humans and mosquitoes \cite%
{yang10}. For tuberculosis model considering drug-sensitive and resistant
strains, there is not $f\left( \varkappa \right) $, but $s^{\ast }$ is
solution of a second degree polynomial \cite{yang3}.

From equation (\ref{rinv}), let us assume (or approximate) that $f\left(
s^{\ast },s_{y}^{\ast },s_{o}^{\ast }\right) =s_{y}^{\ast }+s_{o}^{\ast }$.
Then, we can define the effective reproduction number $Re$ as%
\begin{equation}
R_{e}=R_{r}\left( s_{y}+s_{o}\right) ,  \label{Refe}
\end{equation}%
which depends on time, and when attains steady state ($R_{e}=1$), we have $%
s^{\ast }=1/R_{r}$.

When a mechanism of protection of susceptible persons is introduced in a
population, the basic reproduction number $R_{0}$ is reduced to $R_{r}$, the
reduced reproduction number. The protection of susceptible persons is done
or by vaccine (not yet available), or isolation (or quarantine). The
isolation was described by the isolation rate of susceptible persons $\eta
_{2j}$, with $j=y,o$. When $\eta _{2j}=0$, the fraction of young persons and
elders are, from equation (\ref{trivial}),%
\begin{equation}
\left\{ 
\begin{array}{l}
\bar{s}_{y}^{0}=\displaystyle\frac{\phi }{\phi +\varphi } \\ 
\bar{q}_{y}^{0}=0 \\ 
\bar{s}_{o}^{0}=\displaystyle\frac{\varphi }{\phi +\varphi } \\ 
\bar{q}_{o}^{0}=0,%
\end{array}%
\right.  \label{trivial0}
\end{equation}%
with $\bar{s}_{y}^{0}+\bar{s}_{0}^{0}=1$, and the reduced reproduction
number $R_{r}$ becomes $R_{0}$, with%
\begin{equation}
R_{0}=R_{0y}\bar{s}_{y}^{0}+R_{0o}\bar{s}_{o}^{0},  \label{Rr0}
\end{equation}%
where $R_{0y}$ and $R_{0o}$ are given by equation (\ref{Rr1}).

The basic\ partial reproduction number $R_{0y}^{1}\bar{s}_{y}^{0}$ (or $%
R_{0y}^{2}\bar{s}_{o}^{0}$) is the secondary cases produced by one case of
asymptomatic individual (or pre-diseased individual) in a completely
susceptible young persons without control; and the partial basic
reproduction number $R_{0o}^{1}\bar{s}_{o}^{0}$ (or $R_{0o}^{2}\bar{s}%
_{o}^{0}$) is the secondary cases produced by one case of asymptomatic
individual (or pre-diseased individual) in a completely susceptible elder
persons without control. If all parameters are equal, and $\psi =1$, then%
\[
R_{0}=\left[ pR_{0}^{1}+\left( 1-p\right) R_{0}^{2}\right] , 
\]%
where $R_{0}^{1}=R_{0y}^{1}+R_{0o}^{1}$ and $R_{0}^{2}=R_{0y}^{2}+R_{0o}^{2}$
are the basic partial reproduction numbers due to asymptomatic and
pre-diseased persons.

The global stability follows method proposed in \cite{shuai}. Let the vector
of variables be $x=\left( e_{y},a_{y},d_{1y},e_{o},a_{o},d_{1o}\right) $,
vectors $f$ and $v$, by equations (\ref{fv1}) and (\ref{fv2}), and matrices $%
F$ and $V$ evaluated from $f$ and $v$ at trivial equilibrium $P^{0}$
(omitted here). Vector $g$, constructed as%
\[
g^{T}=\left( F-V\right) x^{T}-f^{T}-v^{T}, 
\]%
results in%
\[
g^{T}=\left( 
\begin{array}{c}
\lambda \left( s_{y}^{0}-s_{y}\right) -e_{y}\left( \alpha _{y}d_{2y}+\alpha
_{o}d_{2o}\right) \\ 
-a_{y}\left( \alpha _{y}d_{2y}+\alpha _{o}d_{2o}\right) \\ 
-d_{1y}\left( \alpha _{y}d_{2y}+\alpha _{o}d_{2o}\right) \\ 
\lambda \psi \left( s_{o}^{0}-s_{o}\right) -e_{o}\left( \alpha
_{y}d_{2y}+\alpha _{o}d_{2o}\right) \\ 
-a_{o}\left( \alpha _{y}d_{2y}+\alpha _{o}d_{2o}\right) \\ 
-d_{1o}\left( \alpha _{y}d_{2y}+\alpha _{o}d_{2o}\right)%
\end{array}%
\right) , 
\]%
and $g^{T}\geq 0$ if $s_{y}^{0}\geq s_{y}$, $s_{o}^{0}\geq s_{o}$ and $%
\alpha _{y}=\alpha _{o}=0$.

Let $v_{l}=\left( z_{1},z_{2},z_{3},z_{4},z_{5},z_{6}\right) $ be the left
eigenvector satisfying $v_{l}V^{-1}F=\rho v_{l}$, where $\rho =\sqrt{R_{r}}$%
, and%
\[
V^{-1}F=\left[ 
\begin{array}{cccccc}
0 & \frac{\beta _{1y}s_{y}^{0}}{\sigma _{y}+\phi } & \frac{\beta
_{2y}s_{y}^{0}}{\sigma _{y}+\phi } & 0 & \frac{\beta _{1o}s_{y}^{0}}{\sigma
_{y}+\phi } & \frac{\beta _{2o}s_{y}^{0}}{\sigma _{y}+\phi } \\ 
\displaystyle\frac{p_{y}\sigma _{y}}{\gamma _{y}+\eta _{y}+\chi _{y}+\phi }
& 0 & 0 & 0 & 0 & 0 \\ 
\frac{\left( 1-p_{y}\right) \sigma _{y}}{\gamma _{1y}+\eta _{1y}+\phi } & 0
& 0 & 0 & 0 & 0 \\ 
0 & \frac{\beta _{1y}\psi s_{o}^{0}}{\sigma _{o}+\phi } & \frac{\beta
_{2y}\psi s_{o}^{0}}{\sigma _{o}+\phi } & 0 & \frac{\beta _{1o}\psi s_{o}^{0}%
}{\sigma _{o}+\phi } & \frac{\beta _{2o}\psi s_{o}^{0}}{\sigma _{o}+\phi }
\\ 
0 & 0 & 0 & \frac{p_{o}\sigma _{o}}{\gamma _{o}+\eta _{o}+\chi _{o}+\phi } & 
0 & 0 \\ 
0 & 0 & 0 & \frac{\left( 1-p_{o}\right) \sigma _{o}}{\gamma _{1o}+\eta
_{1o}+\phi } & 0 & 0%
\end{array}%
\right] . 
\]%
This vector is%
\[
v_{l}=\left( \displaystyle\frac{\sigma _{y}+\phi }{\rho \beta _{2y}s_{y}^{0}}%
R_{ry},\frac{\beta _{1y}}{\beta _{2y}},1,\frac{\sigma _{o}+\phi }{\rho \beta
_{2y}s_{o}^{0}\psi }R_{ro},\frac{\beta _{1o}}{\beta _{2y}},\frac{\beta _{2o}%
}{\beta _{2y}}\right) , 
\]%
and Lyapunov function $L$, constructed as $L=v_{l}V^{-1}x^{T}$, is%
\[
\begin{array}{ccl}
L & = & \displaystyle\frac{z_{1}}{\sigma _{y}+\phi }e_{y}+\frac{z_{2}}{%
\gamma _{y}+\eta _{y}+\chi _{y}+\phi }a_{y}+\frac{1}{\gamma _{1y}+\eta
_{1y}+\phi }d_{1y}+\frac{z_{4}}{\sigma _{o}+\phi }e_{o}+ \\ 
&  & \displaystyle\frac{z_{5}}{\gamma _{o}+\eta _{o}+\chi _{o}+\phi }a_{o}+%
\frac{z_{6}}{\gamma _{1o}+\eta _{1o}+\phi }d_{1o}\geq 0%
\end{array}%
\]%
always, and%
\[
\begin{array}{ccl}
\frac{d}{dt}L & = & -\left( 1-\rho \right) \frac{\sigma _{y}+\phi }{\rho
\beta _{2y}s_{y}^{0}}R_{ry}e_{y}-\left( 1-\rho \right) \frac{\sigma
_{o}+\phi }{\rho \beta _{2y}s_{o}^{0}\psi }R_{ro}e_{o}-\frac{1}{\rho \beta
_{2y}}\lambda \left[ \frac{Rry}{s_{y}^{0}}\left( s_{y}^{0}-\rho s_{y}\right)
+\frac{Rro}{s_{o}^{0}}\left( s_{0}^{0}-\rho s_{0}\right) \right] \\ 
&  & +e_{y}\left( \alpha _{y}d_{2y}+\alpha _{o}d_{2o}\right) +a_{y}\left(
\alpha _{y}d_{2y}+\alpha _{o}d_{2o}\right) +d_{1y}\left( \alpha
_{y}d_{2y}+\alpha _{o}d_{2o}\right) + \\ 
&  & e_{o}\left( \alpha _{y}d_{2y}+\alpha _{o}d_{2o}\right) +a_{o}\left(
\alpha _{y}d_{2y}+\alpha _{o}d_{2o}\right) +d_{1o}\left( \alpha
_{y}d_{2y}+\alpha _{o}d_{2o}\right) \leq 0%
\end{array}%
\]%
only if $\rho <1$, $s_{y}^{0}\geq s_{y}$, $s_{o}^{0}\geq s_{o}$ and $\alpha
_{y}=\alpha _{o}=0$ ($\left( \sigma _{j}+\phi \right) /\left( \sigma
_{j}+\phi s_{y}^{0}\right) >1$).

Hence, the method proposed in \cite{shuai} is valid only for $\alpha
_{y}=\alpha _{o}=0$, in which case $P^{0}$ is globally stable if $%
s_{y}^{0}\geq s_{y}$, $s_{o}^{0}\geq s_{o}$ and $\rho =\sqrt{R_{r}}\leq 1$.

\end{document}